\documentclass{aa}
\usepackage{graphicx}
\usepackage{natbib}
\usepackage{txfonts}
\newcommand{\h}{$^{\rm h}$}
\newcommand{\m}{$^{\rm m}$}

\begin{document}
   \title{Size and properties of the narrow-line region in Seyfert-2 galaxies
from spatially-resolved optical spectroscopy 
\thanks{Based on observations made with ESO Telescopes at
   the Cerro Paranal Observatory under programme ID 72.B-0144 and the La Silla Observatory under programme ID 073.B-0013}}

   \author{Nicola Bennert\inst{1,2}
          \and
           Bruno Jungwiert\inst{2,3,4} \and Stefanie Komossa\inst{5} \and
          Martin Haas\inst{1} \and Rolf Chini\inst{1}
          }

   \offprints{Nicola Bennert}
\institute{Astronomisches Institut Ruhr-Universit\"at Bochum,
              Universit\"atsstrasse 150, D-44780 Bochum, Germany;
\email{haas@astro.rub.de}, \email{chini@astro.rub.de}
\and
Institute of Geophysics and Planetary Physics, University of California, Riverside, CA 95521, \email{nicola.bennert@ucr.edu}
\and
Astronomical Institute, Academy of Sciences of the Czech Republic,
Bo{\v c}n\'\i\ II 1401, 141 31 Prague 4, Czech Republic, 
\email{bruno@ig.cas.cz}
          \and
             CRAL-Observatoire de Lyon, 9 avenue Charles Andr{\'e}, F-69561 
Saint-Genis-Laval cedex, France
\and
   Max-Planck Institut f\"ur extraterrestrische Physik,
              Giessenbachstrasse 1, D-85748 Garching, Germany \email{skomossa@xray.mpe.mpg.de}
}

   \date{Received 30 March 2006/ Accepted 14 June 2006}

% \abstract{}{}{}{}{} 
% 5 {} token are mandatory

   \abstract{While [\ion{O}{iii}] narrow-band imaging is commonly 
used to measure the size
of the narrow-line regions (NLRs) in active galactic nuclei (AGNs), it can be contaminated by emission
from surrounding starbursts. Recently, we have shown 
that long-slit spectroscopy provides a valuable alternative
approach to probe the size in terms of AGN photoionisation. 
Moreover, several parameters of the NLR can be directly accessed.}
{We here apply the same methods developed and described for the Seyfert-2 galaxy NGC\,1386
to study the NLR of five other Seyfert-2 galaxies 
by using high-sensitivity spatially-resolved optical spectroscopy 
obtained at the VLT and the NTT. }
{We probe the AGN-photoionisation of the NLR and thus, its ``real'' size
using diagnostic line-ratio diagrams.
We derive physical properties of the 
NLR such as reddening, ionisation parameter,
electron density, and velocity as a function of distance from the nucleus.
}
{For NGC\,5643, the diagnostic diagrams unveil a similar transition between line ratios falling in the
AGN regime and those typical for \ion{H}{ii} regions as found for NGC\,1386,
thus determining the size of the NLR. For the other four objects, all measured line ratios fall in the AGN regime.
In almost all cases,
both electron density and ionisation parameter decrease with radius.
Deviations from this general behaviour (such as a secondary peak) seen in both the ionisation parameter and
electron density can be interpreted as signs of shocks from the
interaction of a radio jet and the NLR gas.
In several objects, the gaseous velocity distribution is characteristic for
rotational motion in an (inclined) emission-line disk in the centre.
We compare our results to those of NGC\,1386 and show that the latter 
can be considered as prototypical
also for this larger sample. We discuss our findings in detail for each object.}
{}

\keywords{Galaxies: active --
          Galaxies: nuclei --
          Galaxies: Seyfert}
\titlerunning{The NLR in Seyfert-2 galaxies}
\authorrunning{N. Bennert et\,al.}

   \maketitle
%
%________________________________________________________________

\section{Introduction}
The luminous central engine in active galaxies, most likely an accreting
supermassive black hole (BH), ionises the surrounding interstellar medium
leading to the broad-line region (BLR) in the inner part and
the narrow-line region (NLR) further out.
Among the central AGN components, the NLR has the advantage
to be directly accessible via imaging and spatially resolved spectroscopy, at
least for nearby AGNs.

[\ion{O}{iii}]\,$\lambda$5007\AA~(hereafter [\ion{O}{iii}])
narrow-band imaging is commonly used to study the NLRs of active
galaxies. However, this emission can be contaminated by contributions from
star formation, shock-ionised gas or tidal tails, resulting in an apparent
increase of the NLR. In addition, the measured size depends on the depth of the images:
When comparing
ground based [\ion{O}{iii}] images of Seyfert galaxies
from \citet{mul96} with the HST snapshot survey of \citet{sch03},
the latter reveal, on average, six times smaller NLR sizes, probably due to 
the 15 to 20 times lower sensitivity.
These considerations question the definition of the ``NLR size'' from 
[\ion{O}{iii}] imaging alone.

Spatially resolved long-slit spectroscopy
is a valuable alternative approach as it can directly probe
the size in terms of AGN photoionisation and discriminate
the stellar or shock-ionised contribution.
In addition, several physical parameters of the NLR such as reddening, ionisation parameter,
electron density, and velocity can be directly
accessed and analysed as a function of distance from the nucleus.

In \citet{ben05} and \citet{ben06a} (hereafter paper I), we describe methods developed to 
probe the AGN-photoionisation of the NLR and thus, its ``real'' size
as well as to derive physical conditions within the NLR
of the nearby Seyfert-2 galaxy NGC\,1386:
We use the galaxy itself for subtracting the stellar template, applying
reddening corrections to fit the stellar template to the spectra of the
NLR. From spatially resolved spectral diagnostics,
we find a transition between central line ratios falling into the
AGN regime and outer ones in the \ion{H}{ii}-region regime. 
Applying \texttt{CLOUDY} photoionisation models \citep{fer98}, we show that the observed distinction
between \ion{H}{ii}-like and AGN-like ratios in NGC\,1386 represents a true
difference in ionisation source and cannot
be explained by variations of physical parameters such as
ionisation parameter, electron density or metallicity. We 
interpret it as a real border between the NLR,
i.e. the central AGN-photoionised region, and 
surrounding \ion{H}{ii} regions.
We find that both the electron density and the ionisation parameter decrease with
radius. The differences between the reddening distributions 
determined from the continuum slope and the Balmer
decrement argue in favour of dust intrinsic to the NLR clouds with 
varying column density along the
line of sight. The NLR and stellar velocity fields are similar and indicate that
the NLR gas is distributed in a disk rather than a sphere.

Here, we apply the same methods
to a larger sample of five Seyfert--2 galaxies 
to probe the size of the NLR. 
We derive physical properties such as reddening, ionisation parameter,
electron density, and velocity, and discuss their
variations with distance from the nucleus. 
In our discussion, we include the results for NGC\,1386 from
paper I.
A detailed comparison
of our results with literature data is given for each object [see also \citet{ben05}].

A similar study was carried out for six Seyfert-1 galaxies. The results and the comparison
with the Seyfert-2 galaxies presented here will be summarised in \citet{ben06c}.

\section{Observations}
The spectra were taken in the wavelength range $\sim$$\lambda$3700-7000\,\AA~to cover
the emission lines suited for specific diagnoses:
H$\alpha$ and H$\beta$
(normaliser, reddening indicator);
[\ion{O}{ii}]\,$\lambda$3727\,\AA/[\ion{O}{iii}]\,$\lambda$5007\,\AA~(sensitive to ionisation parameter); [\ion{O}{iii}]\,$\lambda$$\lambda$4363,5007\,\AA~(temperature sensitive); [\ion{O}{i}]\,$\lambda$6300\,\AA, [\ion{N}{ii}]\,$\lambda$6583\,\AA,
 [\ion{S}{ii}]\,$\lambda$$\lambda$6716,6731\,\AA~(AGN classifier);
 [\ion{S}{ii}]\,$\lambda$6716\,\AA/[\ion{S}{ii}]\,$\lambda$6731\,\AA~(sensitive to electron density).

Relevant information on the sample and observations is summarised in
Tables~\ref{objsy1} and~\ref{obssy1}.

\subsection{VLT/FORS1}
\label{longvlt}
The high signal-to-noise ratio (S/N) long-slit spectra of three Seyfert-2 galaxies
(NGC\,3281, NGC\,5643, and NGC\,1386)
described here were obtained using  
FORS1\footnote{FOcal Reducer/low
  dispersion Spectrograph} attached to the Cassegrain focus of UT1 at
the VLT on the 25$^{\rm th}$ of February 2004.~Observations 
were made in the spectral range 3050-8300\,\AA~through 
the nucleus of each galaxy with exposure times of 1800\,s with a
typical seeing of $\sim$1-2\arcsec. The spatial
resolution element is 0\farcs2 pix$^{-1}$.

\begin{table*}
\begin{minipage}{180mm}
 \caption[]{\label{objsy1} Properties of the sample\footnote{Unless stated otherwise, the properties were taken from
the NASA/IPAC Extragalactic Database (NED).}}
\begin{center}
\begin{tabular}{lcccccc}
\\[-2.3ex]
\hline
\hline\\[-2.3ex]
&  IC\,5063 & NGC\,7212  & ESO\,362-G008 & NGC\,3281 & NGC\,5643 & {NGC\,1386}\\[0.25ex]
\hline\\[-2.3ex]
altern. name & ESO\,187-G023 & MCG\,+02-56-011 & MCG\,-06-12-009& ESO\,375-G055 & ESO\,272-G016 & ESO\,385--G035 \\
$\alpha$ (J2000) & 20\h52\m02\fs3  & 22\h07\m01\fs3 & 
05\h11\m09\fs1& 10\h31\m52\fs1 &
14\h32\m40\fs8& 03\h36\m46\fs2 \\
$\delta$ (J2000) & -57\degr04\arcmin08\arcsec & +10\degr13\arcmin52\arcsec & 
-34\degr23\arcmin36\arcsec & -34\degr51\arcmin13\arcsec &
 -44\degr10\arcmin29\arcsec & --35\degr59\arcmin57\arcsec \\
i. (\degr)\footnote{Host galaxy inclination [\citet{vau91}; RC3]} & 51 & 66 & 70 & 66 & 31 & 77\\
p.a. (\degr)\footnote{Position angle of host galaxy major axis (RC3); for NGC\,5643, it was taken from \citet{mor85} as it is not
given in RC3.} & 116 & 42 & 167 & 140 & 128 & 25\\
$v_{\rm hel}$ (km\,s$^{-1}$)& 3402$\pm$6 & 7984$\pm$21 & 4785$\pm$24& 3200$\pm$22 & 1199$\pm$5 & 868$\pm$5\\
$v_{\rm 3k}$ (km\,s$^{-1}$)\footnote{Velocity relative to the 3K background using the NED velocity calculator} & 3276 & 7638 & 4809& 3523 & 1400& 774 \\
dist. (Mpc)\footnote{Distance $D$ in Mpc, using $v_{\rm 3K}$ and $H_0$ = 71\,km\,s$^{-1}$\,Mpc$^{-1}$} & 47 & 110 & 67 & 50 & 20& 11\\
lin. scale (pc/\arcsec)\footnote{Linear scale $d$ using distance $D$ and $d$ = 4.848 $\cdot$ 10$^{-6}$
$\cdot$ $D$}& 220 & 504 & 321 &  236 & 95 & 52\\
morphology & SA(s)0+: & Sab & Sa & SAB(rs+)a & SAB(rs)c & SB(s)0+ \\
AGN Type & Sy2 & Sy2 & Sy2 & Sy2 & Sy2& Sy2\\
$E(B-V)_G$ (mag)\footnote{Foreground Milky Way reddening used for reddening correction \citep{sch98}} &  0.061 & 0.072 &  0.032 & 0.096 & 0.169& 0.012 \\
$M_B$ (mag) &  12.89 & 14.8  & 13.6& 12.7 & 10.74& 12.09 \\[0.1ex]
\hline\\[-2.3ex]
\end{tabular}
\end{center}
\end{minipage}
\end{table*}

The slit width was chosen according
to the seeing corresponding to 0\farcs7-1\farcs3 on the sky. These slit
widths project to a spectral resolution of $\sim$8-14\,\AA~($\sim$450-770
km\,s$^{-1}$) as is confirmed by the full-width-at-half-maximum (FWHM)
of wavelength calibration lines as well as
of the [\ion{O}{i}]\,$\lambda$5577\,\AA~night-sky line.
The length of the slit used corresponds to 6\farcm8 on the sky. The
slit was orientated along the position angle (p.a.)~of the 
maximum extent of the high excitation
gas observed in narrow-band images centred on [\ion{O}{iii}]
from \citet{sch03} in the cases of NGC\,3281 and NGC\,1386.
For NGC\,5643, the p.a.~from the HST [\ion{O}{iii}] image of \citet{sim97} was
planned to be used. However, by mistake, the observed p.a.~differ by 24\degr~to
the p.a.~of the major [\ion{O}{iii}] 
extension (66\degr~instead of 90\degr). Fortunately,
this galaxy was observed in a different
observing run at the NTT 
at a p.a.~of 90\degr~by Christian Leipski,
who kindly provided us
with these data (see next Section for details).
We nevertheless used the high S/N VLT data to derive the stellar template of
the galaxy itself and apply it to the NTT data.

\begin{figure*}
\begin{center}
\includegraphics[width=5.2cm,angle=0]{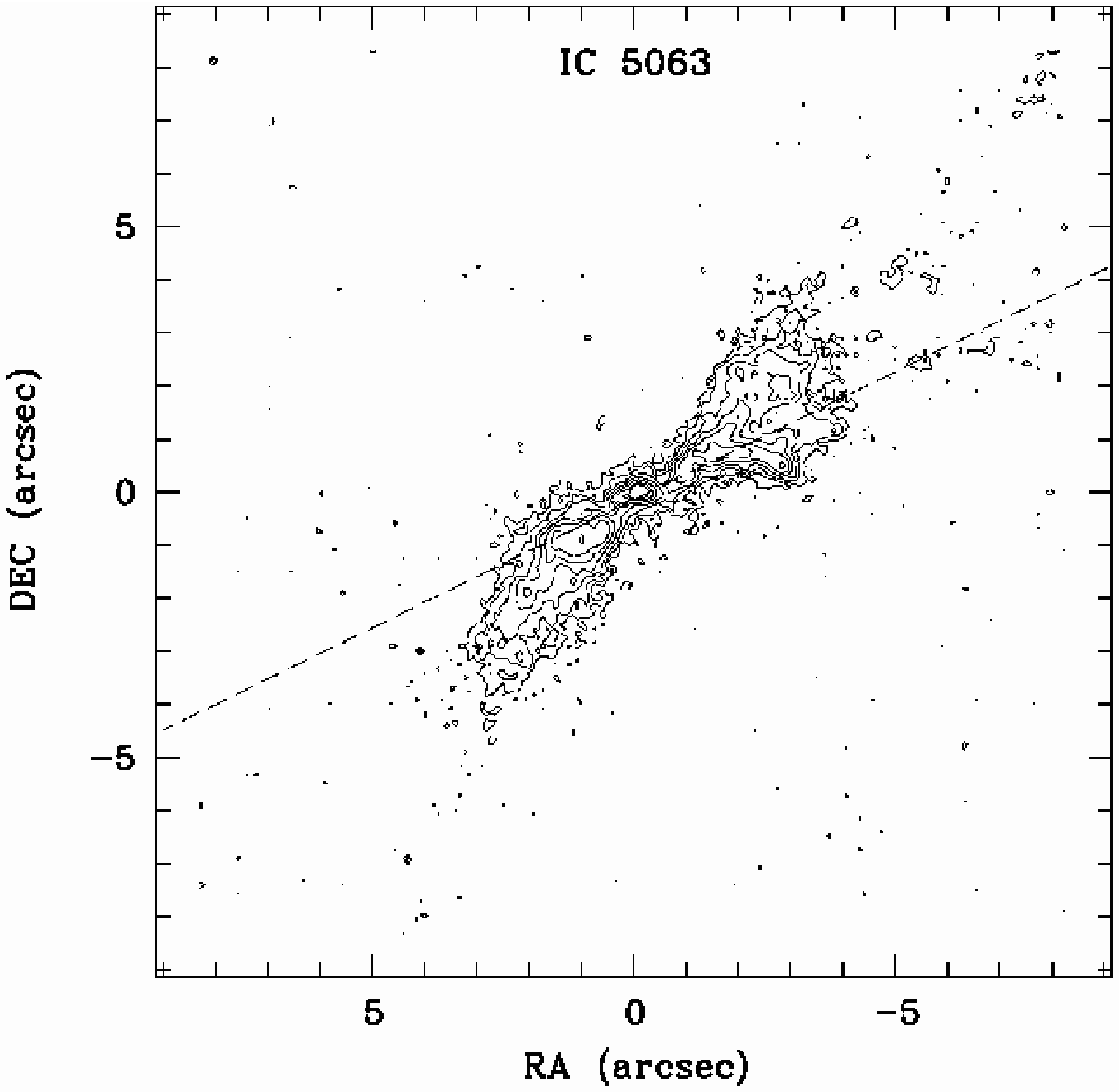}
\includegraphics[width=5.2cm,angle=0]{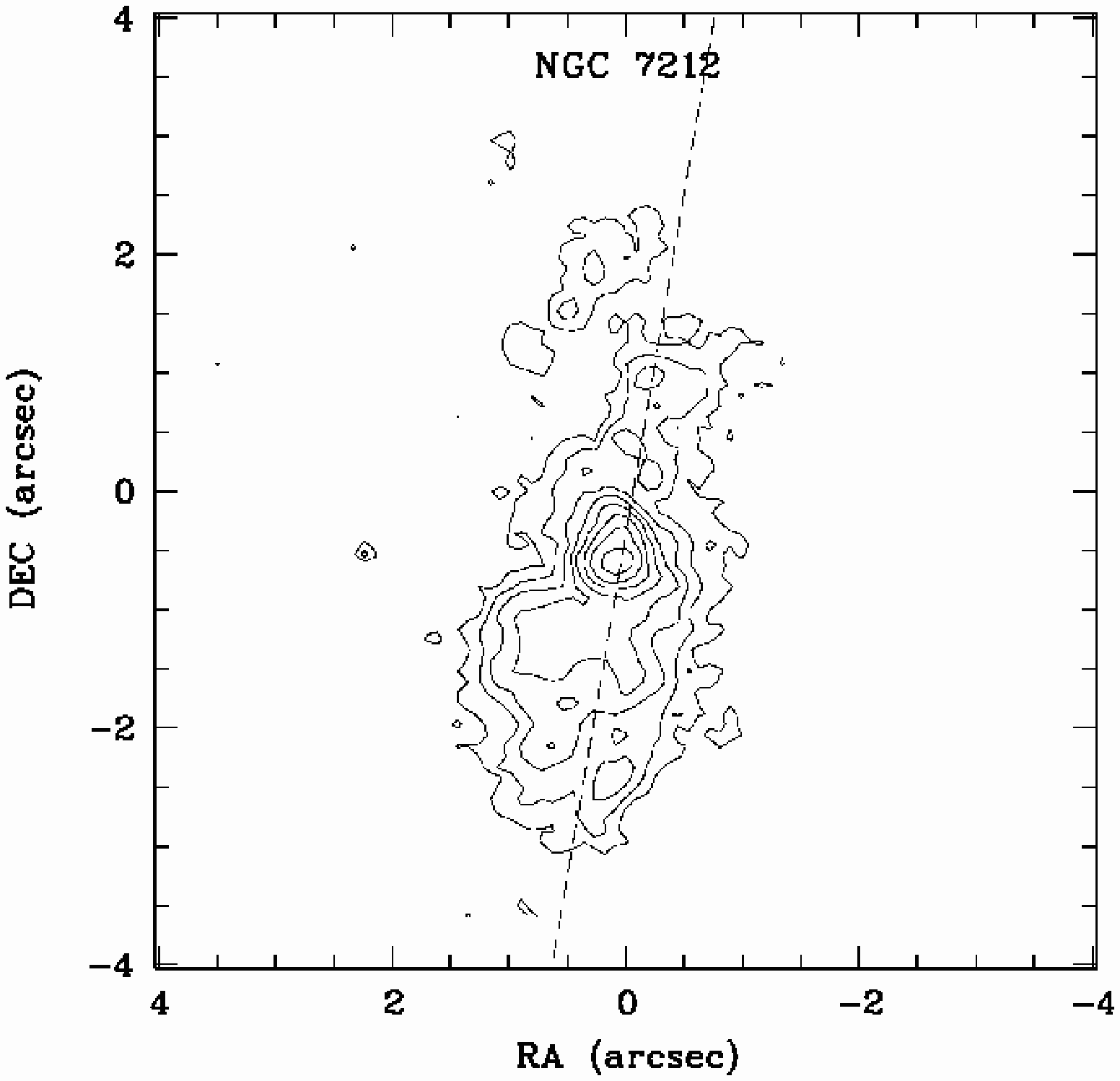}
\includegraphics[width=5.2cm,angle=0]{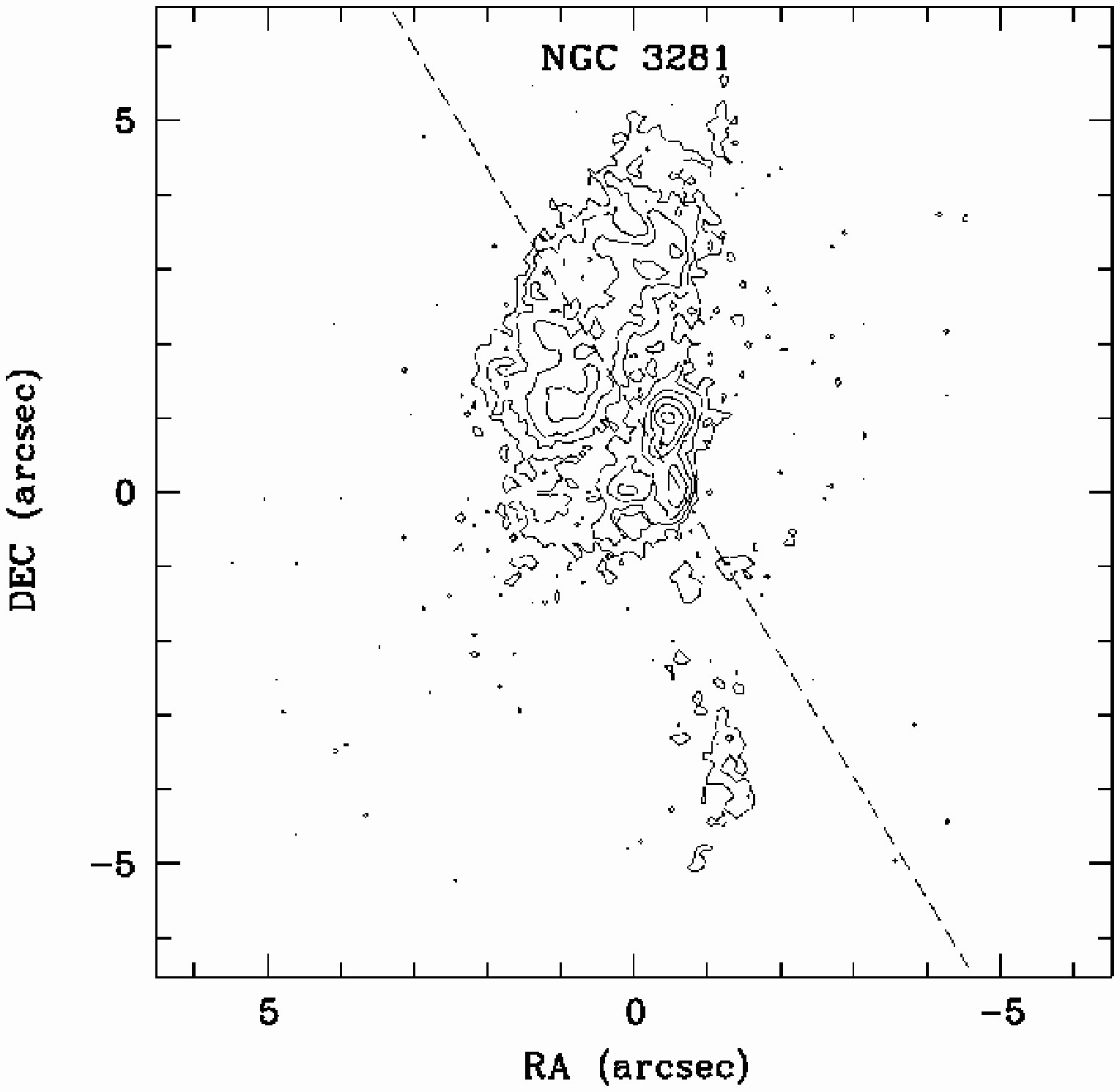}\\*[0.5cm]
\includegraphics[width=5.2cm,angle=0]{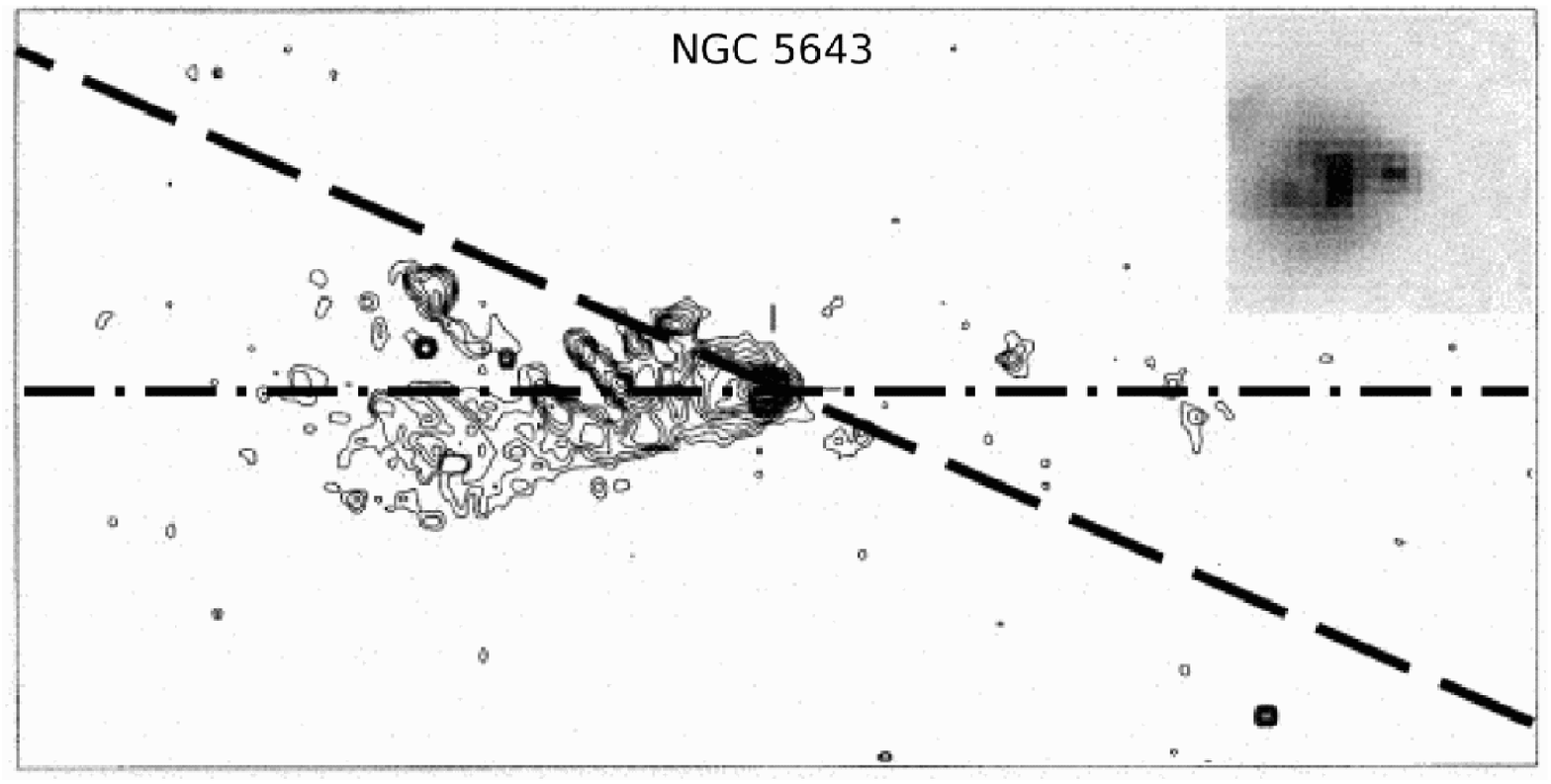}
\includegraphics[width=5.2cm,angle=0]{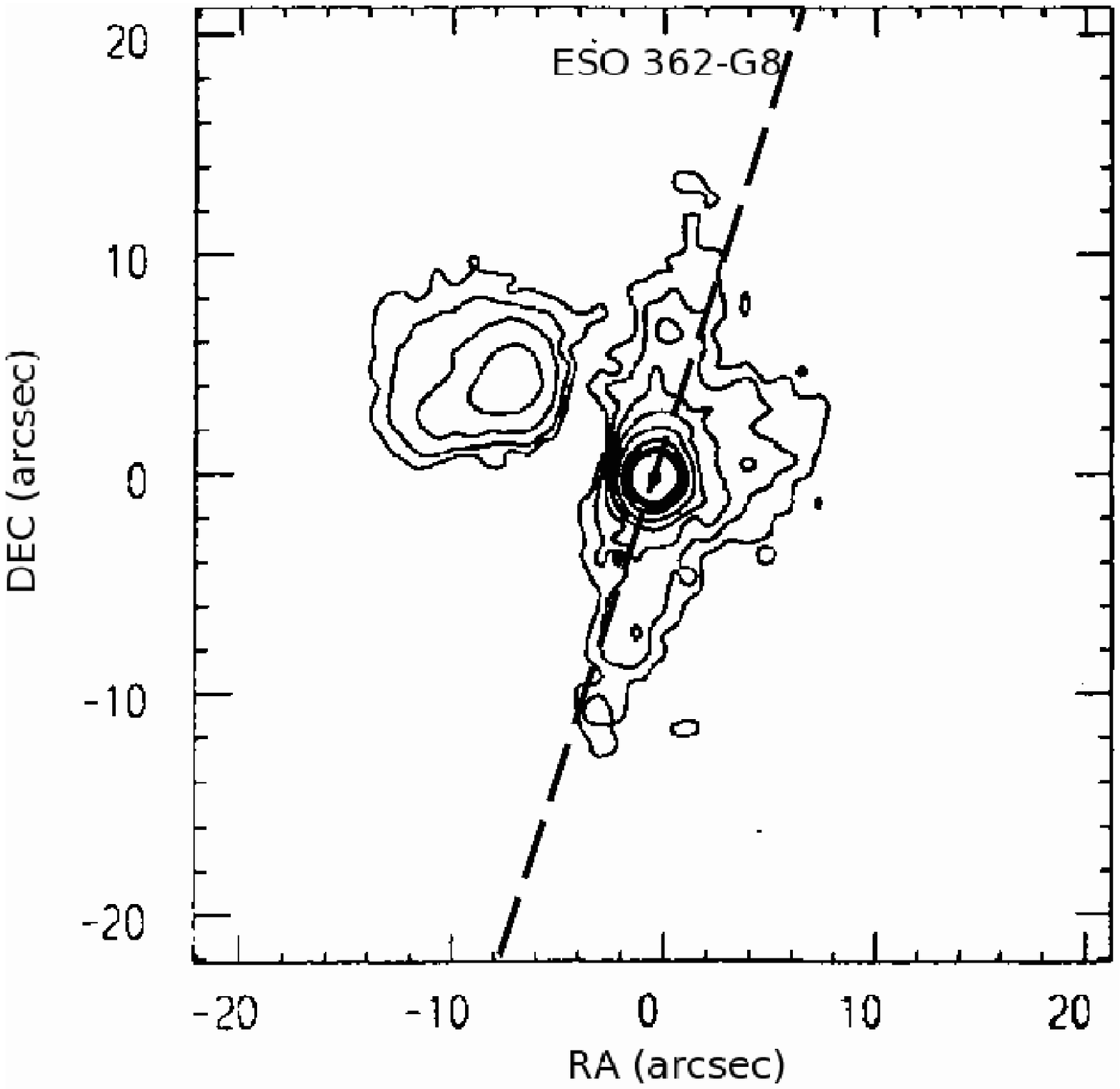}
\end{center}
\caption[]{\label{galaxies2} \small 
HST [\ion{O}{iii}] images of the Seyfert-2 galaxies
IC\,5063, NGC\,7212, and
NGC\,3281 (WF chip: $\sim$0\farcs1\,pix$^{-1}$) taken from \citet{sch03}.
Contours start at the 3$\sigma$ level above the background 
(\citet{sch03}, their Table 2) and increase in
powers of 2 times 3$\sigma$ (3$\sigma$ $\times$ $2^n$).
The HST [\ion{O}{iii}] image of NGC\,5643  was
taken from \citet{sim97} and has the dimensions 30\arcsec~$\times$ 15\arcsec. The
inset shows the central 1\arcsec~$\times$
1\arcsec. The direction
of the long slit used in our VLT observations is presented by the long-dashed
black line while the p.a.~of the long slit used in the NTT observations by
Christian Leipski is indicated by the dashed-dotted black line. 
The groundbased [\ion{O}{iii}] image of ESO\,362-G008 was taken from \citet{mul96}.
The position of the long slit is shown as dashed line.
North is up, east to the left.
}
\end{figure*}

As UT1 (FORS1) is equipped with an atmospheric dispersion corrector (ADC),
there was no need to consider the effects of atmospheric diffraction.
The [\ion{O}{iii}] images with the slit position overlaid are shown in
Fig.~\ref{galaxies2}.

\begin{table*}
\begin{minipage}{180mm}
 \caption[]{\label{obssy1} Observations of the sample}
\begin{center}
\begin{tabular}{lcccccc}
\\[-2.3ex]
\hline
\hline\\[-2.3ex]
&  IC\,5063 & NGC\,7212 & ESO\,362-G008 & NGC\,5643 & NGC\,3281 & NGC\,1386\\
\hline\\[-2.3ex]
telescope & NTT & NTT & NTT & NTT/VLT & VLT &VLT\\
date (beg.) &  15/16-Sep-04 & 17-Sep-04 &  14/15/16-Sep-04 & 21-Apr-04/25-Feb-04 & 25-Feb-04 & 25-Feb-04 \\[0.25ex]
exp. time blue (s)\footnote{Total integration time. At the VLT, 
the blue and red spectral range were covered in one exposure.} & 6000 & 2732 & 5400 & 3600/1800 & 1800 & 1800 \\
exp. time  red (s)$^a$ & 3600 & 1800 & 5400 & 3600/1800 & 1800 & 1800 \\
seeing & $<$ 1\arcsec & $<$ 1\arcsec & $<$ 1\arcsec & 1\arcsec/$\sim$2\arcsec&  $\sim$1\farcs5& $\sim$1\arcsec \\
slit width & 1\arcsec & 1\arcsec & 1\arcsec & 0\farcs7/1\farcs3 & 0\farcs7 & 0\farcs7 \\
FWHM$_{\rm instr}$ (km\,s$^{-1}$) & 250 & 250 & 250 & 90/770 &  460& 460  \\
p.a.~(deg)\footnote{Position angle of the slit} & 115 & 170 &  163 & 90/66 &  31 & 5 \\
hel. corr. (km\,s$^{-1}$)\footnote{This heliocentric correction was added to the measured radial
velocities.} & 0 & -12 & +29 & -9/+11 & -4  & --29\\
average (pixel)\footnote{Number of pixel rows which were averaged} & 3 & 3 & 3 & 3/9 & 7 & 5 \\
scale\footnote{Formal spatial resolution of final extracted spectra} & 1\farcs1 $\times$ 1\arcsec & 1\farcs1 $\times$ 1\arcsec & 1\farcs1 $\times$ 1\arcsec & 1\farcs1 $\times$ 1\arcsec/1\farcs8~$\times$ 1\farcs3& 1\farcs4~$\times$ 0\farcs7 & 1\arcsec~$\times$ 0\farcs7 \\[0.1ex]
\hline\\[-2.3ex]
\end{tabular}
\end{center}
\end{minipage}
\end{table*}

\subsection{NTT/EMMI}
\label{longntt}
High S/N long-slit spectra of another
three Seyfert-2 galaxies 
(IC\,5063, NGC\,7212, and ESO\,362-G008)
were obtained using EMMI\footnote{ESO Multi-Mode Instrument}
attached to the Nasmyth B focus of the NTT
from the 14$^{\rm th}$ to the 17$^{\rm th}$ of September 2004.
Medium dispersion spectroscopy was performed in the observing
modes
REMD\footnote{REd Medium Dispersion spectroscopy} and BLMD\footnote{BLue Medium
Dispersion spectroscopy} in the red and blue wavelength range, respectively.
The spatial
resolution element is 0\farcs37 pix$^{-1}$
in the blue and 0\farcs33 pix$^{-1}$ in the red.
Observations were made in the spectral range 
3650-5350\,\AA~(blue) and 4540-7060\,\AA~(red)
through 
the nucleus of each galaxy.
All four nights were photometric with a
typical seeing of $\sim$0.5-1\arcsec. 
Thus, for all objects, a slit width of 1\arcsec~was chosen.
It projects to a spectral resolution of  
$\sim$4\,\AA~($\sim$250\,km\,s$^{-1}$ 
at [\ion{O}{iii}]) as is confirmed by the FWHM
of wavelength calibration lines as well as
of the [\ion{O}{i}]\,$\lambda$5577\,\AA~night-sky line. The
slit (5\farcm5) was orientated along the p.a.~of the maximum extent of the  [\ion{O}{iii}] emission taken
from \citet{mul96} for ESO\,362-G008, and from
\citet{sch03} for IC\,5063 and NGC\,7212 (Fig.~\ref{galaxies2}).

Additionally, NGC\,5643 was observed with EMMI at the NTT in REMD mode (spatial resolution
  $\sim$0\farcs33 pix$^{-1}$) on April 21$^{\rm
  st}$ 2004 by Christian Leipski. 
The seeing was $\sim$1\arcsec~and the slit corresponds to
0\farcs7. NGC\,5643
  was observed at a low airmass ($<$ 1.1) with a total integration time of
  3600\,s
 in both the blue (4650-5450\,\AA) and the red wavelength
  range (6500-7250\,\AA). While the spectral resolution is high
  ($\sim$1.5\,\AA~$\simeq$ 90\,km\,s$^{-1}$ at [\ion{O}{iii}]), the spectral
  range does not cover the [\ion{O}{ii}]\,$\lambda$3727\,\AA~and 
the [\ion{O}{i}]\,$\lambda$6300\,\AA~lines. Therefore,
  we can neither derive the ionisation parameter from the ratio
  [\ion{O}{ii}]/[\ion{O}{iii}] nor the second diagnostic diagrams which
  involves the [\ion{O}{i}] line.

At the NTT, no ADC is available. As we aimed to observe along the major extension of 
the [\ion{O}{iii}] emission, we could not observe along the parallactic
angle. Thus, several
precautions were taken: (i) A blue filter was used to centre
the objects for observations in the blue wavelength range,
and vice versa for the red.
(ii) All objects were observed
at airmasses smaller than 1.3 so that the atmospheric differential refraction is 
smaller than the slit width in both the blue
and red wavelength range \citep{fil82}.

\section{Reduction and Analysis}
\subsection{Data reduction}
\label{data}
Standard reduction including bias subtraction, flat-field correction,
and cosmic-ray rejection
was performed using the ESO
\texttt{MIDAS}\footnote{Munich Image Data Analysis System,
trade mark of ESO} software (version Nov. 99).
Night-sky spectra at 1\arcmin-3\arcmin~distance 
on both sides of any notable galaxy emission
were interpolated in the region of the galactic spectrum and subtracted
in each case.  

Wavelength calibration was achieved by rebinning the
spectra to a scale of 2.65\,\AA\,pix$^{-1}$ for
the VLT spectra. For the NTT data, a scale of 1.84\,\AA\,pix$^{-1}$ for the blue
and 1.58\,\AA\,pix$^{-1}$ for the red wavelength range was obtained.
For NGC\,5643, the spectra taken by Christian Leipski have a
higher resolution, corresponding to a scale of 0.4\,\AA\,pix$^{-1}$.
The curve of \citet{tug97} was used to correct for atmospheric extinction.
The spectra were flux calibrated using the standard star CD-32\degr9927
in case of the VLT data and LTT\,7379 for the NTT data (and 
LTT\,3684 for NGC\,5643, respectively).
Foreground Milky Way reddening was corrected using values from
\citet{sch98}, as listed in NED, and the extinction law from \citet{sav79}.
Forbidden-line wavelengths were taken from \citet{bow60}. Heliocentric
corrections as given in Table~\ref{obssy1}
were added to the observed velocities. 

\subsection{Extraction of spatially resolved spectra}
\label{longextra}
In the spatial direction perpendicular to the dispersion axis,
several pixel rows were extracted (see below).
Each pixel row corresponds to 1-2\arcsec~along the slit direction.
In the case of the EMMI spectra, the spectra taken with the red chip and its
slightly higher spatial resolution (0\farcs33 pix$^{-1}$
compared to 0\farcs37 pix$^{-1}$
in the blue) were rebinned to the resolution of the blue chip.

We choose the spectrum with the maximum intensity of the continuum
as ``photometric centre'' (``zero'' on the spatial scale).
It coincides with the highest emission-line fluxes in H$\alpha$ and [\ion{O}{iii}].
In the following, we also refer to it as 
``central spectrum''.
Note that this optical
centre needs not to coincide with the position of the AGN, 
which may be hidden by dust.
In the case of the FORS1 spectra, between
three and nine pixel rows were averaged according
to the seeing to enhance the S/N without losing any spatial information.
Table~\ref{obssy1} lists detailed information on each galaxy. We averaged 3
pixel rows of all NTT spectra, corresponding to 1\farcs1.
Line intensities and ratios refer to windows
of 1 square arcsecond size.
On average, we could measure [\ion{O}{iii}] emission at a S/N $>$ 3
out to $r \sim$ 13\arcsec~distance from the nucleus, 
ranging from a spatial coverage of 4\arcsec~(ESO\,362-G008) to 20\arcsec~(IC\,5063), 
plus extended \ion{H}{ii} regions in some galaxies
[e.g.~in NGC\,5643 out to 75\arcsec~($\sim$7\,kpc) from the centre]. Line ratios at a S/N $>$ 3
were measured out to an average distance of $r \sim$ 8\arcsec.

The angular distances were transformed to linear distances
at the galaxies (Table~\ref{objsy1}). 
As the linear distances are projected
distances and depend on the (uncertain) distance
to the galaxy,
we instead use in our figures the angular scale on the x-axis and give a scale bar 
as a measure of the corresponding linear scale.

\subsection{Subtracting the stellar population}
\label{stellarpop}
As discussed in paper I, removing the contribution of the
stellar population is one of the first and most critical steps in the
analysis of AGN emission-line spectra. 

Here, we apply the same methods
to subtract the stellar absorption lines by using
a template derived from the galaxy itself.
The template was scaled to the NLR spectra by normalisation in
the red ($\sim$5400-5700\,\AA)
justified by the fact that the slope at $\lambda \ge 5400$\,\AA~does 
not change significantly for different stellar populations \citep{bic86}.
Note that we also chose this range as it does not cover any strong NLR emission lines.
To allow for a possible reddening difference 
of the template and each NLR spectrum
due to different dust amounts in different galactic regions (or dust
intrinsic to the NLR),
we applied a reddening correction to the template
by fitting the continuum slope of the template to the spectra of the NLR
[MIDAS command ``extinct/long''
with extinction-law from \citet{sav79}].
Moreover, to take into account
a mismatch in redshifts between the stellar template obtained
from the outer portions of the galaxy
and the underlying stellar absorption lines in
the inner spectra, we 
corrected the redshift of the stellar template 
to the redshift of the inner spectra as measured by fitting a
Gaussian to the absorption profile of \ion{Ca}{ii} K.

\begin{figure*}  
\begin{center}
\includegraphics[width=6.2cm,angle=-90]{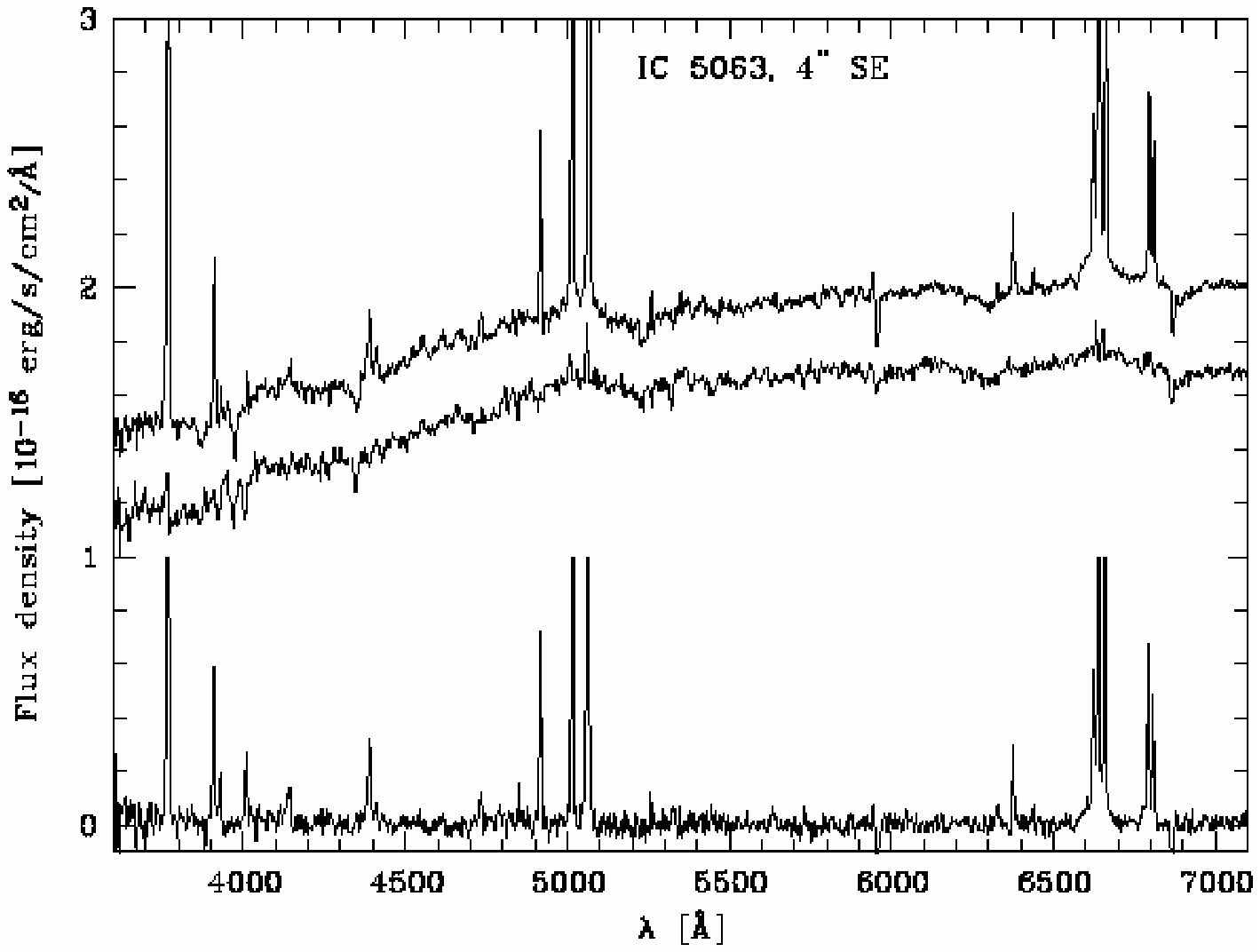}
\includegraphics[width=6.2cm,angle=-90]{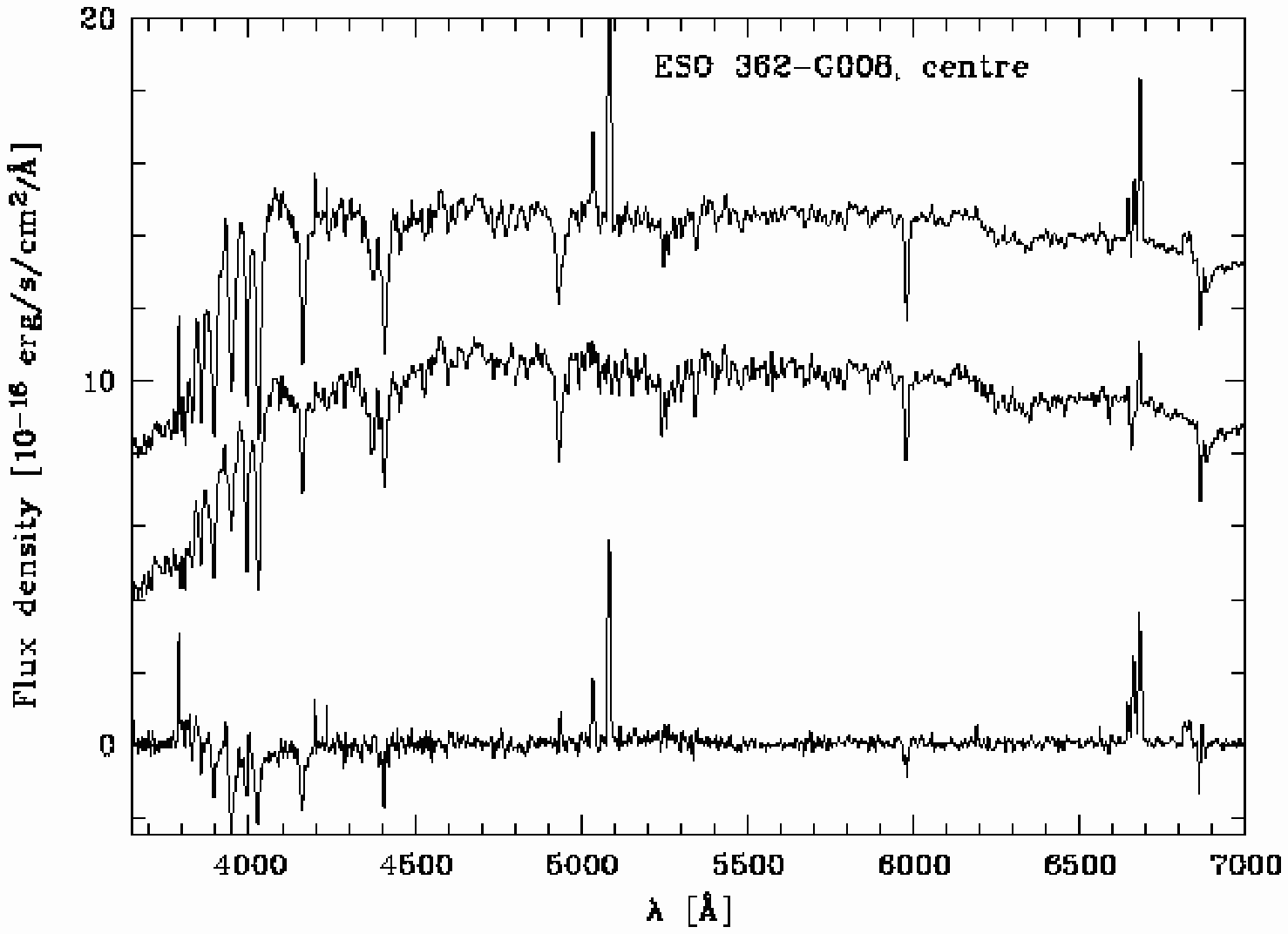}
\includegraphics[width=6.2cm,angle=-90]{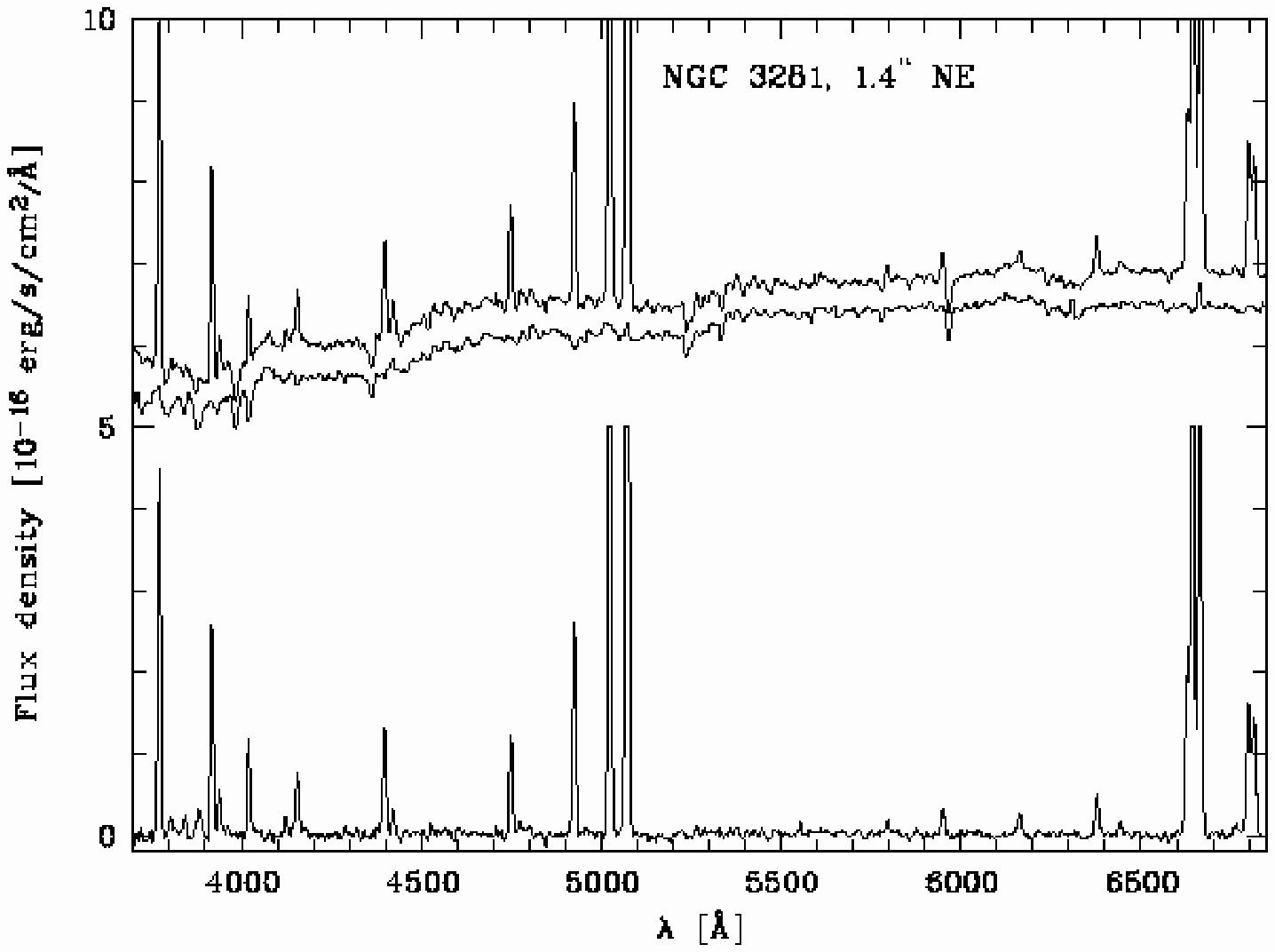}
\includegraphics[width=6.2cm,angle=-90]{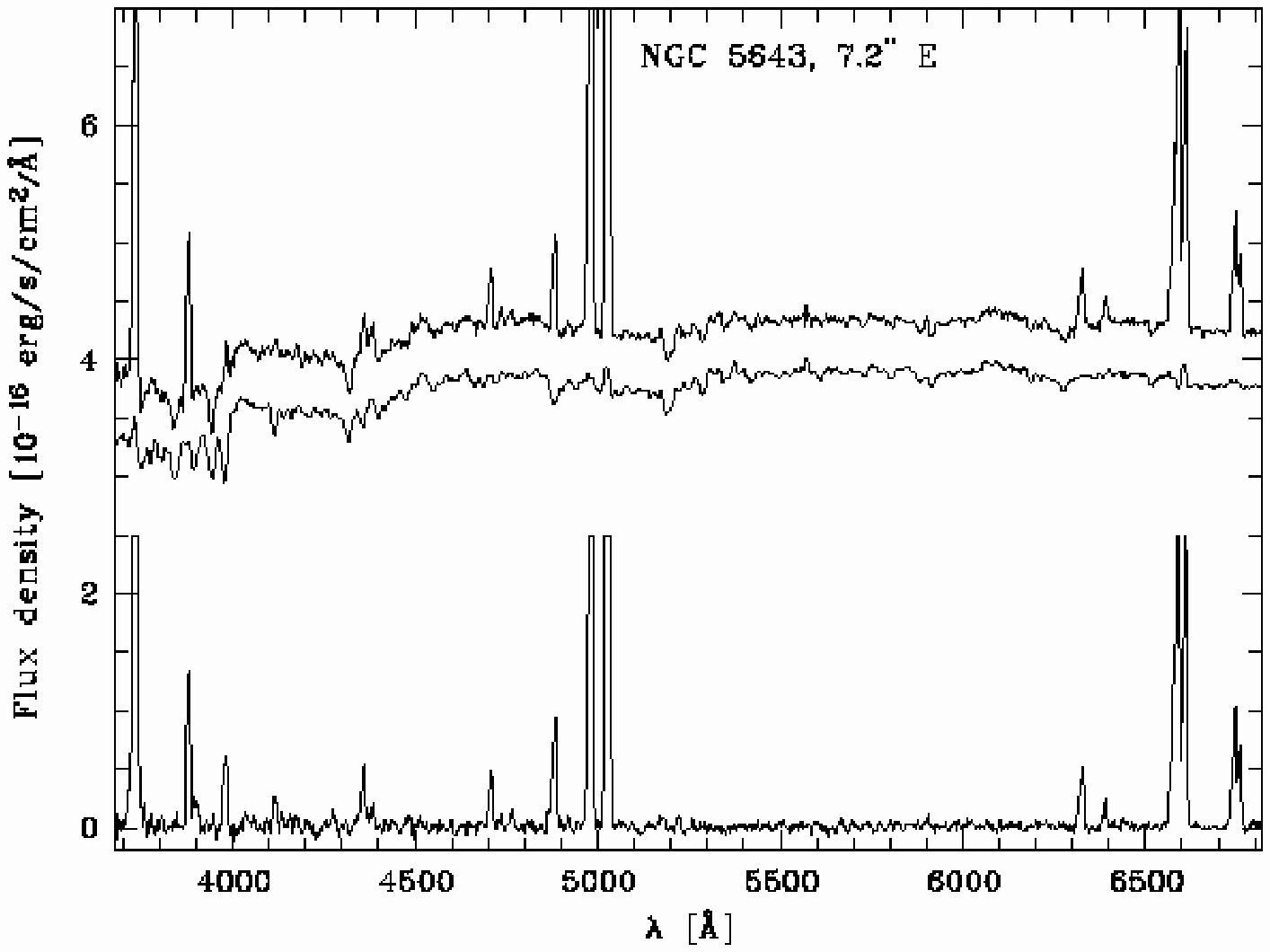}
\includegraphics[width=6.2cm,angle=-90]{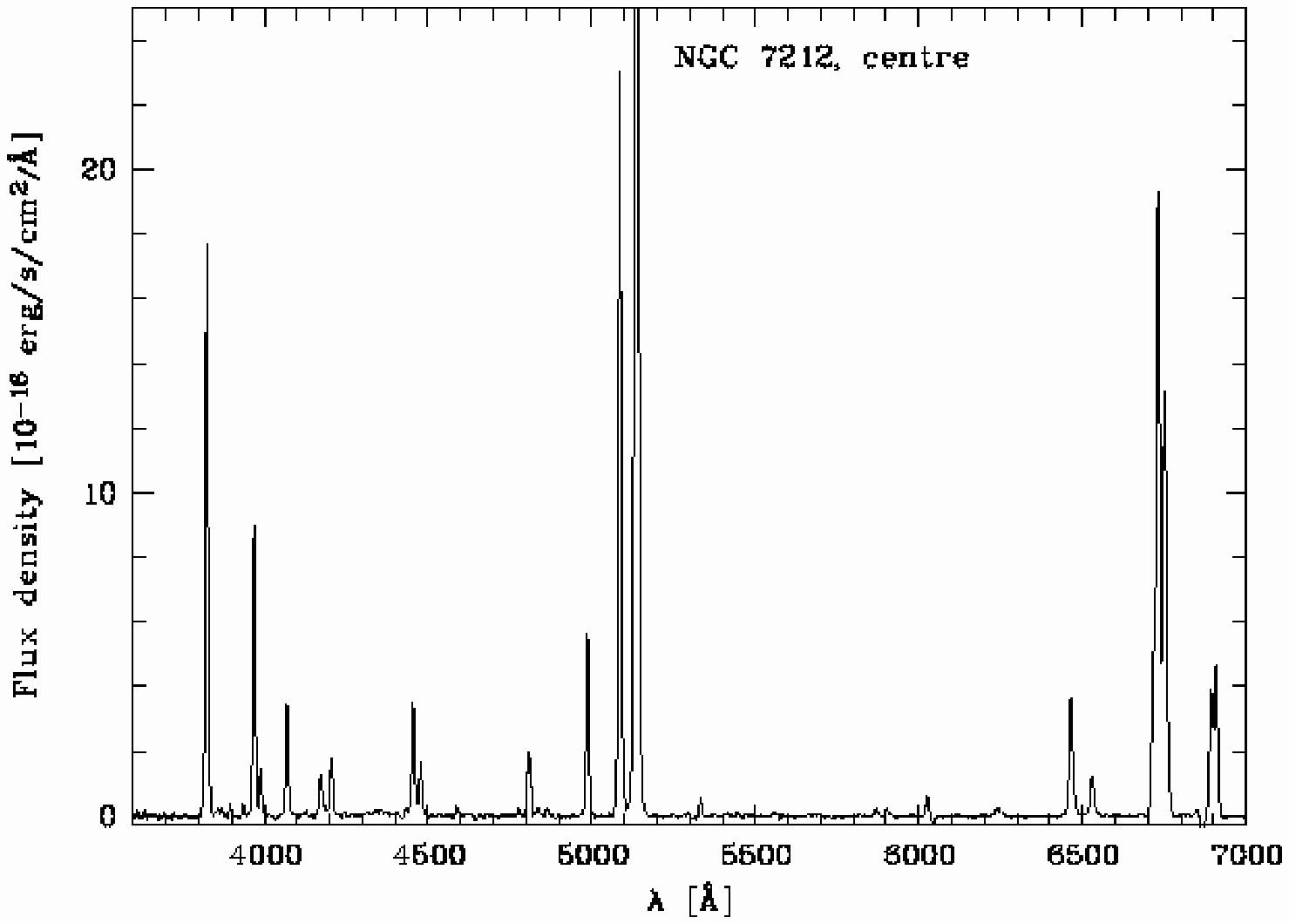}
\caption[Template subtraction]
{\label{figtemplate} \small
 Template subtraction for 
IC\,5063, ESO\,362-G008,
NGC\,3281, and NGC\,5643. 
The observed (upper), the template (middle) and
the template-subtracted spectra (lower spectrum) are shown.
In these plots, both upper spectra are shifted
vertically by an arbitrary amount.
Strong emission lines are truncated in the difference spectrum.
The template matches the stellar absorption lines seen in the NLR spectrum
fairly well, with the exception of ESO\,362-G008 for which strong Balmer
  absorption lines remain. 
For NGC\,7212, no absorption features are distinguishable and thus
no template was subtracted.
}
\end{center}
\end{figure*}

The procedure described above was generally applied (Table~\ref{tabletemplate}; Fig.~\ref{figtemplate}) with the
exception of NGC\,7212 as no stellar absorption lines were seen in the
spectra. For IC\,5063, the \ion{Na}{i} D absorption line was used to
estimate the redshift difference between template and inner spectra as 
this is the only strong absorption line that can be followed throughout the region
of interest.
For NGC\,5643, there is no absorption line within the spectral range
that can be measured at a sufficient S/N. Therefore, 
the [\ion{O}{iii}] emission line was used to estimate the difference in velocities.

To probe the quality of the match between the stellar template and each
observed NLR spectrum, we concentrated on stellar absorption lines
which were not contaminated by emission throughout the NLR. Two strong
lines suited for this purpose are \ion{Ca}{ii} K and \ion{Na}{i} D.
While the residuum of \ion{Ca}{ii} K absorption in the resulting spectrum
after subtraction of the stellar template is within the noise levels,
\ion{Na}{i} D absorption
remains in some objects (e.g.~NGC\,3281 and IC\,5063).
However, at least part of the \ion{Na}{i} D absorption line may originate from
interstellar absorption and can be strong in the regions of high reddening
\citep{bic86}. Thus, we additionally checked that there are no remaining
absorption features of the G band at 4300\AA~and
\ion{Mg}{i}$\lambda$5176\AA. Indeed, any remaining putative absorption
is within the noise limit, confirming our results from the \ion{Ca}{ii} K
line.

The lack of significant other stellar absorption lines after
subtraction of the template confirms that the stellar population does not change
much throughout the NLR in the observed objects. 
One exception is ESO\,362-G008:
Here, the stellar population seems to change
very rapidly in the innermost regions and the equivalent widths (EW)
of the underlying Balmer absorption lines get significantly 
larger towards the centre. It implies the existence of a very young stellar
population close to the AGN. It was not possible to derive a template
spectrum neither from the outer part of the galaxy which fitted the stellar
absorption lines throughout the NLR nor from the inner part due to
the ``contamination'' by emission lines. Close to the centre, the Balmer emission
lines start to fill the underlying absorption trough making it impossible
to correct for the stellar absorption.
We decided to use an outer template as first approach. Our results indicate
that for H$\beta$ and H$\alpha$ the correction was not
sufficient. 

\begin{table} 
\begin{minipage}{80mm}
\caption[]{\label{tabletemplate} Template subtraction details\footnote{The template distance from the 
nucleus is given with the direction in brackets as well as the
size (diameter) of the region that was averaged to obtain the template.
The spectra were median-filtered over three pixels to increase the S/N.}}
\begin{center}
\begin{tabular}{lcc}
\\[-2.3ex]
\hline
\hline\\[-2.3ex]
\multicolumn{1}{c}{Galaxy} & Distance & Size\\[0.25ex]
\hline\\[-2.3ex]
IC\,5063 & 23\arcsec~(NW) & 7\arcsec\\ 
ESO\,362-G008 & 6\arcsec~(SE) &  1\arcsec \\[0.1ex]
NGC\,3281 & 11\arcsec~(SW) &  \hspace*{0.2cm}1\farcs5 \\
NGC\,5643\footnote{Template taken from VLT spectra (p.a.~66\degr) due to higher S/N; no reddening correction applied.} & 26\arcsec~(SW) & 2\arcsec \\
NGC\,1386 & 18\arcsec~(N) & 5\arcsec\\
\hline\\[-2.3ex]
\end{tabular}
\end{center}
\end{minipage}
\end{table}

\subsection{Emission-line fluxes and reddening}
The fluxes of the pure emission-line spectra
were measured as a function
of distance from the nucleus by integrating along a Gaussian
fit to the line profile.
The fit routine ``\texttt{fit/spec}'' \citep{rou92} was used for this purpose.

The uncertainties in deriving the fluxes were mostly caused by  the
placement of the continuum and were thus estimated as the product of the
FWHM of the line and the root-mean square
deviation of the local continuum fluxes. Gaussian error propagation was used
to calculate the errors of subsequent parameters such as line ratios,
ionisation parameter, etc. The resulting errors are in the range of $\sim$1-15\%.
Note that we did not take into account uncertainties from stellar absorption
correction and the quality of the Gaussian fits which was very good given
  the low spectral resolution of our data.

The spectra were dereddened using the recombination
value for the intensity ratio H$\alpha$/H$\beta$ = 2.87
(a typical value for $T_e$ = 10000 K, \citet{ost89}, Table 4.2) and an average
reddening curve (\citet{ost89}, Table 7.2).
Note that in the following, only those spectra are used 
which have emission-line fluxes exceeding the S/N ratio of 3.

For  NGC\,5643 and IC\,5063,
a simple
Gaussian was not sufficient to fit the observed narrow-line profiles.
The profiles revealed substructure such
as blue- or redshifted subpeaks and shoulders. 
This is a well known phenomenon for Seyfert 
galaxies [e.g.~\citet{whi85,vei90}] and has recently also been found for quasars
\citep{lei06a}. It is commonly interpreted as outflow 
and/or interaction with a radio jet. 
The pronounced substructure was limited to the central arcseconds 
(22\arcsec~for NGC\,5643 and 9\arcsec~for IC\,5063) and we used
three Gaussians (a central one as well as a blueshifted and a redshifted one)
to fit the narrow emission lines with the strengths varying spatially to
closely approximate the total line flux.  
For NGC\,5643, we were able to study the profiles and their spatial variation
in greater detail, given the high resolution of these spectra (Appendix~\ref{ngc5643}).

\section{Results and Discussion}
In the following, 
we include the results for the Seyfert 2
NGC\,1386 from paper I in the Tables presented here for comparison.
We do not show the corresponding figures for NGC\,1386 but
refer the reader to paper I.

\subsection{Nuclear spectra}
\begin{table*}
\begin{minipage}{180mm}
 \caption[]{\label{lineratio1} Observed and reddening-corrected 
line intensity ratios relative to H$\beta$\footnote{All narrow emission line ratios were derived from the nuclear
spectra. After reddening correction, other Balmer line-ratios such as H$\gamma$/H$\beta$ and
H$\delta$/H$\beta$ are consistent with the recombination values within the
errors. The uncertainties are in the range of $\sim$1-15\%.}}
\begin{center}
\begin{tabular}{lcccccccccccc}
\\[-2.3ex]
\hline
\hline\\[-2.3ex]
\multicolumn{1}{c}{Line} 
& \multicolumn{2}{c}{\rm IC\,5063} 
& \multicolumn{2}{c}{\rm NGC\,7212} & \multicolumn{2}{c}{\rm ESO\,362-G008}
& \multicolumn{2}{c}{\rm NGC\,3281} & \multicolumn{2}{c}{\rm NGC\,5643}& \multicolumn{2}{c}{\rm NGC\,1386} 
\\
& \multicolumn{1}{c}{$F_{\rm obs}$} & \multicolumn{1}{c}{$F_{\rm dered}$}
& \multicolumn{1}{c}{$F_{\rm obs}$} & \multicolumn{1}{c}{$F_{\rm dered}$}
& \multicolumn{1}{c}{$F_{\rm obs}$} & \multicolumn{1}{c}{$F_{\rm dered}$} 
& \multicolumn{1}{c}{$F_{\rm obs}$} & \multicolumn{1}{c}{$F_{\rm dered}$}
& \multicolumn{1}{c}{$F_{\rm obs}$} & \multicolumn{1}{c}{$F_{\rm dered}$}
& \multicolumn{1}{c}{$F_{\rm obs}$} & \multicolumn{1}{c}{$F_{\rm dered}$}\\[0.25ex]
\hline\\[-2.3ex]
$[\ion{O}{ii}]\,\lambda3727$\,\AA &
1.48 & 2.96 & 2.03 & 2.63 & 1.48 & 2.61& 1.80 & 2.62 & --\footnote{Not covered by wavelength range} & --$^b$ & 1.81 & 2.63 \\*[0.01cm]
$[\ion{Ne}{iii}]\,\lambda3869$\,\AA 
& 0.47 & 0.86 & 0.98 & 1.22 & --\footnote{Underlying absorption lines} & --$^c$& 0.73 & 1.01 & --$^b$ & --$^b$& 0.77 & 1.07 \\*[0.01cm]
$[\ion{Ne}{iii}]\,\lambda3967$\,\AA 
& 0.18 & 0.31 & 0.4 & 0.49& --$^c$ & --$^c$& 0.26 & 0.35 & --$^b$ & --$^b$& 0.48 & 0.64 \\*[0.01cm]
$[\ion{O}{iii}]\,\lambda$4363\,\AA 
& 0.11 & 0.15 & 0.20 & 0.23 & --$^c$ & --$^c$&  0.12 & 0.14 & --$^b$ & --$^b$& 0.19 & 0.23 \\*[0.01cm]
$\ion{He}{ii}\,\lambda$4686\,\AA &
0.14 & 0.15 & 0.23 & 0.24 & --$^c$ & --$^c$& 0.36 & 0.38 & --$^b$ & --$^b$ & 0.46 & 0.48 \\*[0.01cm]
$[\ion{O}{iii}]\,\lambda$5007\,\AA &
8.88 & 8.03 & \hspace*{-0.2cm}12.52 & \hspace*{-0.2cm}12.06 & \hspace*{-0.2cm}11.57 & \hspace*{-0.2cm}10.65&  9.21 & 8.72 & \hspace*{-0.2cm}12.38 & \hspace*{-0.2cm}11.40 & \hspace*{-0.2cm}11.34 & \hspace*{-0.2cm}10.73 \\*[0.01cm]
$[\ion{Fe}{vii}]\,\lambda$5721\,\AA
& -- & -- & -- & -- & -- & -- & 0.08 & 0.06 & --$^b$ & --$^b$& 0.29 & 0.22 \\*[0.01cm]
$[\ion{Fe}{vii}]\,\lambda$6087\,\AA 
& 0.14 & 0.07 & 0.06 & 0.04 & -- & --& 0.17 & 0.12 & --$^b$ & --$^b$& 0.44 & 0.31 \\*[0.01cm]
$[\ion{O}{i}]\,\lambda$6300\,\AA  &
0.71 & 0.32 & 0.76 & 0.56 & 0.44 & 0.23&  0.46 & 0.30 & --$^b$ & --$^b$& 0.46 & 0.30\\*[0.01cm]
$[\ion{Fe}{x}]\,\lambda$6375\,\AA &
0.04 & 0.02 & 0.04 & 0.03 & -- & --& 0.02 & 0.01 & --$^b$ & --$^b$ & 0.07 & 0.05 \\*[0.01cm]
H$\alpha$& 7.10 & 2.87 & 4.04 & 2.87
& 6.06 & 2.87 &   4.71 & 2.87 & 6.03 & 2.87 &  4.70 & 2.87 \\*[0.01cm]
$[\ion{N}{ii}]\,\lambda$6583\,\AA  & 4.27 &
1.71 & 2.82 & 1.99 & 6.78 & 3.18& 4.02 & 2.44 & 6.30 & 2.98 & 5.60 & 3.41 \\*[0.01cm]
$[\ion{S}{ii}]\,\lambda$6716\,\AA  & 1.51 &
0.58 & 0.80 & 0.56 & 1.19 & 0.54& 1.31 & 0.78 & 1.74 & 0.80 & 1.04 & 0.62 \\*[0.01cm]
$[\ion{S}{ii}]\,\lambda$6731\,\AA  & 1.49
& 0.57 & 0.98 & 0.63 & 1.53 & 0.70& 1.24 & 0.74 & 1.94 & 0.89 & 1.29 & 0.77 \\*[0.01cm]
\hline\\[-2.3ex]
\end{tabular}
\end{center}
\end{minipage}
\end{table*}

\begin{table*}
\begin{minipage}{180mm}
\caption[]{\label{result} Reddening-corrected H$\beta$ luminosity and 
results from dereddened line ratios of the nuclear spectra.}
\begin{center}
\begin{tabular}{lcccccc}
\\[-2.3ex]
\hline
\hline\\[-2.3ex]
& \rm{IC\,5063} & \rm{NGC\,7212} & \rm{ESO\,362-G008}&  \rm{NGC\,3281} & \rm{NGC\,5643} &  \rm{NGC\,1386} \\[0.25ex]
\hline\\[-2.3ex]
$F_{\rm H\beta}$ (10$^{-14}$\,erg\,s$^{-1}$\,cm$^{-2}$) & 54$\pm$1 & 12$\pm$0.5 & 11.5$\pm$2 & 36$\pm$1 & 10$\pm$0.5 & 99$\pm$7\\
$L_{\rm H\beta}$ (10$^{39}$\,erg\,s$^{-1}$) & \hspace{-2mm}140$\pm$3     & \hspace{-2mm}169$\pm$5         & 65$\pm$9&  \hspace{-5.5mm} 107$\pm$4  & 4.8$\pm$0.3     & \hspace{-5mm}14.2$\pm$1       \\
$T_{\rm e, obs}$ (K)                                     & 13865$\pm$1800             & 14635$\pm$1500                 & \hspace{+4mm}--\footnote{Underlying absorption lines}&  \hspace{-4mm}13715$\pm$440 & \hspace{+2mm}--\footnote{Not covered by wavelength range} & \hspace{-2mm}15650$\pm$1500    \\
$E_{B - V}$ (mag)\footnote{Note that this central value is not necessarily representative for the reddening within the NLR; for more details on reddening see Table~\ref{reddening}}                           & \hspace{+2mm}0.89$\pm$0.01 & \hspace{+2mm}0.33$\pm$0.01     & \hspace{+2mm}0.73$\pm$0.05&  0.49$\pm$0.01              & 0.73$\pm$0.02 & 0.48$\pm$0.02         \\
$n_{\rm e, obs}$ (cm$^{-3}$)                     & 635$\pm$30                 & \hspace{-0.5mm}1420$\pm$50\footnote{[\ion{S}{ii}]\,$\lambda$6716\,\AA~is slightly truncated by telluric absorption bands.} & \hspace{+2mm}1420 (1710)$\pm$70\footnote{Using $T_e$ = 10000\,K and, in brackets, $<$$T_{e}$$>_{\rm 4 Sy2s}$ $\sim$
14465, respectively}&  \hspace{-2mm}540$\pm$40    & 860 (1035)$\pm$30$^e$& \hspace{-4mm}1545$\pm$50  \\
$U_{\rm log (n_e) = 3, obs}$ (10$^{-3}$)         & \hspace{+2mm}2.15$\pm$0.01 & \hspace{+2mm}3.27$\pm$0.02     & 2.76$\pm$0.1&  \hspace{+2mm}2.3$\pm$0.02  & \hspace{+2mm} --$^b$ & 2.83$\pm$0.01    \\[0.1ex]
\hline\\[-2.3ex]
\end{tabular}
\end{center}
\end{minipage}
\end{table*}

Table~\ref{lineratio1} lists the observed and
reddening-corrected line-intensity ratios relative to H$\beta$ from the
nuclear spectrum (uncorrected for slit losses). 
For pairs of lines 
([\ion{O}{iii}], [\ion{O}{i}], and [\ion{N}{ii}]) with a fixed
line ratio ($\sim$3:1), only the brighter line is shown.
Emission-line ratios of the strongest lines as a function of distance
from the centre can be found online for each individual galaxy (including
NGC\,1386).

In Table~\ref{result}, we give the reddening-corrected H$\beta$ luminosity
and summarise the results from dereddened line ratios such as the electron
temperature $T_{\rm e, obs}$\footnote{Derived from the [\ion{O}{iii}]($\lambda$4959\,\AA+$\lambda$5007\,\AA)/$\lambda$4363\,\AA~emission-line ratio}, 
the reddening value $E_{B - V}$, 
the electron density $n_{\rm e, obs}$, and the ionisation
parameter $U$$_{\rm obs}$ for the nuclear spectra of all objects.
The parameters represent an average over the central several hundred
parsecs.

The temperature was, in most objects, only
determined for the nuclear spectrum due to the faintness of the involved 
[\ion{O}{iii}]\,$\lambda$4363\,\AA~emission line in the outer spectra.
In some objects, we were able to derive the electron temperature in the inner
few arcseconds (NGC\,3281, NGC\,7212, and IC\,5063) where it stays roughly
constant within the errors or scatters without showing a clear dependency on radius. 
The central temperature was used to apply a correction to the electron density.
In those cases in which no temperature was measured we used $T_e = 10000$\,K or
an average temperature derived from the other galaxies instead. 
The other values (reddening, electron density, and ionisation parameter) were
determined throughout the NLR and we discuss each of them in turn.

\subsection{Reddening distribution}
\begin{figure*}
\begin{center}
\includegraphics[width=6.2cm,angle=-90]{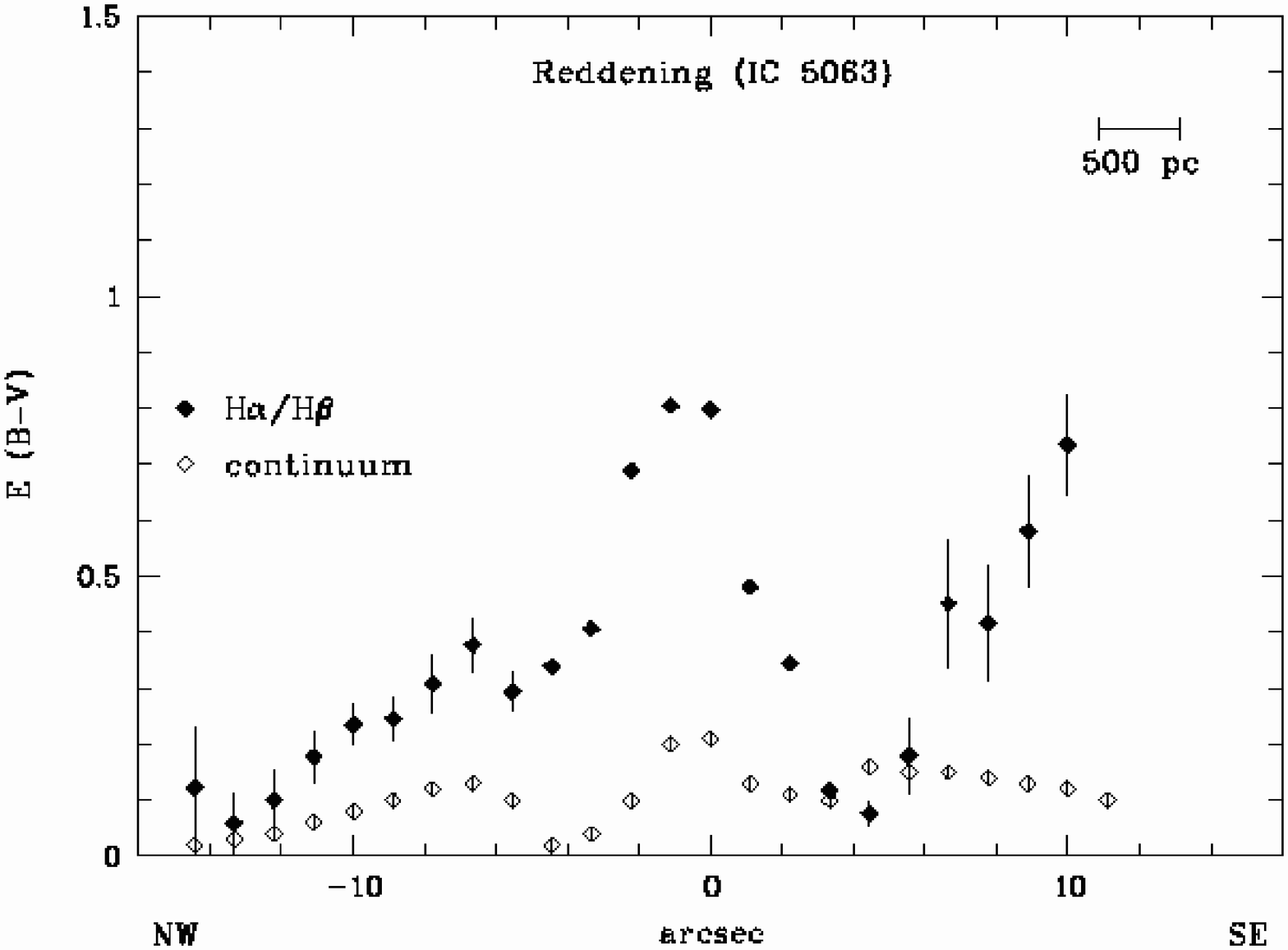}
\includegraphics[width=6.2cm,angle=-90]{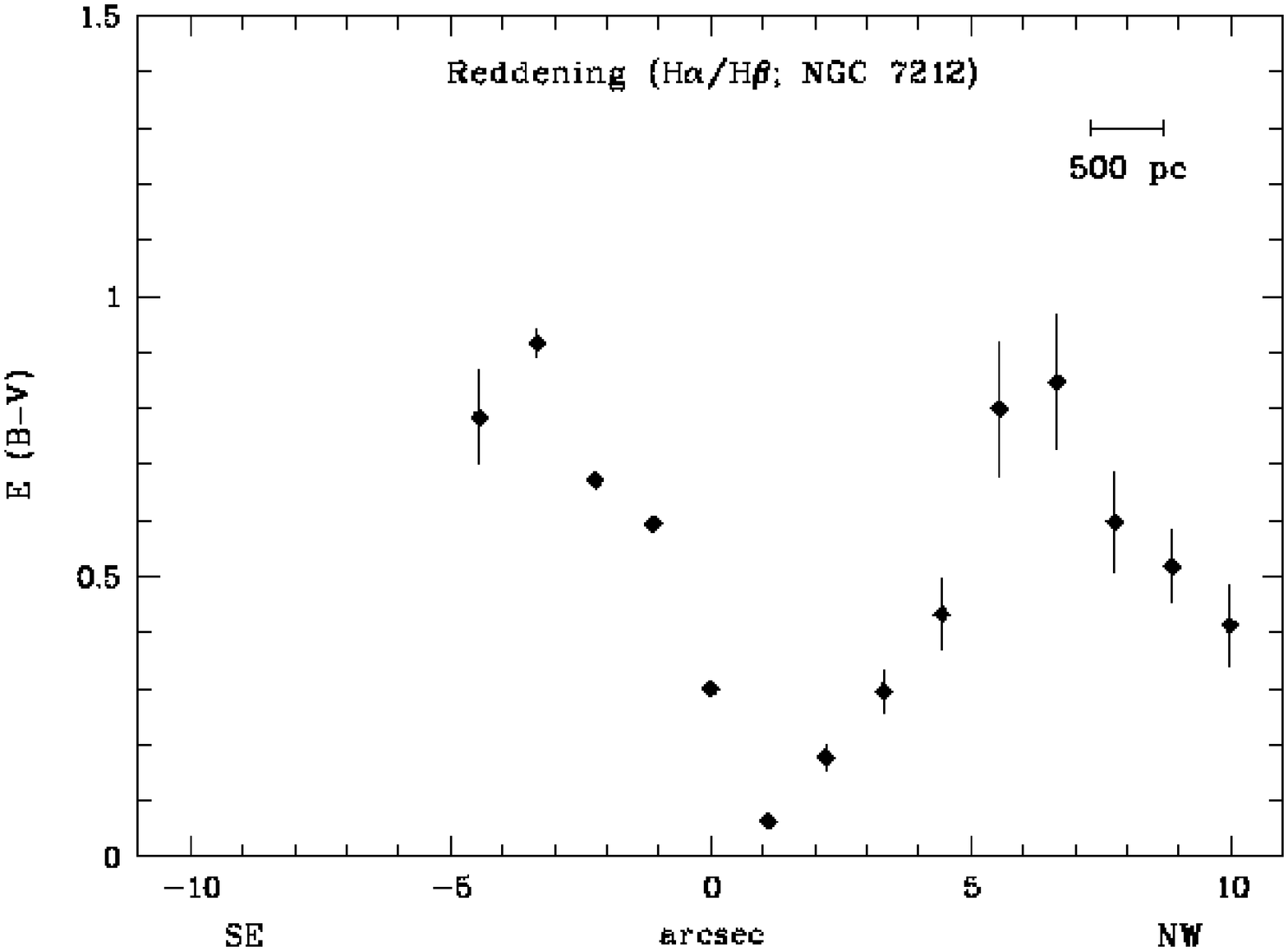}
\includegraphics[width=6.2cm,angle=-90]{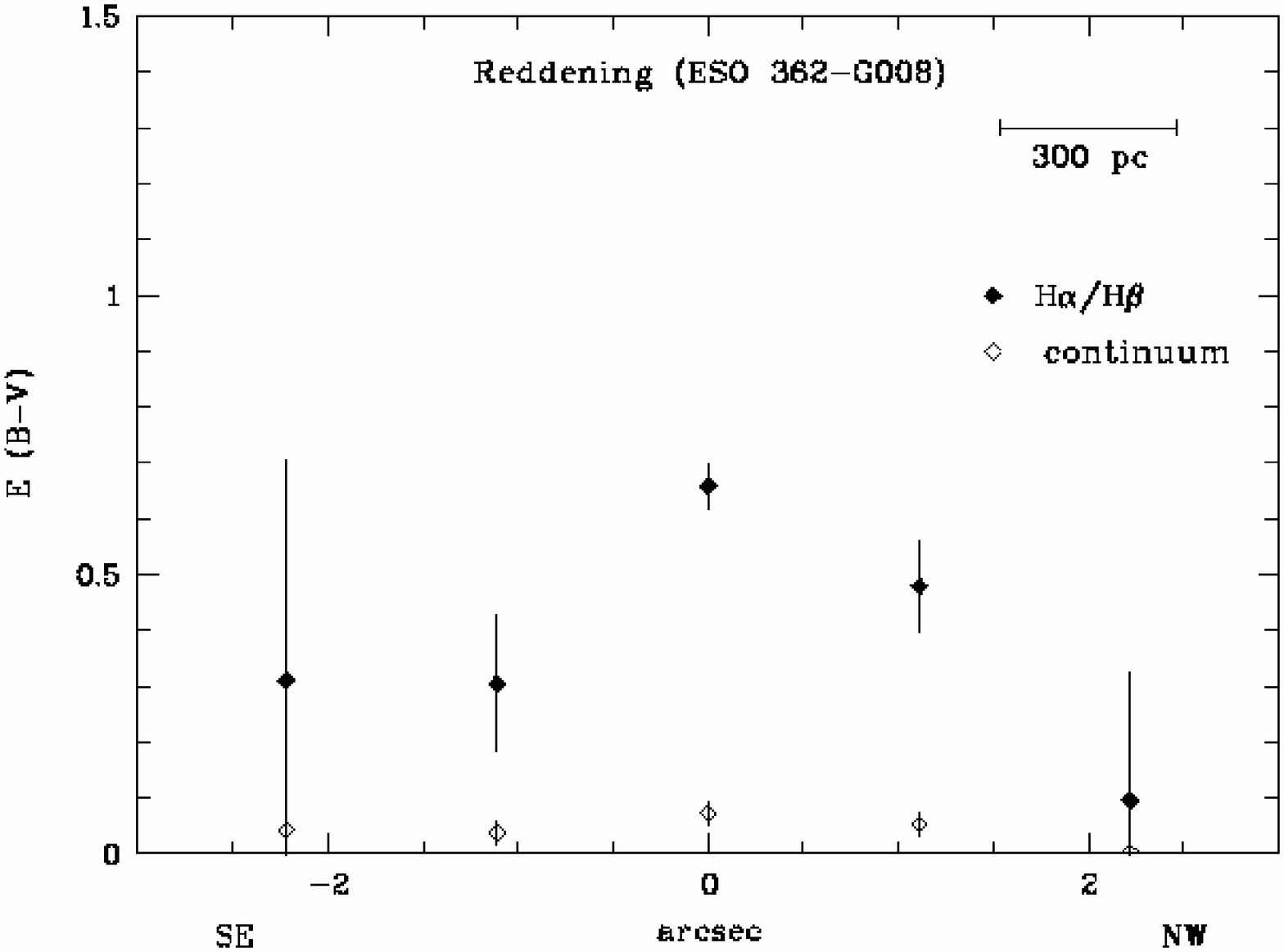}
\includegraphics[width=6.2cm,angle=-90]{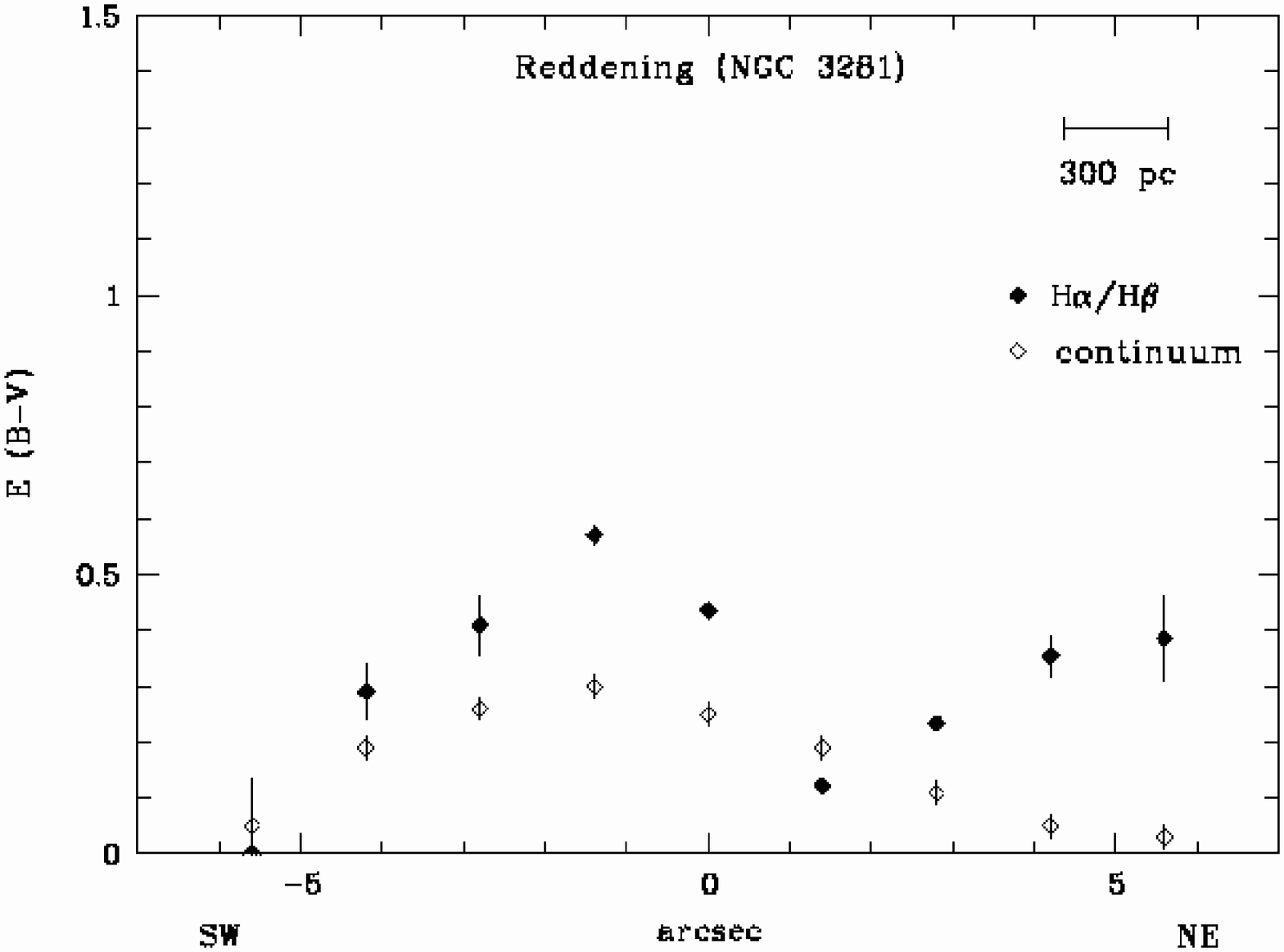}
\includegraphics[width=6.2cm,angle=-90]{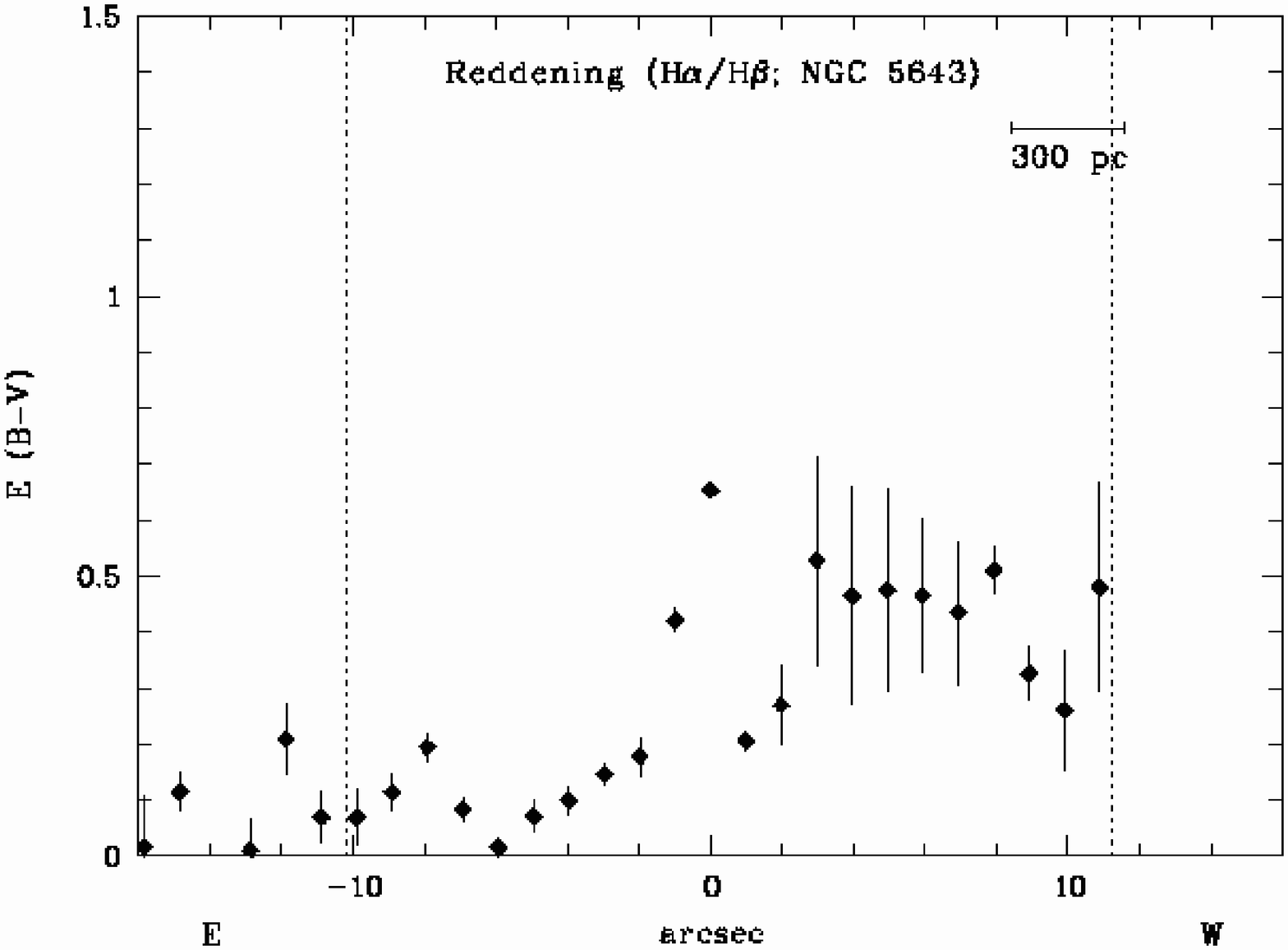}
\caption[]{\label{reddening2} \small
Reddening distributions of the Seyfert-2 galaxies 
IC\,5063, NGC\,7212, ESO\,362-G008, NGC\,3281, 
and NGC\,5643. 
The reddening was derived from the
recombination value of the narrow H$\alpha$/H$\beta$ emission-line ratio (filled symbols).
For some objects, we are able to also show  
the reddening distribution of the continuum of the central spectra relative to
the stellar template as determined during template subtraction (open symbols).
For NGC\,5643, the edge of the NLR as determined from the
  diagnostic diagrams is indicated by dotted lines.}
\end{center}
\end{figure*}

Two different measures of reddening across the galaxy were derived:~(i) 
the reddening of the continuum slope in the
central parts with respect
to the template derived in the outer parts of the galaxy;
(ii) the reddening distribution obtained from the recombination
value for the (narrow-line) intensity ratio H$\alpha$/H$\beta$.
The first reddening value was only derived for those objects for which a
 stellar template correction was applied.
The reddening distributions are shown in
Fig.~\ref{reddening2}. 
While the nuclear reddening is given in Table~\ref{result}, 
we give in Table~\ref{reddening} the highest reddening value within the NLR, 
the distance from the centre at which it occurs as well as the global 
reddening
(as derived by summing the H$\alpha$ and H$\beta$ flux within the NLR).

As the match between the absorption lines of the stellar template and
those seen in the spectra is quite close for all
spectra, we believe that reddening by dust 
is the cause of the spatially varying continuum slope
and not an intrinsically redder stellar
population in the central part.

We cannot compare the absolute values of $E_{(B - V)}$ directly
as the reddening determined from the continuum
slope is a value relative to the template. 
The dispersion $\Delta E_{(B - V)}$ is significantly
smaller than that obtained from the Balmer decrement. This can be due to
extinction by foreground dust in e.g.~the host galaxy which affects both the
template and the central spectra and thus do not reflect in the relative
reddening value. However, often both reddening values are distributed differently
indicating that the stellar population and the NLR are suffering different dust
extinctions. As discussed in paper I, 
the most probable explanation is dust intrinsic to the NLR
clouds with a varying column density along the line-of-sight. 

As in paper I, we use the reddening distribution determined from the 
narrow H$\alpha$/H$\beta$ emission-line ratio to correct for the intrinsic
reddening of the NLR itself 
as these lines originate in the NLR and thus give a better estimate for
the reddening within the NLR than the one determined from the continuum slope.

In some cases (e.g.~NGC\,5643, IC\,5063), the reddening is highest in the
centre and decreases with distance from the nucleus. But more often, the
distributions show deviations from such a general trend. 

For ESO\,362-G008, the
reddening correction $E_{(B - V)}$ needed to fit the continuum of the stellar
template to that of the observed central spectra was set arbitrarily to zero in
Fig.~\ref{reddening2} for comparison. Note that in this object, the difference in slope
between the continuum of the NLR and that of the stellar template may 
not entirely
be due to dust but may at least partially
be due to the change in stellar population: The continuum in  the centre
appears to be bluer than the template, reflecting the very young stellar population already
mentioned earlier. 

\begin{table}
\begin{minipage}{80mm}
 \caption[]
{\label{reddening} Maximum and global reddening within the NLR}
\begin{center}
\begin{tabular}{lccc}
\\[-2.3ex]
\hline
\hline\\[-2.3ex]
\multicolumn{1}{c}{Galaxy} & max. $E_{B - V}$\footnote{Highest reddening value within the NLR} & Distance\footnote{Distance from the centre at which highest reddening occurs} & global $E_{B - V}$\footnote{Derived by adding the H$\alpha$ and H$\beta$ flux within the NLR}\\
& (mag) & (\arcsec) & (mag) \\[0.25ex]
\hline\\[-2.3ex]
IC\,5063      & 0.80$\pm$0.01 & -1.11 & 0.39$\pm$0.05\\
NGC\,7212     & 0.92$\pm$0.03 & -3.33 & 0.56$\pm$0.07\\
ESO\,362-G008 & 0.66$\pm$0.04 & 0 & 0.39$\pm$0.09\\
NGC\,3281     & 0.57$\pm$0.02 & -1.4 & 0.33$\pm$0.05\\
NGC\,5643     & 0.65$\pm$0.01 & 0  & 0.30$\pm$0.04\\
NGC\,1386     & 0.88$\pm$0.2  & -4 & 0.48$\pm$0.06\\[0.1ex]
\hline\\[-2.3ex]
\end{tabular}
\end{center}
\end{minipage}
\end{table}

\subsection{Spatially resolved spectral diagnostics}
\label{2ddiag}
As in paper I, we use
diagnostic line-ratio diagrams
of the three types pioneered by \citet{bal81}
to not only distinguish
bet\-ween emission-line object classes (e.g. Seyfert galaxies, LINERs, Starbursts,
transition objects), but 
to  probe the ``real'' NLR size, i.e.~the central region which
is photoionised by the AGN, and to discriminate the contribution from
starbursts.

\begin{figure*}
\begin{center}
\includegraphics[width=6.2cm,angle=-90]{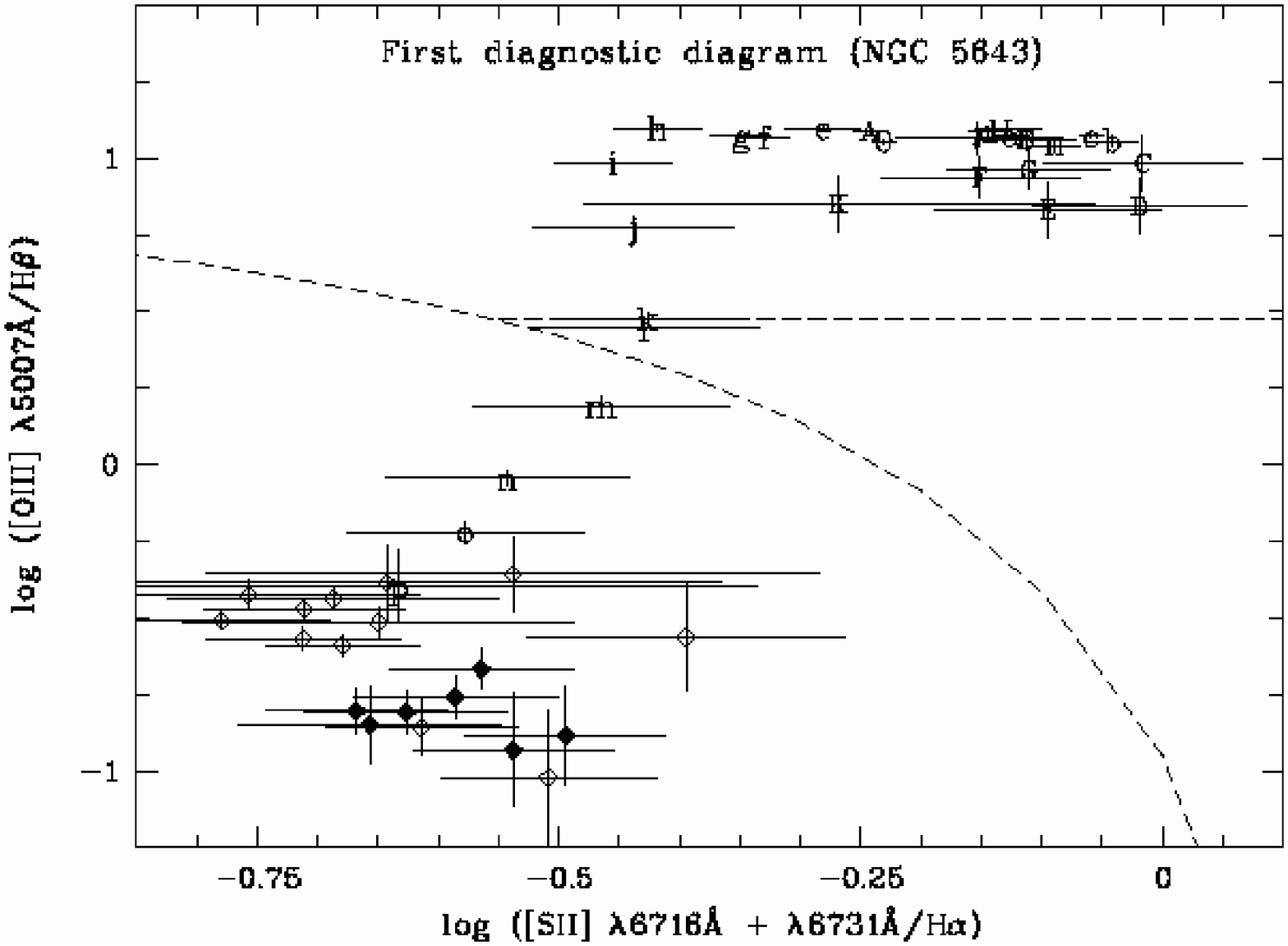}
\includegraphics[width=6.2cm,angle=-90]{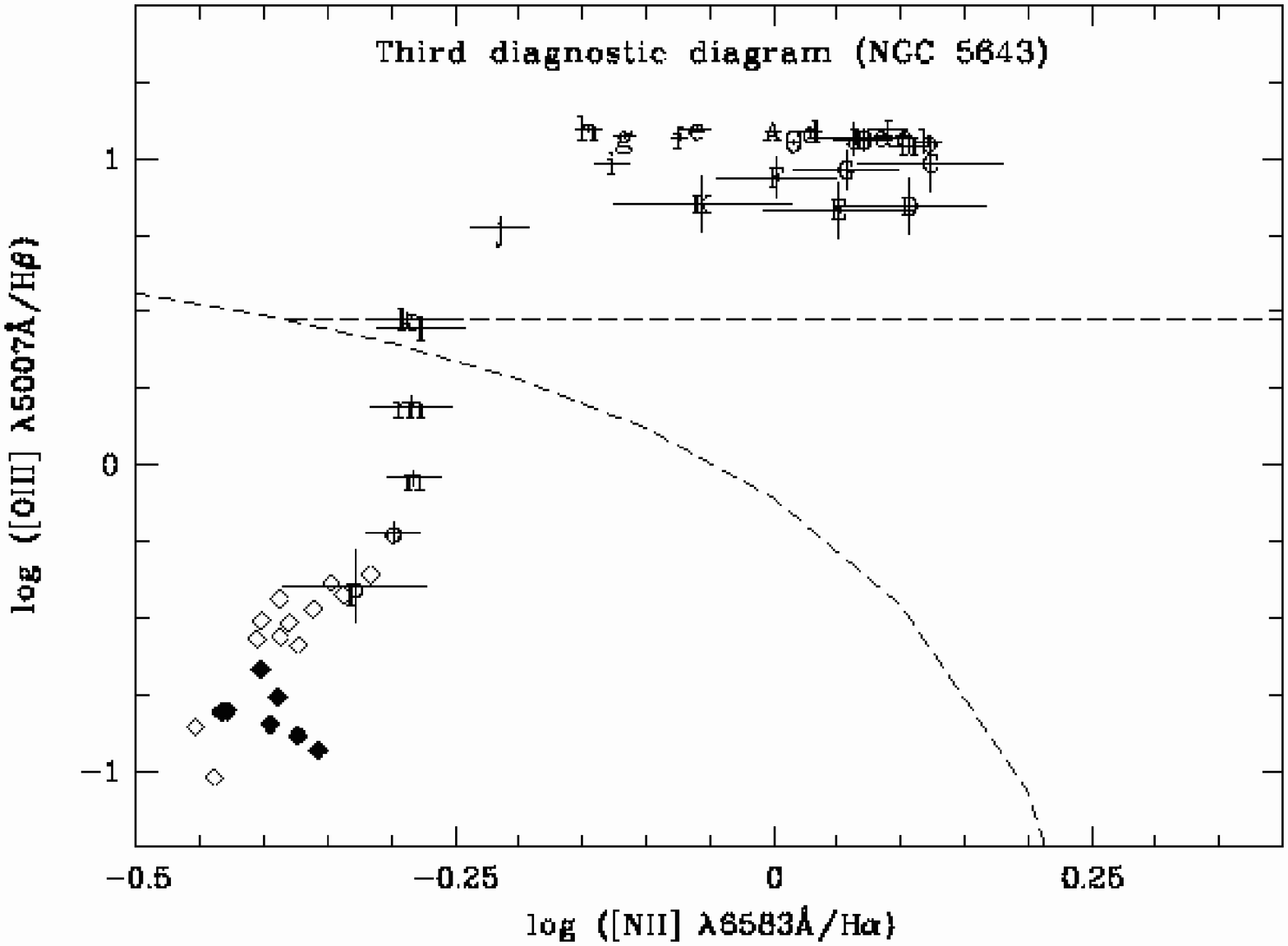}
\includegraphics[width=6.2cm,angle=-90]{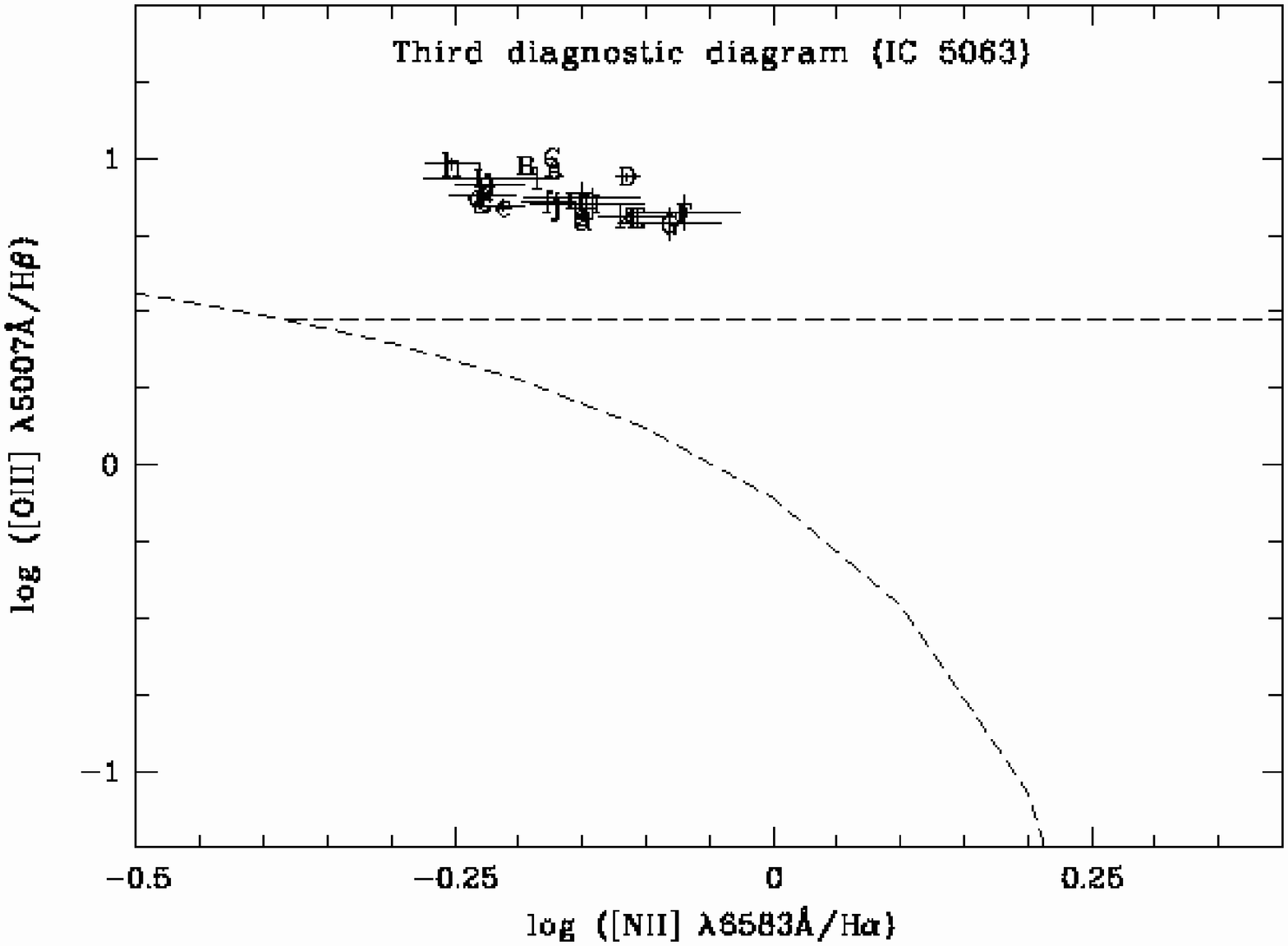}
\includegraphics[width=6.2cm,angle=-90]{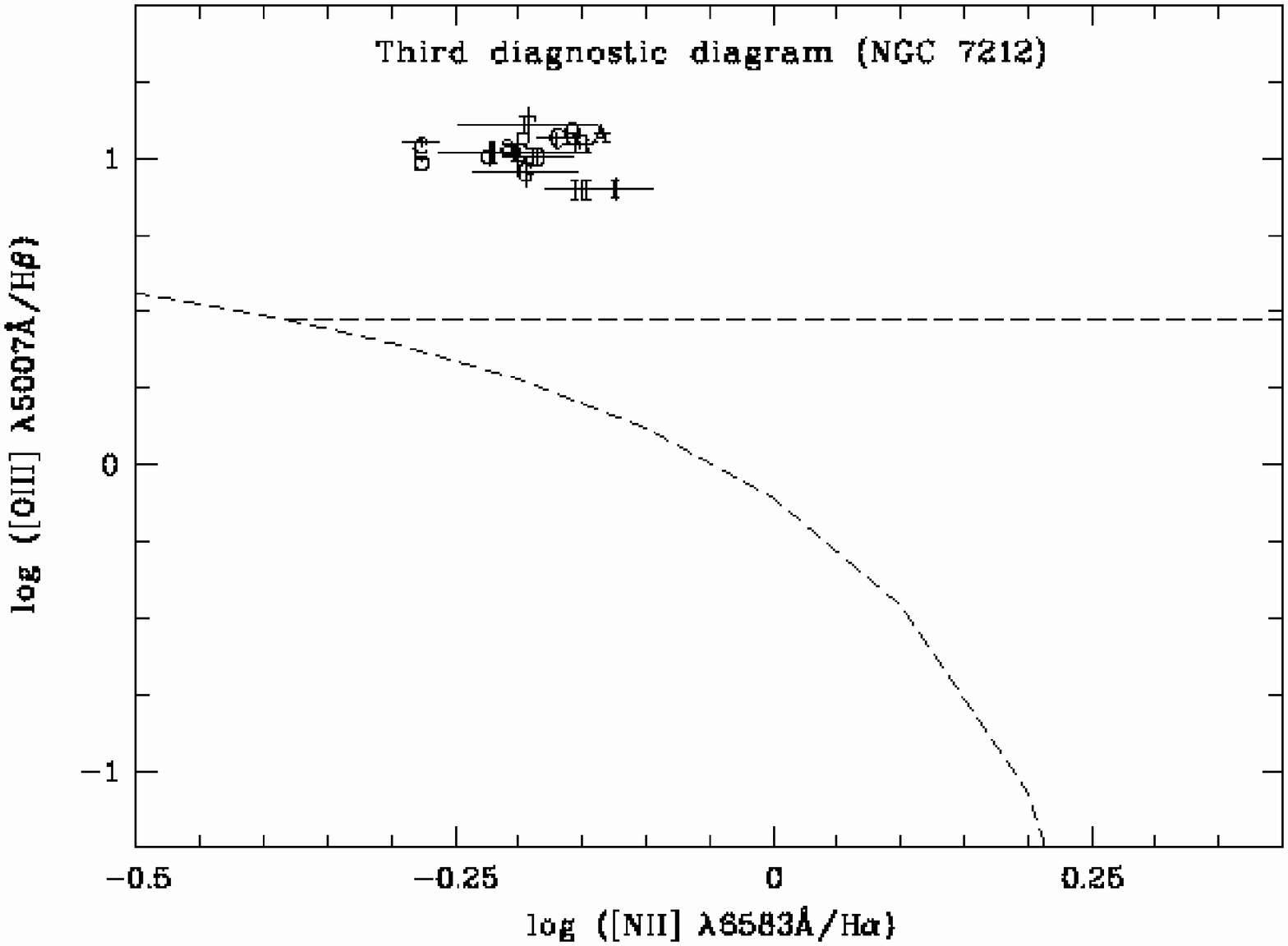}
\includegraphics[width=6.2cm,angle=-90]{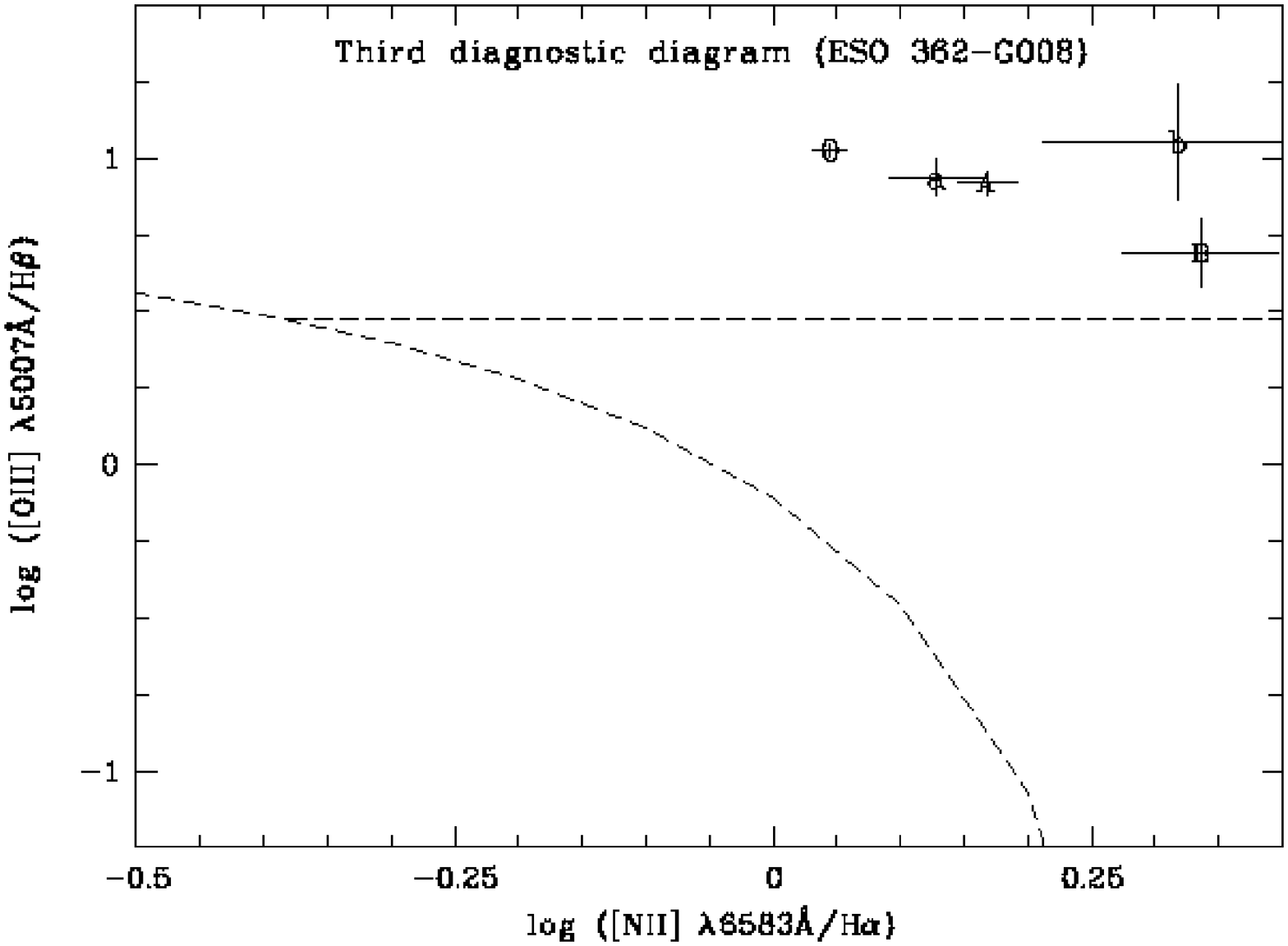}
\includegraphics[width=6.2cm,angle=-90]{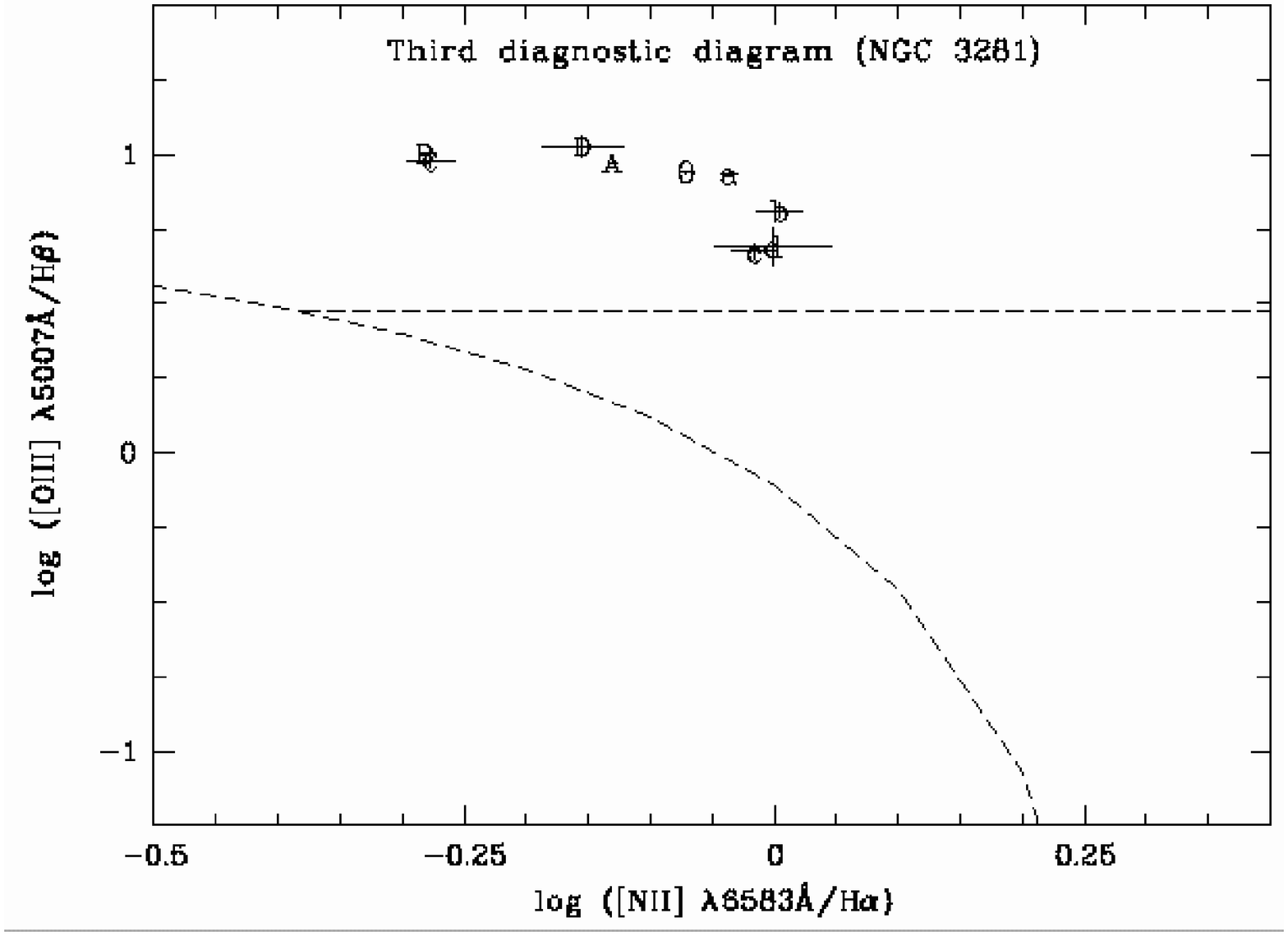}
\caption[]
{\label{diag1} \small
Diagnostic diagrams for
spatially-resolved emission-line ratios in NGC\,5643 (first and third diagnostic diagram) as
well as in IC\,5063,
NGC\,7212, ESO\,362-G008, and NGC\,3281 
(due to similarity, only third diagnostic diagram is shown).}
\end{center}
\end{figure*}

The high S/N ratio of our spectra enables us 
to measure line ratios for all three diagrams (``first'': [\ion{O}{iii}]/H$\beta$ versus [\ion{S}{ii}]/H$\alpha$;
``second'': [\ion{O}{iii}]/H$\beta$ versus [\ion{O}{i}]/H$\alpha$; ``third'': 
[\ion{O}{iii}]/H$\beta$ versus [\ion{N}{ii}]/H$\alpha$) out to several arcseconds from the
nucleus. We present typical diagnostic diagrams for all objects
in Fig.~\ref{diag1}. 
The symbols are chosen such that ``O'' refers to
the central spectrum, the small letters mark regions corresponding to ``-''
arcseconds from the nucleus, the capital
ones mark regions corresponding to ``+''
arcseconds from the nucleus (Table~\ref{tablediag}).

As for NGC\,1386 (paper I),
we find a clear transition between line ratios falling in the AGN
regime and those typical for \ion{H}{ii} regions
for NGC\,5643.
The transition is not as sharp as for NGC\,1386, but
the line ratios are gradually changing from
AGN-type to \ion{H}{ii}-like ratios with line ratios of two outer spectra falling in the
corner between LINER, AGN and \ion{H}{ii} regions.
For the other four galaxies, no such transition is observed but all
emission-line ratios  are typical for gas 
ionised by an AGN power-law continuum.

We use the diagnostic diagrams to determine the NLR size. The results are
summarised in Table~\ref{tablediag}. For those objects which show a transition
of emission-line ratios from the central AGN region to \ion{H}{ii} regions,
this method gives a measure of the NLR size without [\ion{O}{iii}] contamination from circumnuclear starbursts:
Although \ion{H}{ii} regions may be
present over the entire emission--line region, 
the AGN ionisation dominates in the innermost arcseconds,
determining the size of the NLR.

For both NGC\,1386 and NGC\,5643, which show a transition of line ratios, 
the determined NLR size is about twice as large
as those measured from the HST snapshot survey of \citet{sch03}, once
again showing the low sensitivity of this survey.
On the other hand, some authors have attributed all [\ion{O}{iii}] emission
to the extended NLR.
While for NGC\,5643, \citet{fra03} classify the [\ion{O}{iii}] line emission
detected out to $r \sim$ 15--20\arcsec~from longslit spectroscopy 
as extended NLR, we here show that [\ion{O}{iii}] emission beyond
$r \sim$ 11\arcsec~originates from \ion{H}{ii} regions.
The same applies for NGC\,1386: \citet{fra03} determine the extended
NLR from the observed [\ion{O}{iii}] line emission to $r \sim$ 10\arcsec
while we can show that [\ion{O}{iii}] emission beyond
$r \sim$ 6\arcsec~is predominantly ionised by surrounding stars.

To conclude, compared to the spatially resolved
spectral diagnostics measuring the ``real'' NLR size, the apparent NLR size determined
by [\ion{O}{iii}] images can be either smaller in case of low sensitivity 
or larger in case of contributions of circumnuclear starbursts. 
A meaningful measure of the NLR size is of great importance to determine
the slope, and thus the origin, of the NLR size-luminosity relation 
\citep{ben02}. Moreover, it will help to probe whether indeed a different
slope for type-1 and type-2 AGNs exist \citep{ben04, ben06b}.

For the remaining four objects, the estimated NLR size is a lower limit.
This points out the limits of the method presented here:
The NLR radius determination depends on the relative brightness of the
AGN and the central starburst component and therefore the NLR
radius also depends on the presence, strength and
distribution of starbursts.
Such a method fails if 
there are no or just weak starforming regions surrounding the central AGN.
In that case, we cannot say for sure whether the extension of the 
detected [\ion{O}{iii}] emission is limited by the 
competition between AGN and starburst luminosity.

\begin{table*}
\begin{minipage}{180mm}
 \caption[]
{\label{tablediag} Results from diagnostic diagrams\footnote{The second column gives 
the distance from the centre to the first spectra
(marked with the letters ``a'' and ``A'' in the diagnostic diagrams). 
In the third column, 
the orientation of the small and capital letters is listed.
The maximum [\ion{O}{iii}] radius (S/N $>$ 3) at the
same p.a.~taken from literature is given in the fourth column.
We also give the
 [\ion{O}{iii}] radius (S/N $>$ 3) observed from our
spectra (column 5). In the sixth column, the radius is given until which we
were able to plot line ratios in the diagnostic diagrams.  In the last column, 
the radius of
the NLR as determined from the diagnostic diagrams is given in \arcsec~and, in
brackets, pc, respectively.
The two objects with a clear transition between
NLR and \ion{H}{ii} region are marked in bold.}}
\begin{center}
\begin{tabular}{lcr@{/}lcccc}
\\[-2.3ex]
\hline
\hline\\[-2.3ex]
\multicolumn{1}{c}{Galaxy} & ``a/A'' & \multicolumn{2}{c}{``a/A''} &  $R_{\rm [OIII]}$ & $R_{\rm [OIII]}$ & $R_{\rm line-ratios}$ & $R_{\rm NLR}$\\
& Distance (\arcsec) & \multicolumn{2}{c}{Orientation} & Literature (\arcsec) & Our Data (\arcsec) & Our data (\arcsec) & Our Data (\arcsec, pc)\\[0.25ex]
\hline\\[-2.3ex]
IC\,5063                 & 1   & NW & SE & 6\footnote{Taken from HST image of \citet{sch03}} & 20      & 13       & \hspace{0.1cm}$>$13 (2860)\\
NGC\,7212                & 1   & SE & NW & 2.4$^b$               & 12      & 10       & \hspace{0.1cm}$>$10 (5040)\\
ESO\,362-G008           & 1   & SE & NW & \hspace{-2mm}11.9\footnote{Taken from groundbased image of \citet{mul96}} & \hspace{+1.5mm}4 & \hspace{+1.5mm}3 & \hspace{-1.5mm}$>$3 (960)\\
NGC\,3281                & 1.4 & SW & NE & 3$^b$              & \hspace{+1.5mm}9   & \hspace{+1.5mm}5          & $>$5 (1180)\\
{\bf NGC\,5643}          & 1   & E & W   & 6\footnote{Taken from HST image of \citet{sim97}} & 16      & 16       & \hspace{+2mm}{\bf 11 (1045)}\\
{\bf NGC\,1386}          & 1   & S & N   & 3$^b$               & 12           & 10  & \hspace{-0mm} {\bf 6 (310)}\\[0.1ex]
\hline\\[-2.3ex]
\end{tabular}
\end{center}
\end{minipage}
\end{table*}

\subsection{Surface-brightness distribution}
\label{longsur}
\begin{figure*}
\begin{center}
\includegraphics[width=6.2cm,angle=-90]{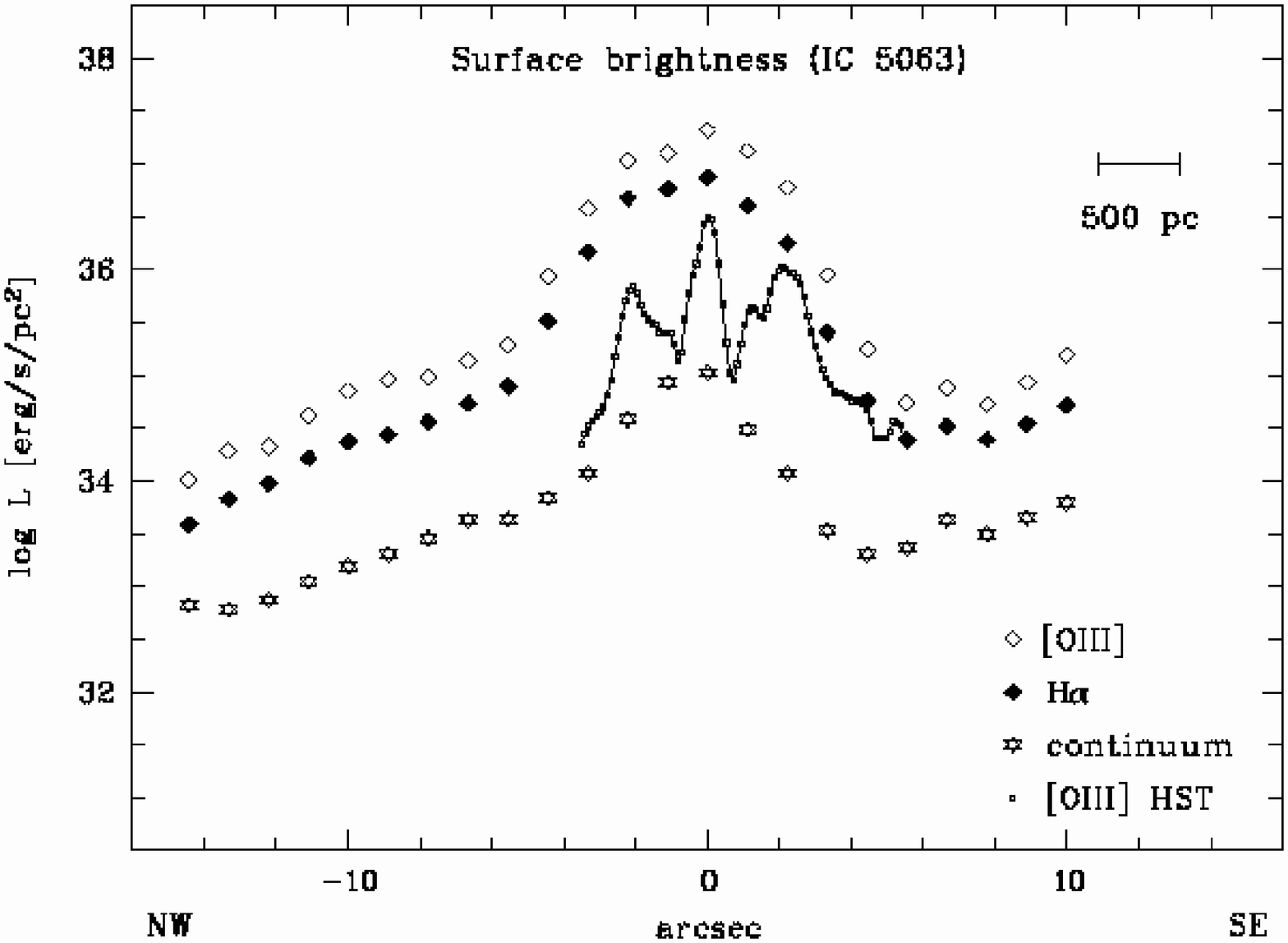}
\includegraphics[width=6.2cm,angle=-90]{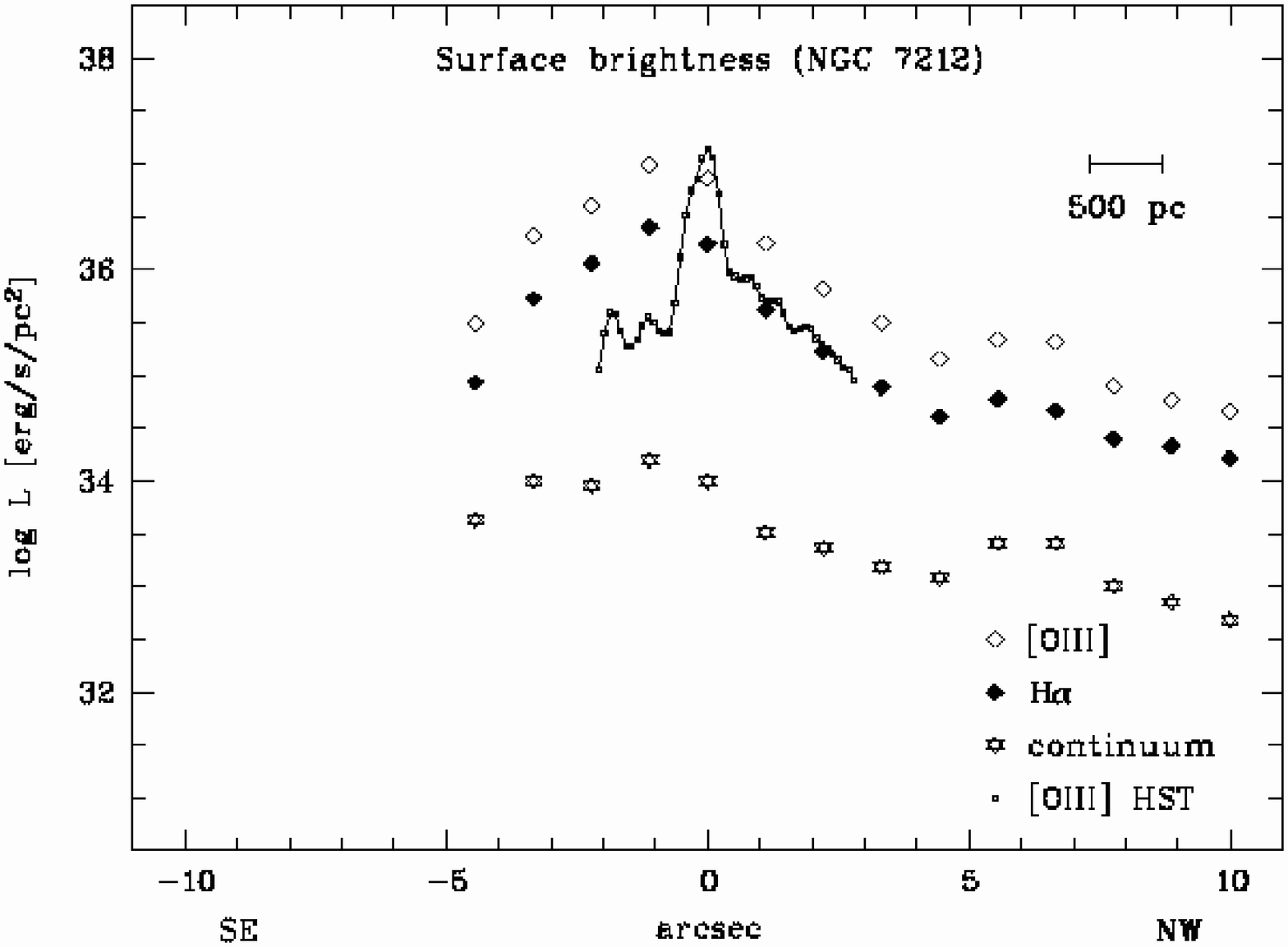}
\includegraphics[width=6.2cm,angle=-90]{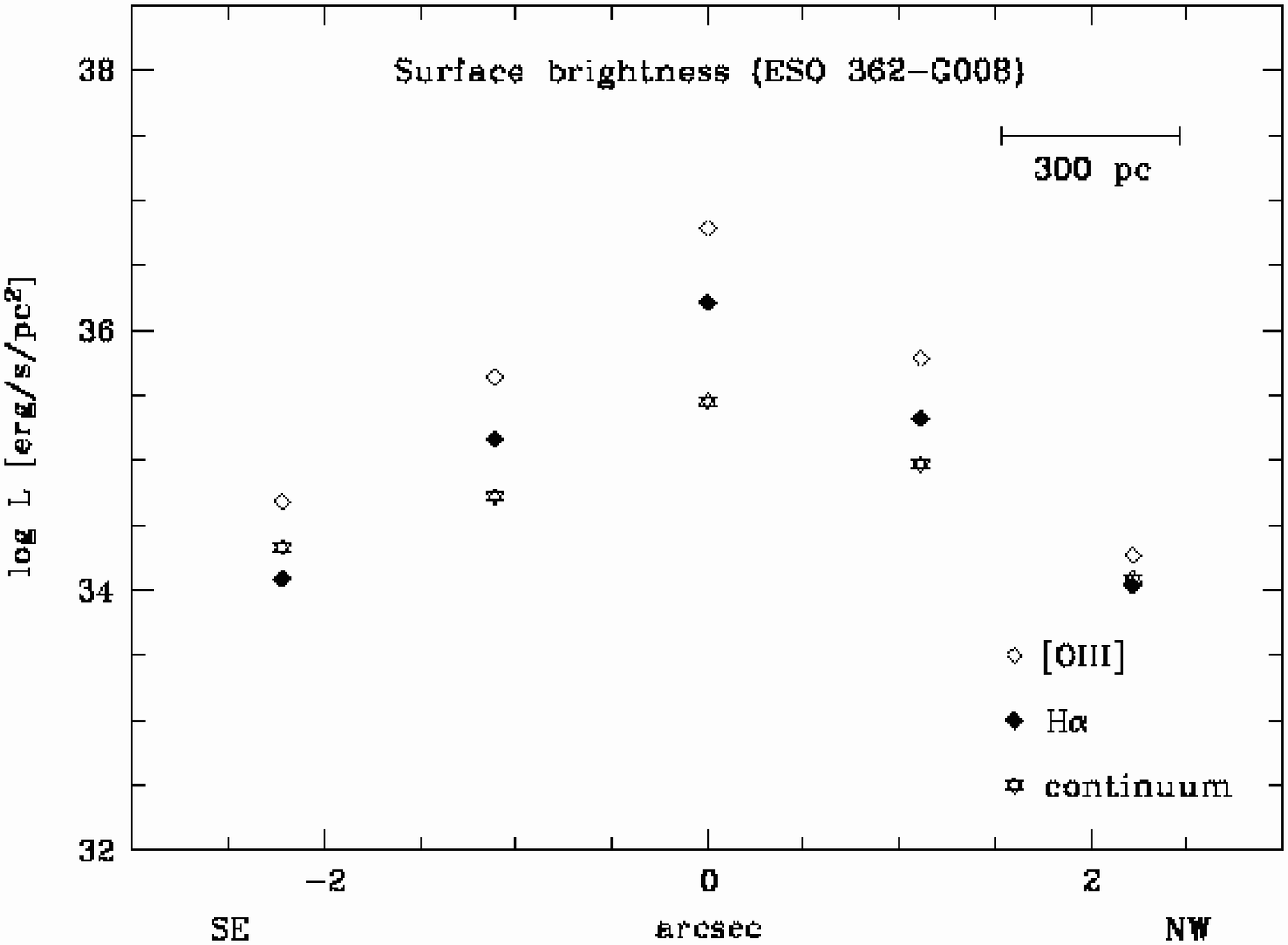}
\includegraphics[width=6.2cm,angle=-90]{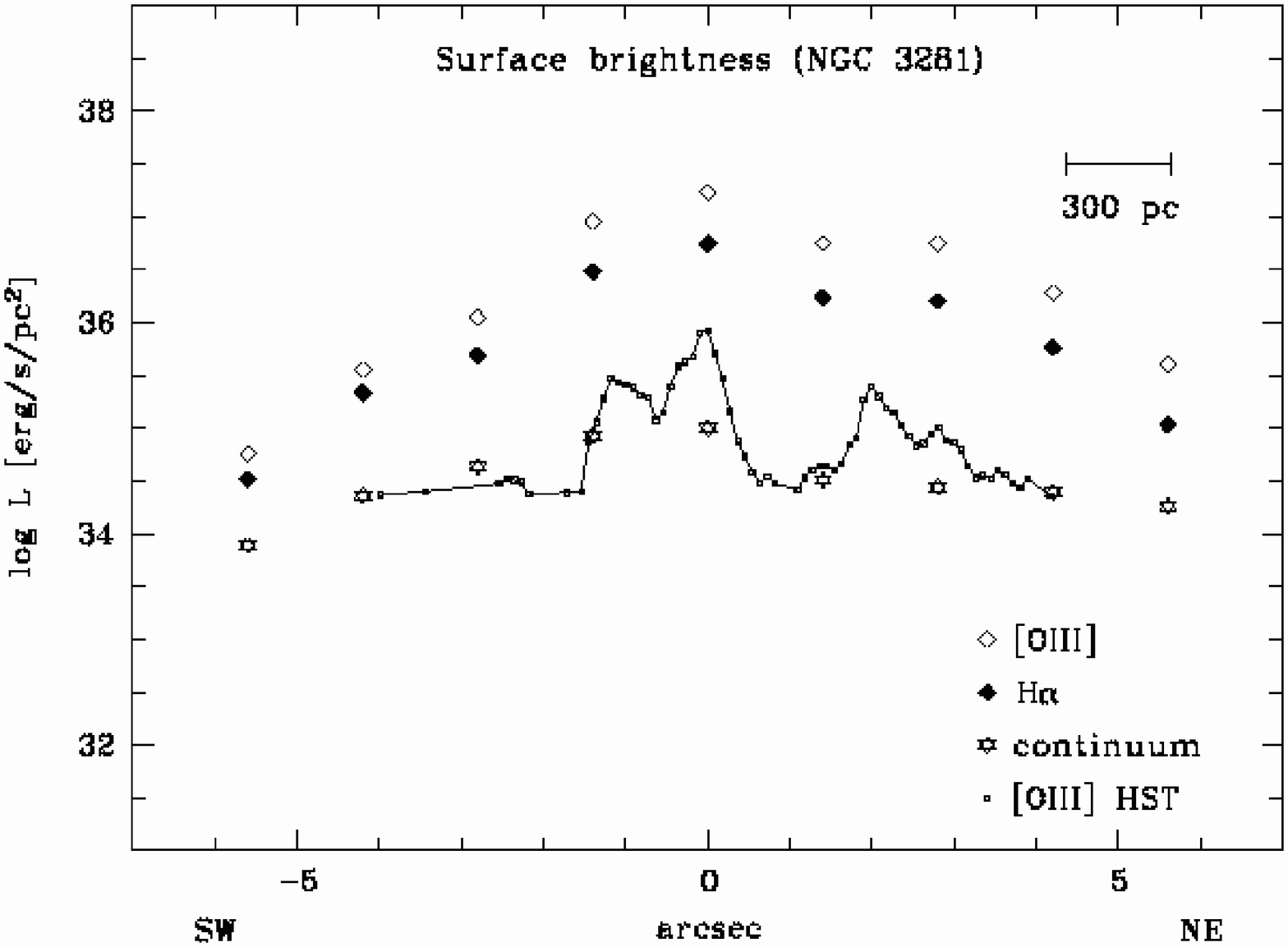}
\includegraphics[width=6.2cm,angle=-90]{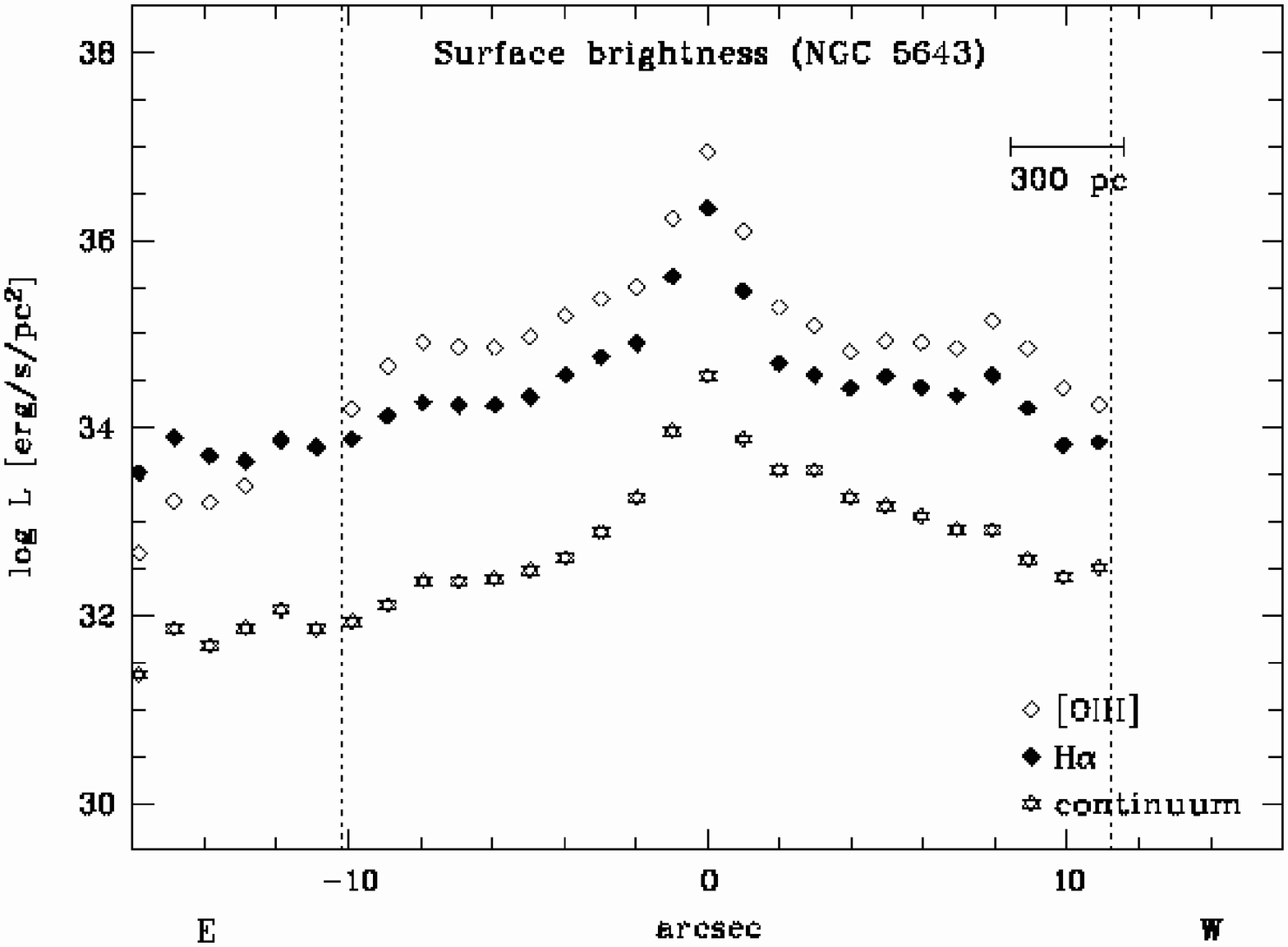}
\caption[]
{\label{lum2} \small
Surface-brightness distributions of  
IC\,5063, NGC\,7212, ESO\,362-G008, NGC\,3281,
and NGC\,5643
in [\ion{O}{iii}] (open diamonds), 
H$\alpha$ (filled diamonds), and continuum (at
  5450-5700\,\AA, stars). The [\ion{O}{iii}] surface-brightness distribution
  from the HST image is shown as small open squares connected by a line
(HST pixel scale $\sim$ 0\farcs1\,pix$^{-1}$). 
Only data points with S/N $>$ 3
 were included. 
Error bars are smaller than the symbol size.
The HST image has a 10 to 20 times higher spatial resolution but
a significantly lower sensitivity, not allowing to measure the outer parts
of the NLR. 
Note that ESO\,362-G008 and
NGC\,5643 are not included in the HST snap-shot survey by \citet{sch03}.
For NGC\,5643, the edge of the NLR as determined from the
  diagnostic diagrams is indicated by dotted lines.
}
\end{center}
\end{figure*}

The spatially varying luminosities in the [\ion{O}{iii}] and H$\alpha$ emission lines as well as
the continuum (at 5450-5700\,\AA) were calculated and divided by the corresponding area in square
  parsecs at the galaxy to allow a comparison among all galaxies in our sample
(Fig.~\ref{lum2}).
The surface-brightness distributions are similar to each other, centrally peaked and 
decreasing with distance from the
nucleus. 

For comparison, the [\ion{O}{iii}] surface-brightness distributions
from the HST images of \citet{sch03} are shown for those objects included in the
  HST snapshot survey. 
They were derived by averaging three vectorplots along the major axis of the NLR
emission. In all objects, they clearly show the
higher spatial resolution of the HST image (0\farcs05 - 0\farcs1
pix$^{-1}$) compared to the 1-2\arcsec~spatial sampling of our spectral data.
However, they also reveal the low sensitivity of the HST images
compared to our spectroscopy: The
[\ion{O}{iii}] emission at a S/N of 3 ends significantly earlier than what can
be seen in our spectral data.
In some cases, the HST [\ion{O}{iii}] surface-brightness distributions reveal several
subpeaks of possibly individual NLR clouds, as can be already seen in the
[\ion{O}{iii}] images (Fig.~\ref{galaxies2}).
These substructures are smoothed out in our $\sim$10-20 times lower spatial resolution
spectra but are nevertheless still visible as a secondary or tertiary peak,
mostly in emission lines.

We fitted a power-law function $L = L_{0} (\frac{R}{R_0})^{\delta}$ (with projected radius $R$) to the
surface-brightness distributions of [\ion{O}{iii}], H$\alpha$, and the continuum. 
The fitting parameters are shown in Table~\ref{fitlum} (with $L_0$ referring to $R_0$ = 100 pc from the nucleus). 
Only data points within the NLR were included
and taken as average from both sides of the nucleus.
The [\ion{O}{iii}] surface brightness falls faster with radius than the
H$\alpha$ surface brightness and also faster than the continuum
($<$$\delta_{\rm [OIII]}$$> \sim -2.24\pm0.2$; 
$<$$\delta_{\rm H\alpha}$$> \sim -2.16\pm0.2$;
$<$$\delta_{\rm cont}$$> \sim -1.19\pm0.1$).

This general trend is comparable to what has been found by \citet{fra03} in their
long-slit spectroscopic study of the NLR of
18 Seyfert-2 galaxies. 
However, they find on average a slightly
steeper  slope for especially [\ion{O}{iii}] ($<$$\delta_{\rm [OIII]}$$> \sim -3.6$; 
$<$$\delta_{\rm H\alpha}$$> \sim -2.6$; $<$$\delta_{\rm cont}$$> \sim -1.4$;
calculated from their Table 2).
This discrepancy can have different reasons.
For one, we extracted spectra at every $\sim$1\arcsec, whereas they
extracted spectra at every $\sim$2\arcsec. 
Secondly, the samples differ and only two objects
are common to both samples.  
Thirdly, \citet{fra03} include all
visible emission in their fit, not concentrating on the NLR alone as they do
not have a quantitative measure of the NLR size. Therefore, they
possibly include [\ion{O}{iii}] emission from surrounding \ion{H}{ii} regions
where the surface brightness drops significantly.
This explanation is supported when comparing the results for 
NGC\,1386 and NGC\,5643, common to both samples:
\citet{fra03} report a steeper slope in the [\ion{O}{iii}] surface-brightness
distribution but include emission out to distances from the nucleus where we
can show from diagnostic diagrams
that this emission can be attributed to surrounding \ion{H}{ii} regions. 

\begin{table*} 
\begin{minipage}{180mm}
\caption[]{\label{fitlum} Fitting parameters of surface-brightness distributions\footnote{A linear least-squares fit was applied with $\log L = \delta \cdot
\log R/R_0 + \log L_{0}$ and $R_0$ = 100 pc from the nucleus.
The number of data points included in the fit is
given in column 2 (= half the number of averaged values from both sides of
the nucleus). 
For NGC\,5643 which shows a transition between line
ratios typical for AGNs and \ion{H}{ii}-regions, only data points within the NLR
were included}. For
ESO\,362-G008, too few data points were available for a fit.}
\begin{center}
\begin{tabular}{lccccccc}
\\[-2.3ex]
\hline
\hline\\[-2.3ex]
\multicolumn{1}{c}{Galaxy} & Data Points & \hspace{+3mm}$\delta_{\rm [OIII]}$ & $\log L_{\rm [OIII], 0}$  & \hspace{+3mm}$\delta_{\rm H\alpha}$ & $\log L_{\rm H\alpha, 0}$  & \hspace{+3mm}$\delta_{\rm cont}$ & $\log L_{\rm cont, 0}$ \\
& & & (erg\,s$^{-1}$\,pc$^{-2}$) & & (erg\,s$^{-2}$\,pc$^{-2}$) & & (erg\,s$^{-2}$\,pc$^{-2}$)\\[0.25ex]
\hline\\[-2.3ex]
IC\,5063           & 9               & $-$2.73$\pm$0.33 & 38.41 & $-$2.75$\pm$0.31 & 38.01 & $-$1.37$\pm$0.18 & 35.2 \\
NGC\,7212          & 4               & $-$2.13$\pm$0.57 & 38.46 & $-$2.09$\pm$0.56 & 37.85 & $-$0.77$\pm$0.26 & 34.57 \\
NGC\,3281          & 4               & $-$2.35$\pm$0.60 & 38.22 & $-$2.34$\pm$0.64 & 37.73 & $-$1.02$\pm$0.18 & 35.34 \\
NGC\,5643          & \hspace{-2mm}10 & $-$1.42$\pm$0.19 & 35.97 & $-$1.34$\pm$0.16 & 35.35 & $-$1.54$\pm$0.11 & 33.94 \\
NGC\,1386          & 6               & $-$2.56$\pm$0.82 & 37.16 & $-$2.27$\pm$0.60 & 36.55 & $-$1.26$\pm$0.48 & 35.46 \\[0.1ex]
\hline\\[-2.3ex]
\end{tabular}
\end{center}
\end{minipage}
\end{table*}

\subsection{Electron-density distribution}
\label{longdens}
Applying the classical methods outlined in \citet{ost89},
we derive the electron density as a function of distance
to the nucleus using the ratio of the 
[\ion{S}{ii}]\,$\lambda$$\lambda$6716,6731\,\AA~pair 
of emission lines.
We used the observed central temperature to 
correct for the dependency of electron
density on temperature\footnote{$n_e ({\rm T}) 
= n_e ({\rm [SII]\,ratio}) \cdot \sqrt{(T/10000)}$}. 
Due to the faintness of the involved
[\ion{O}{iii}]\,$\lambda$4363\,\AA~emission line, we were not able to measure
the temperature in the outer parts.
For those objects for which no temperature was determined, we assumed  $T_e =
10000$\,K.

\begin{figure*}
\begin{center}
\includegraphics[width=6.2cm,angle=-90]{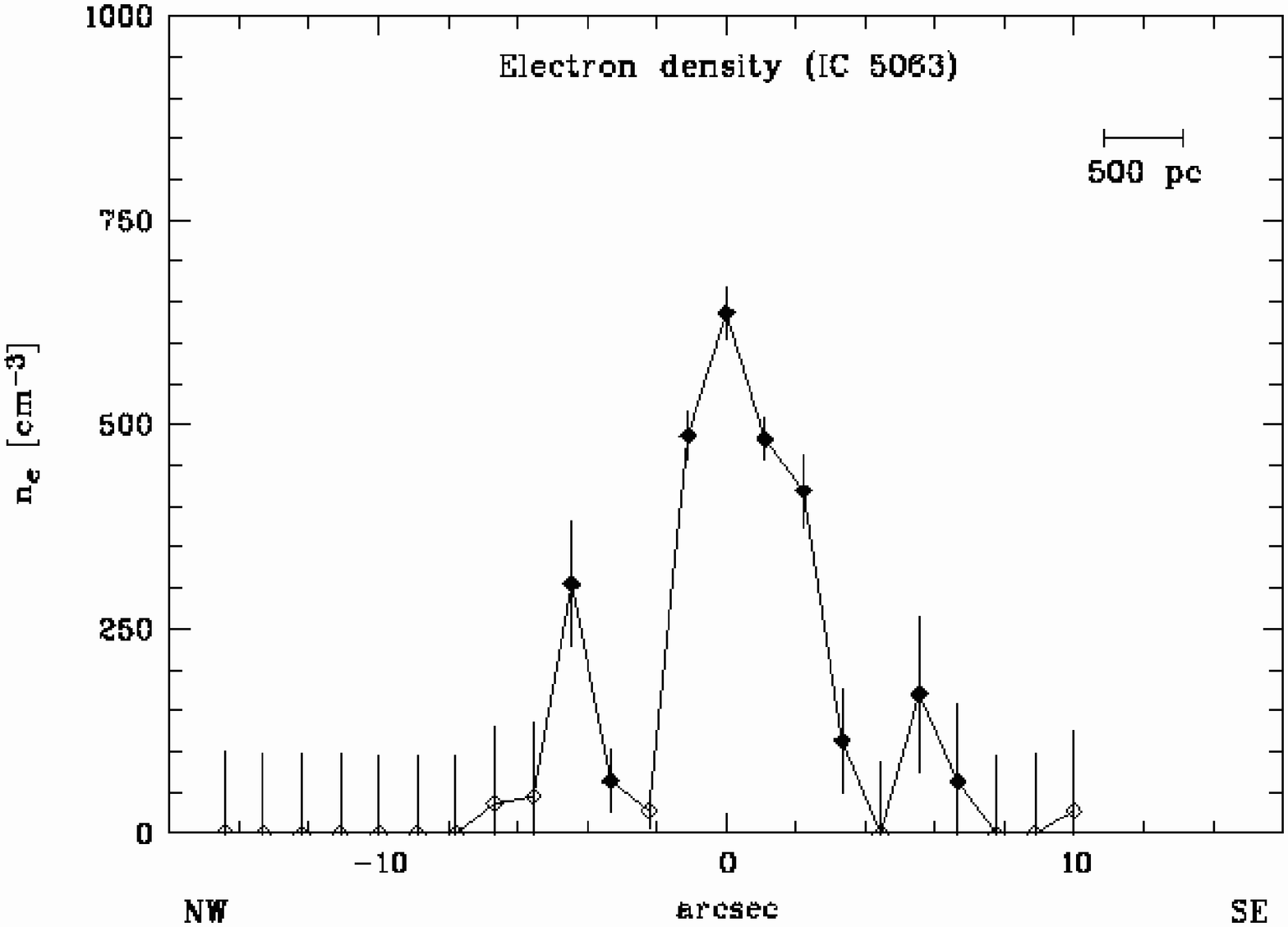}
\includegraphics[width=6.2cm,angle=-90]{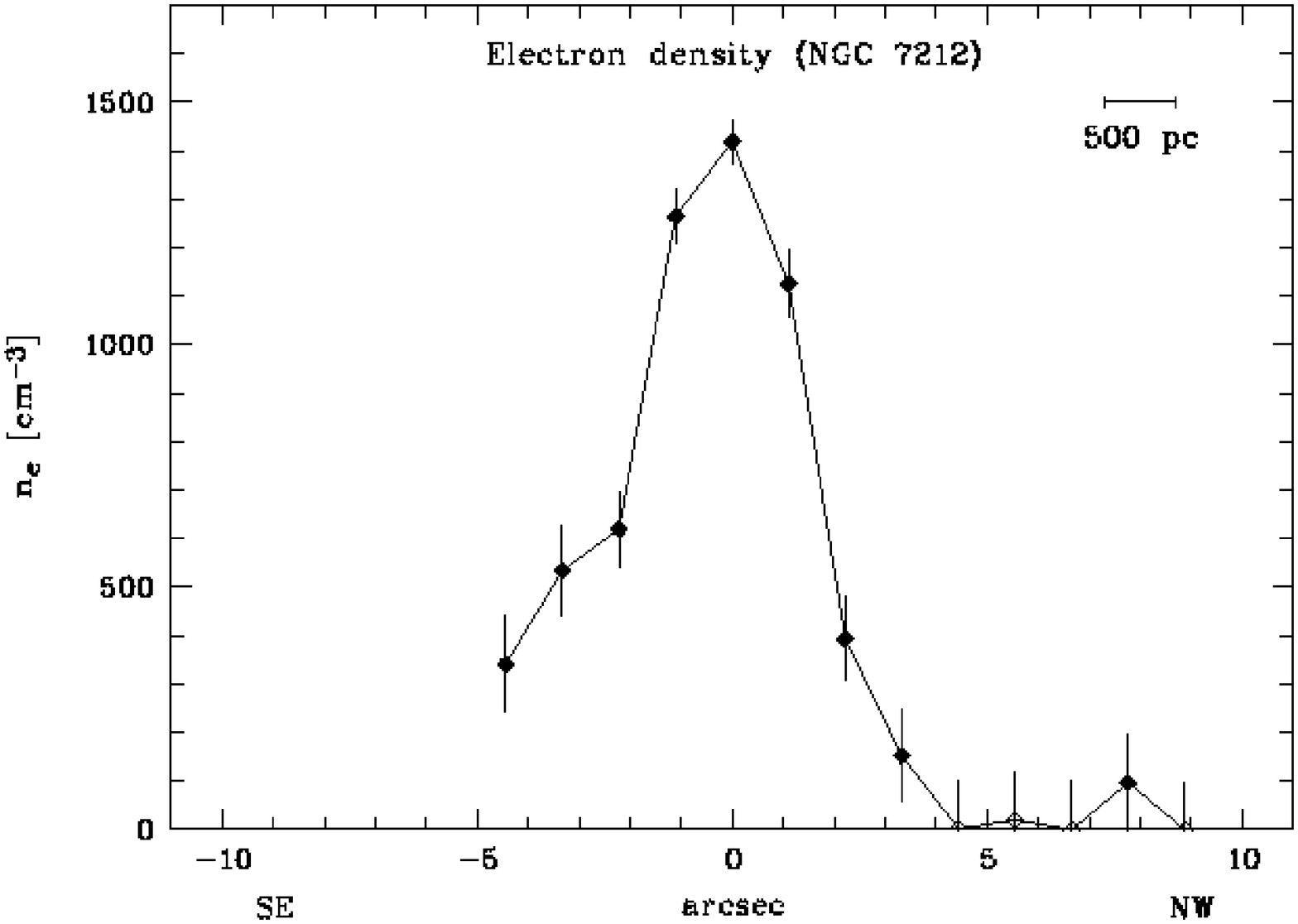}
\includegraphics[width=6.2cm,angle=-90]{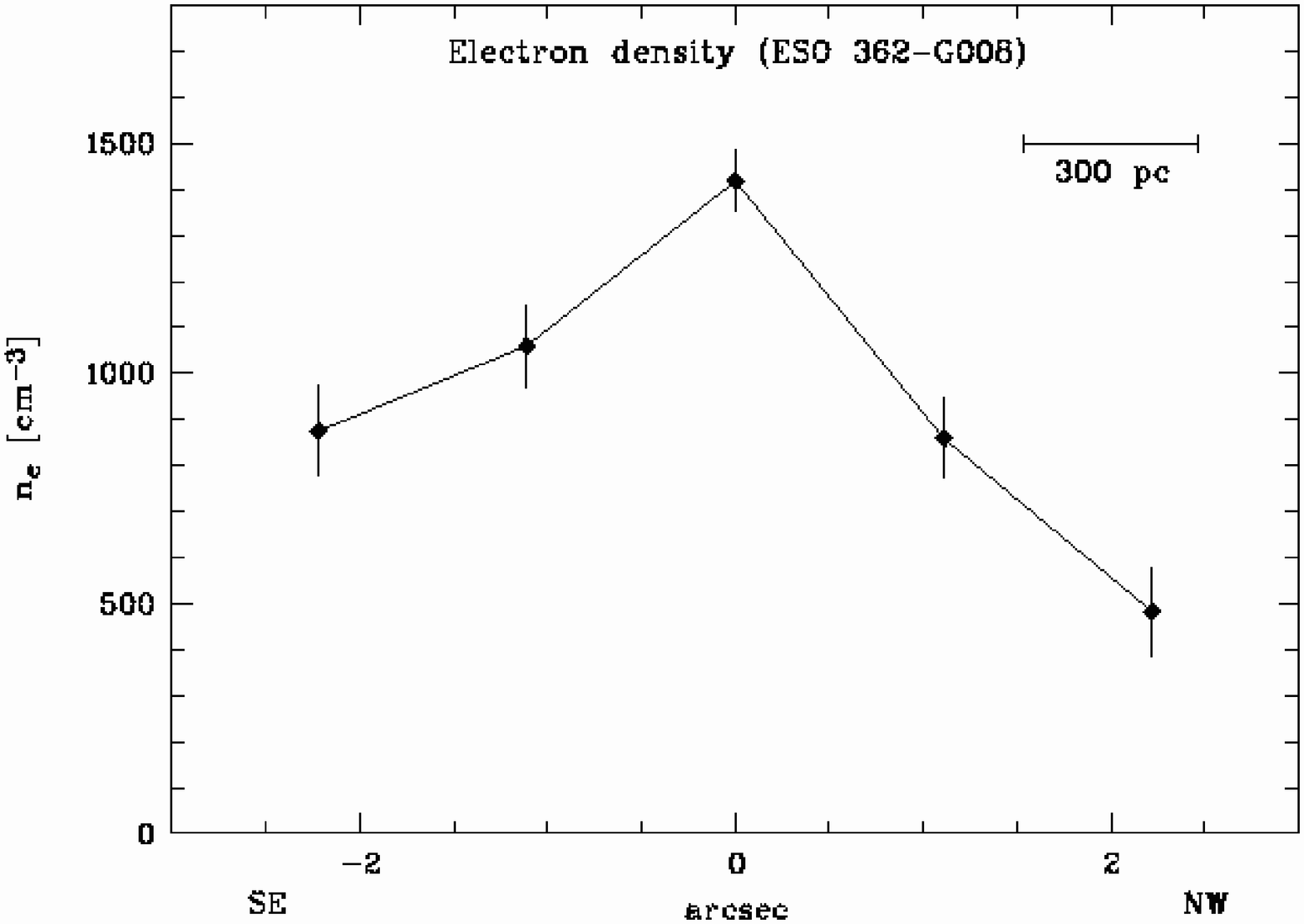}
\includegraphics[width=6.2cm,angle=-90]{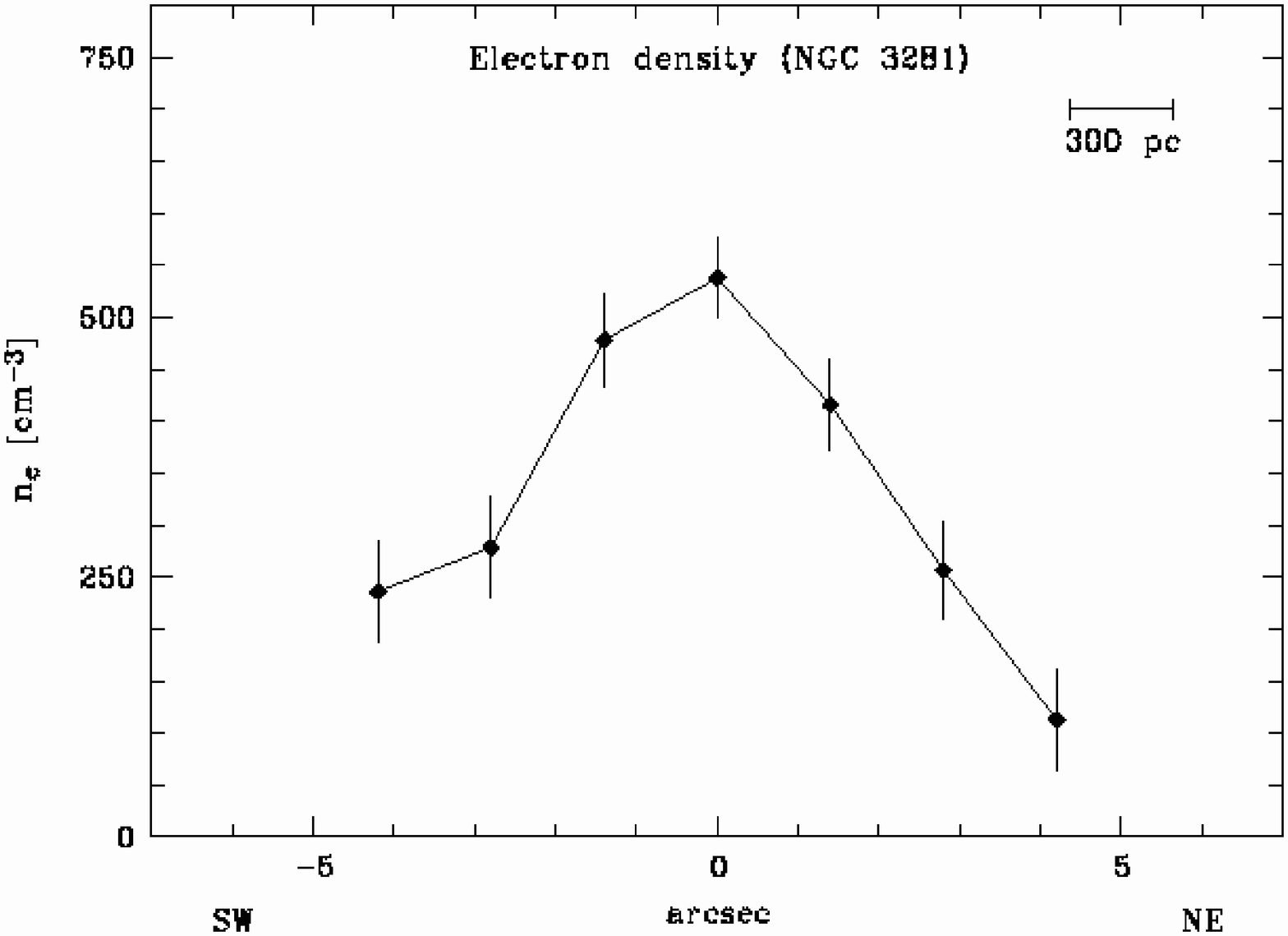}
\includegraphics[width=6.2cm,angle=-90]{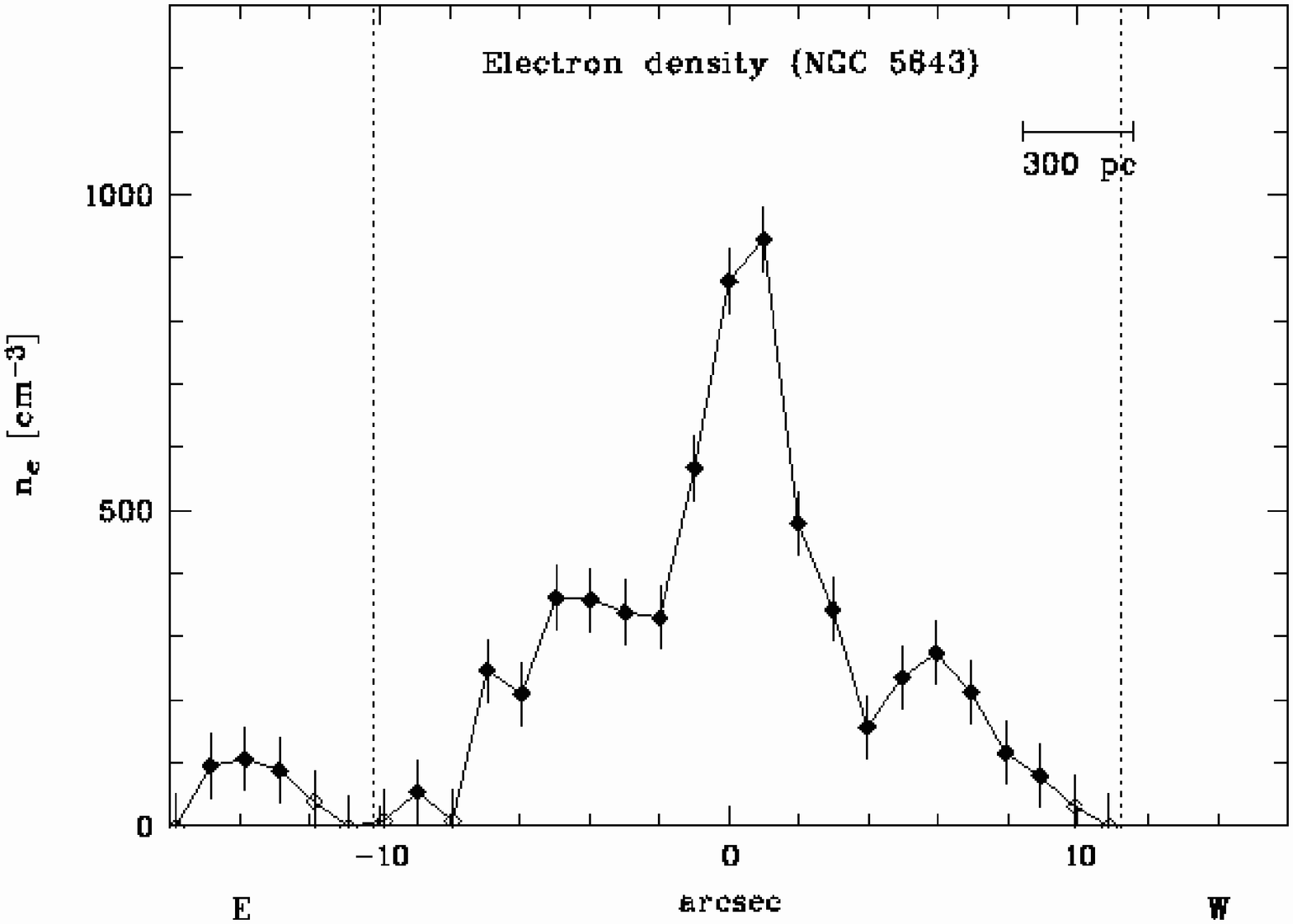}
\caption[]
{\label{density2} \small 
Electron density obtained
from the [\ion{S}{ii}]\,$\lambda$6716\,\AA/$\lambda$6731\,\AA~ratio 
as a function of the distance from the nucleus for IC\,5063, NGC\,7212, 
ESO\,362-G008, NGC\,3281, and NGC\,5643.
Open symbols indicate locations
where $n_{\rm e, obs}$ is in the low-density limit (assumed $\le$ 50\,cm$^{-3}$).
For NGC\,5643, the edge of the NLR as determined from the
  diagnostic diagrams is indicated by dotted lines.
}
\end{center}
\end{figure*}

In all objects, 
the electron density is highest at the nucleus and 
decreases outwards down to the low-density limit
(assumed to be 50\,cm$^{-3}$; Fig.~\ref{density2}). In
some cases, 
it reveals a secondary or tertiary peak on one or both sides of the optical centre.
A characteristic structure with a central peak and two smaller peaks on both
sides of the nucleus can be identified in three objects 
[IC\,5063, NGC\,5643, and NGC\,1386 (paper I)]. 
The outer
peaks are often close to the boundary of the NLR. These density enhancements
may indicate shocks occurring at the edge of the NLR.

In Table~\ref{fitden}, we give the results of fitting a power-law function $n_{\rm e, obs}$
= $n_{e, 0} (\frac{R}{R_0})^{\delta}$ to the observed electron densities 
(with $n_{\rm e, 0}$ at $R_0$ = 100 pc
from the nucleus).
We included only data points within the NLR and averaged the
electron densities from both sides of the nucleus.
 $\delta$ ranges
between -0.8 and -1.3. On average, the density decreases with $R^{-1.14 \pm
  0.1}$.
These results are comparable to those of \citet{fra03} who give power-law fits to
electron densities of 14 Seyfert-2 galaxies.

The temperature can be a function of distance from the central
AGN. Unfortunately, we are not able to determine the temperature dependency on
distance from the nucleus. In those objects where we are able to trace the
electron temperature in the inner few arcseconds, it remains roughly constant.
One may expect that the temperature is decreasing if the AGN is the only heating source.
In that case, correcting with the central temperature overestimates
the electron density in the outer parts. The observed decreasing slope can
therefore not be artificially introduced by a wrong temperature correction.
On the other hand, some authors report an increasing temperature with
distance from the nucleus [e.g.~\citet{ber83}] and explain it with a decrease
in electron density faster than $n_e \propto r^{-2}$. However, the average decrease of
electron density $n_{\rm e, obs}$ 
we observe is with $\delta \sim -1.1$ slower than that.

Note that the critical density for
[\ion{S}{ii}]\,$\lambda\lambda$6716,6731\,\AA~is $\sim$1500\,cm$^{-3}$ and
3900\,cm$^{-3}$, respectively. 
Thus, these lines can only be used to measure the density in an environment
with densities below $\sim$1500\,cm$^{-3}$. For some objects
in which we measure central densities in this regime,
the central density may thus be underestimated.

\begin{table}
\begin{minipage}{80mm}
 \caption[]
{\label{fitden} Fitting parameters of electron-density distribution\footnote{A linear least-squares fit was applied with 
$\log$ $n_{\rm e, obs}$ = $\delta \cdot
\log R/R_0 + \log n_{e, 0}$$_{\rm obs}$. 
$n_{e, 0}$ corresponds to the value at $R_0$ = 100 pc distance
from the centre.
The number of data points included in the fit is
given in column 2 (= half the number of averaged values from both sides of
the nucleus). 
For those objects which show a transition between line
ratios typical for AGNs and \ion{H}{ii}-region like ones in the diagnostic
diagrams, determining the size of the NLR, only data points within the NLR
were included (NGC\,5643 and NGC\,1386).}}
\begin{center}
\begin{tabular}{lccc}
\\[-2.3ex]
\hline
\hline\\[-2.3ex]
\multicolumn{1}{c}{Galaxy} & Data Points & $\delta$ & $\log n_{e, 0}$ (cm$^{-3}$)\\[0.25ex]
\hline\\[-2.3ex]
IC\,5063 & 6   & -1.08$\pm$0.25 & 3.1 \\
NGC\,7212 & 4  & -1.33$\pm$0.15 & 4.1 \\
NGC\,3281 & 3  & -0.84$\pm$0.08 & 3.1  \\
NGC\,5643 & \hspace*{-0.2cm}10 & -1.22$\pm$0.30  & 3.0 \\
NGC\,1386 & 5  & --1.23$\pm$0.09 & 2.4 \\[0.1ex]
\hline\\[-2.3ex]
\end{tabular}
\end{center}
\end{minipage}
\end{table}

\subsection{Ionisation-parameter distribution}
\label{longioni}
The line ratio
[\ion{O}{ii}]$\lambda$3727\,\AA/[\ion{O}{iii}]\,$\lambda$5007\,\AA~can be used
to estimate the value of the ionisation parameter $U$ 
[e.g.~\citet{pen90, kom97}]. Here, we follow the method
described in paper I.

In most objects, the ionisation parameter peaks at the optical nucleus and
decreases with distance. The two exceptions are the Seyfert-2 galaxies
NGC\,3281 and IC\,5063 where the ionisation parameter reaches its maximum
value several arcseconds to one side of the centre. However, the optical
nucleus needs not necessarily coincide with the position of the ionising
source, the AGN, which may also be hidden by dust. We will discuss this
issue when discussing the objects individually.

We fitted a 
power-law function $U_{\log (n_e) = 2, obs}$
= $U_{0} (\frac{R}{R_0})^{\delta}$ to the observed ionisation parameter (with $R_0$ = 100 pc
from the nucleus; Table~\ref{fitioni}). 
We include only data points within the NLR and averaged the
ionisation parameters of both sides of the nucleus.
 $\delta$ ranges
between -0.4 and -0.6. 
Note that we did not measure the slope of the ionisation parameter in the
Seyfert-2 galaxies
NGC\,3281 and IC\,5063 as
the ionisation parameter does not peak in the centre.
We could not determine the ionisation parameter for
NGC\,5643 as the [\ion{O}{ii}]\,$\lambda$3727\AA~emission line
is not covered by the spectral range.

\begin{table} 
\begin{minipage}{80mm}
\caption[]{\label{fitioni} Fitting parameters of ionisation-parameter distribution\footnote{A linear least-squares fit was applied with $\log$$U_{\log (n_e) = 2, obs}$ = $\delta \cdot
\log R/R_0 + \log$$U_{0}$. 
$U_{0}$ corresponds to the value at $R_0$ = 100 pc distance
from the centre. The number of data points included in the fit is
given in column 2 (= half the number of averaged values from both sides of
the nucleus). For NGC\,1386 which shows a transition between line
ratios typical for AGNs and \ion{H}{ii}-regions, only data points within the NLR
were included. For
ESO\,362-G008, too few data points were available for a fit.
For NGC\,3281 and IC\,5063,
no fit was applied as the ionisation parameter does not peak in the centre. For
NGC\,5643, the [\ion{O}{ii}] line was not covered by the observations.
}}
\begin{center}
\begin{tabular}{lccc}
\\[-2.3ex]
\hline
\hline\\[-2.3ex]
\multicolumn{1}{c}{Galaxy} & Data Points & $\delta$ & $\log U_{0}$\\[0.25ex]
\hline\\[-2.3ex]
NGC\,7212 & 4  & -0.43$\pm$0.08 & -2.1  \\
NGC\,1386 & 6  & --0.58$\pm$0.03 & -2.7\\[0.1ex]
\hline\\[-2.3ex]
\end{tabular}
\end{center}
\end{minipage}
\end{table}
\begin{figure*}
\begin{center}
\includegraphics[width=6.2cm,angle=-90]{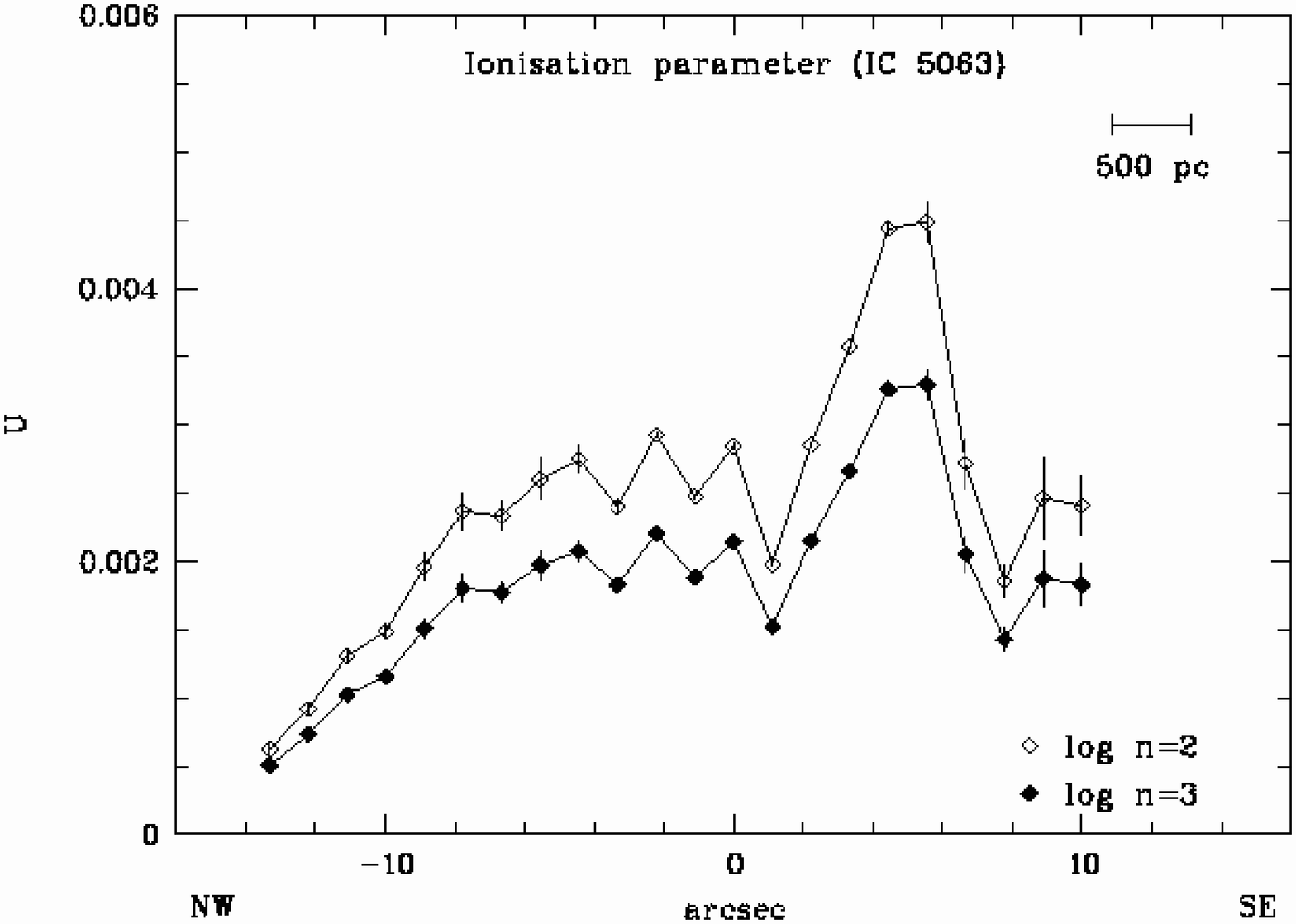}
\includegraphics[width=6.2cm,angle=-90]{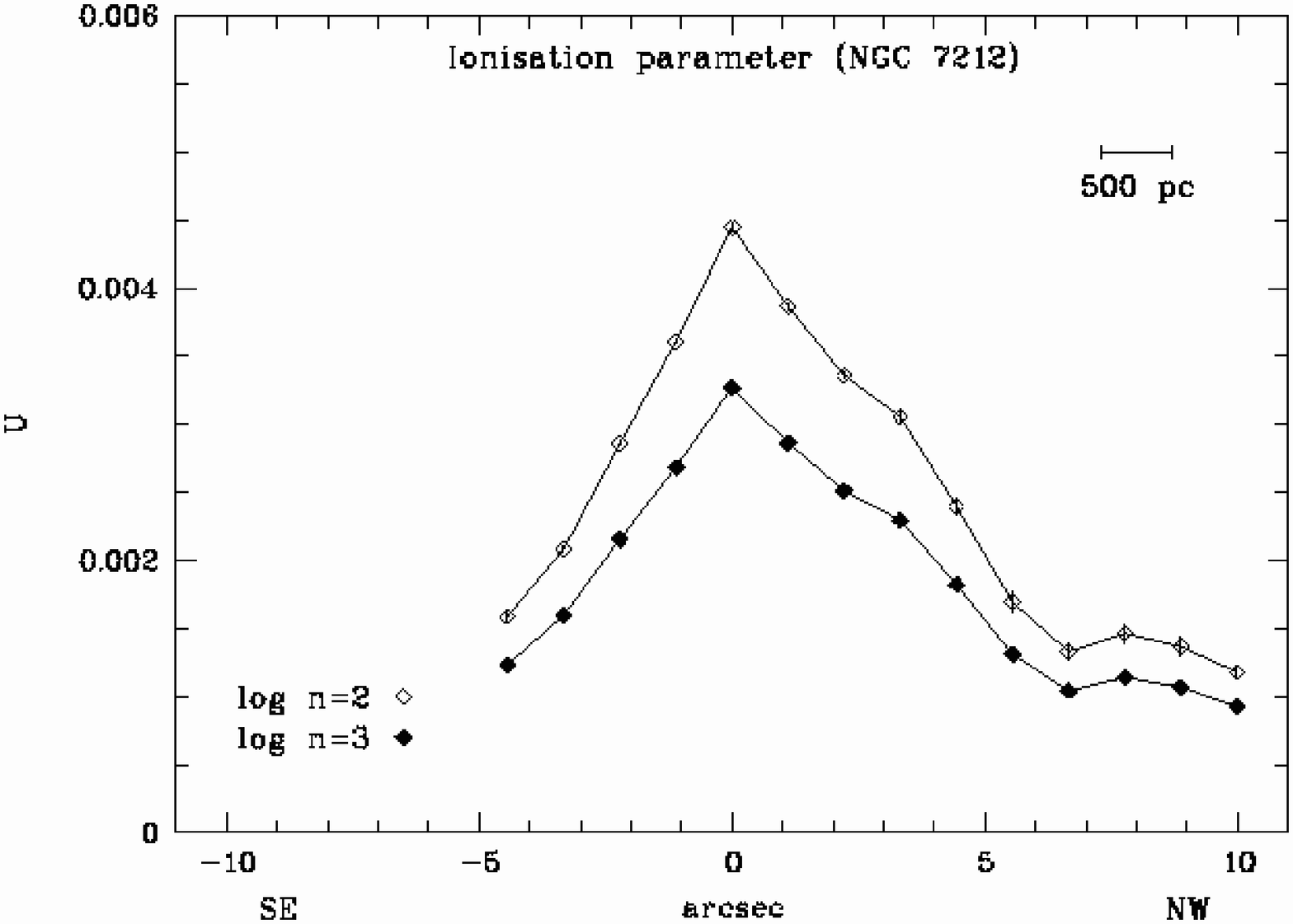}
\includegraphics[width=6.2cm,angle=-90]{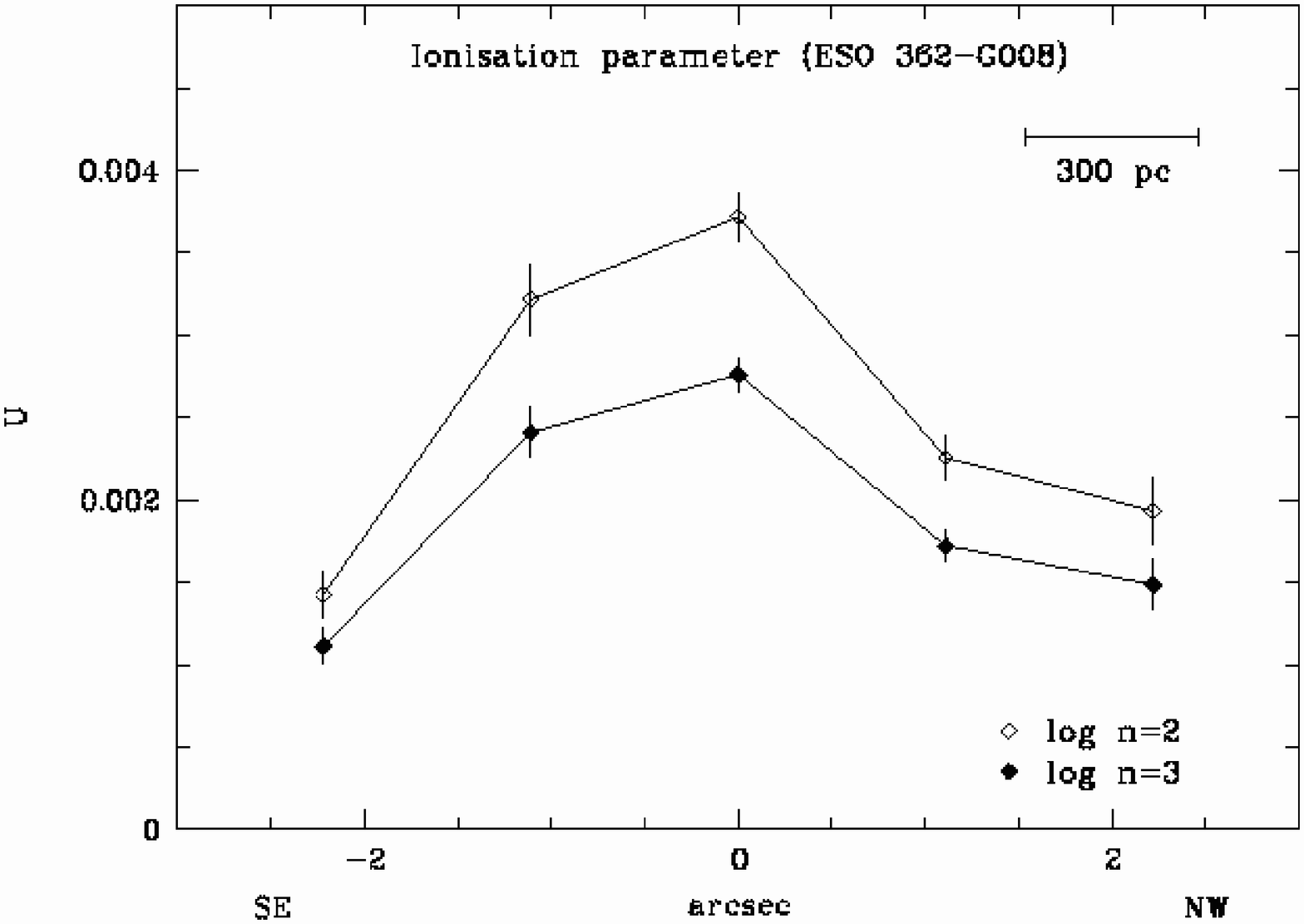}
\includegraphics[width=6.2cm,angle=-90]{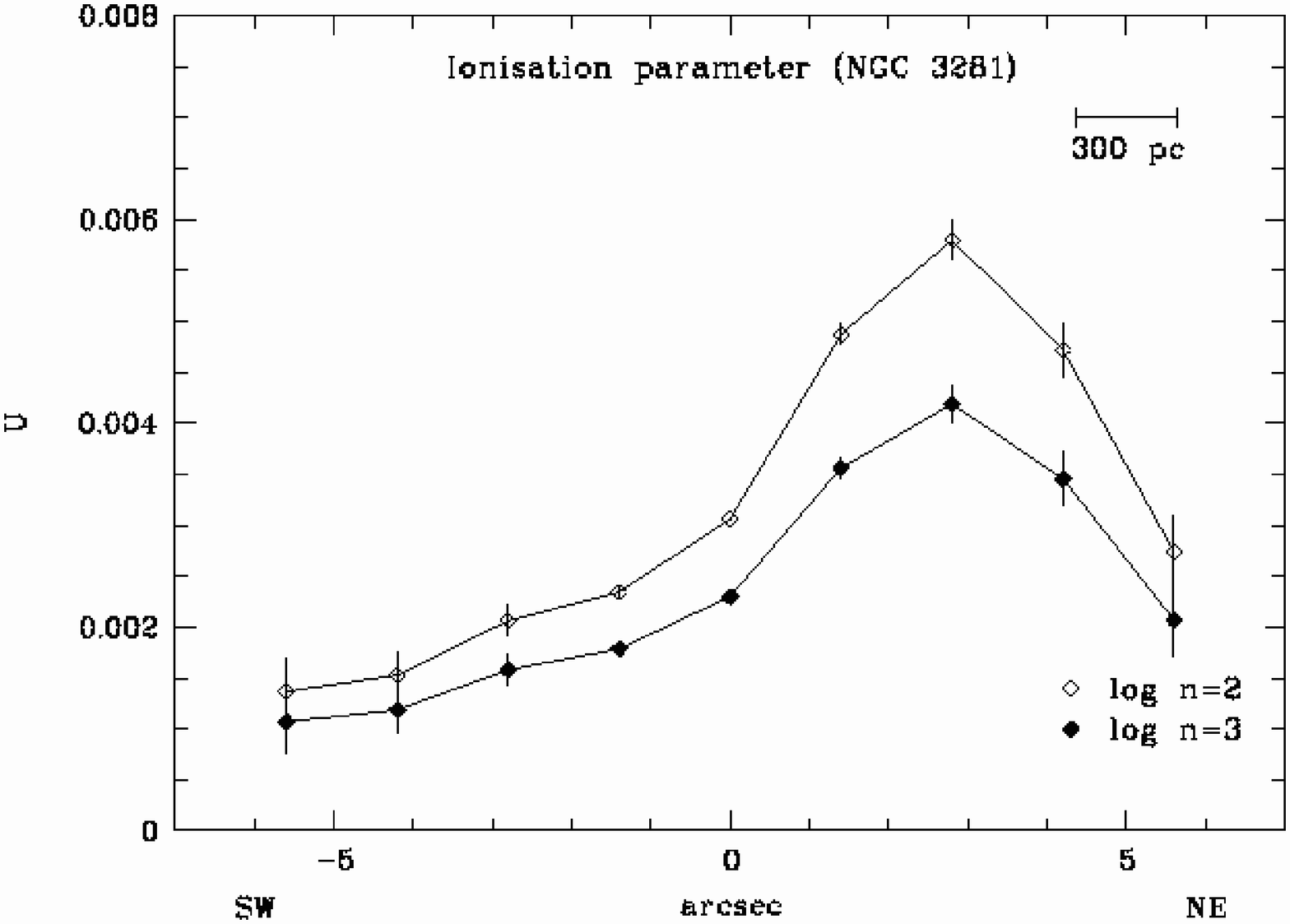}
\caption[]
{\label{ioni2} \small Ionisation parameter
derived from [\ion{O}{ii}]/[\ion{O}{iii}] ratio
as a function of the distance from the nucleus 
for IC\,5063, NGC\,7212, ESO\,362-G008, and NGC\,3281 (open symbols: $n_H$ = 100\,cm$^{-3}$,
filled ones: $n_H$ = 1000\,cm$^{-3}$).
The edge of the NLR as determined from the
  diagnostic diagrams is indicated by dotted lines. Note that for NGC\,5643, the [\ion{O}{ii}]
line was not included in the observed spectral range and we cannot present
the ionisation parameter. 
}
\end{center}
\end{figure*}

\subsection{Velocities}
\label{longvel}
\begin{figure*}
\begin{center}
\includegraphics[width=6.2cm,angle=-90]{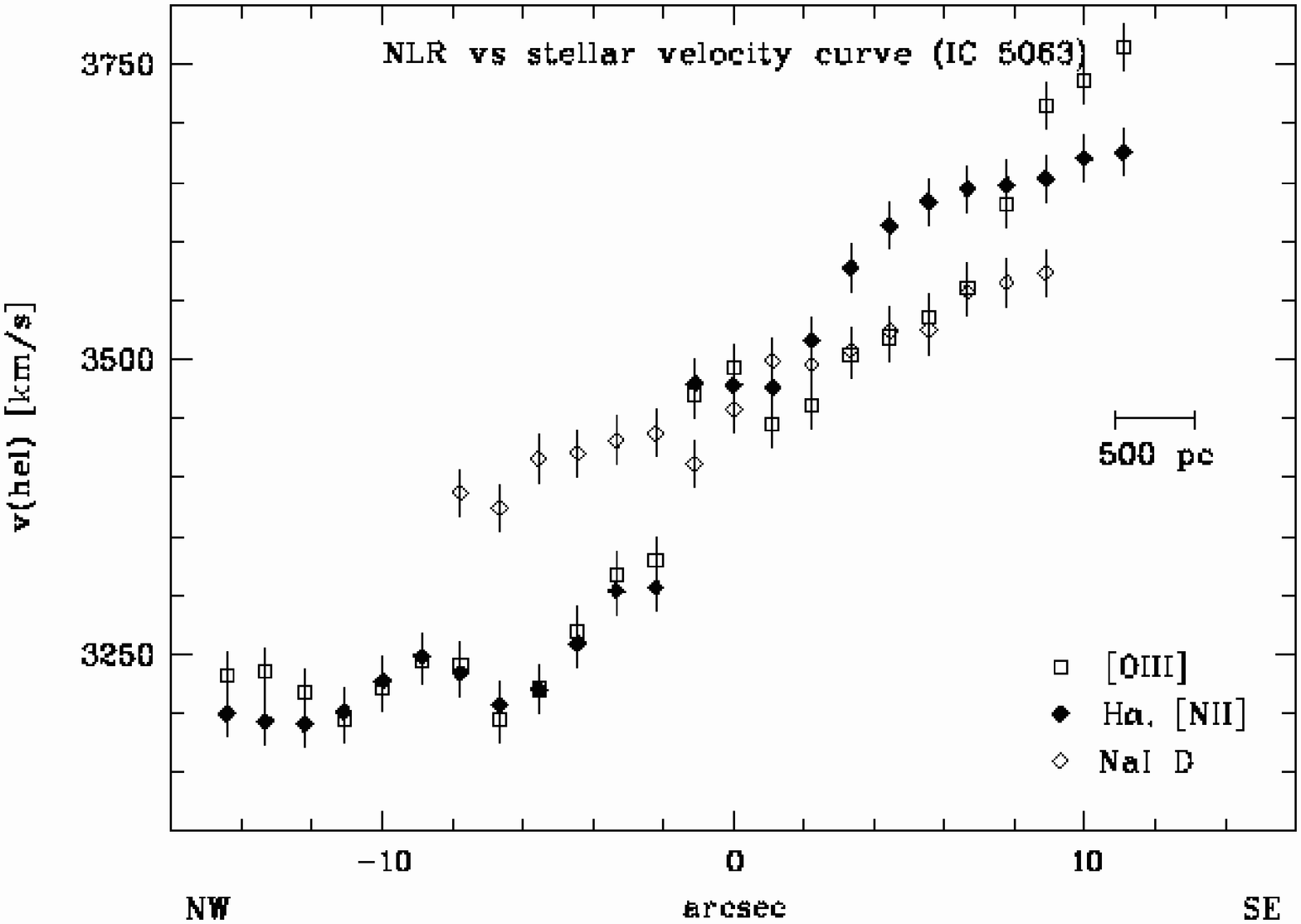}
\includegraphics[width=6.2cm,angle=-90]{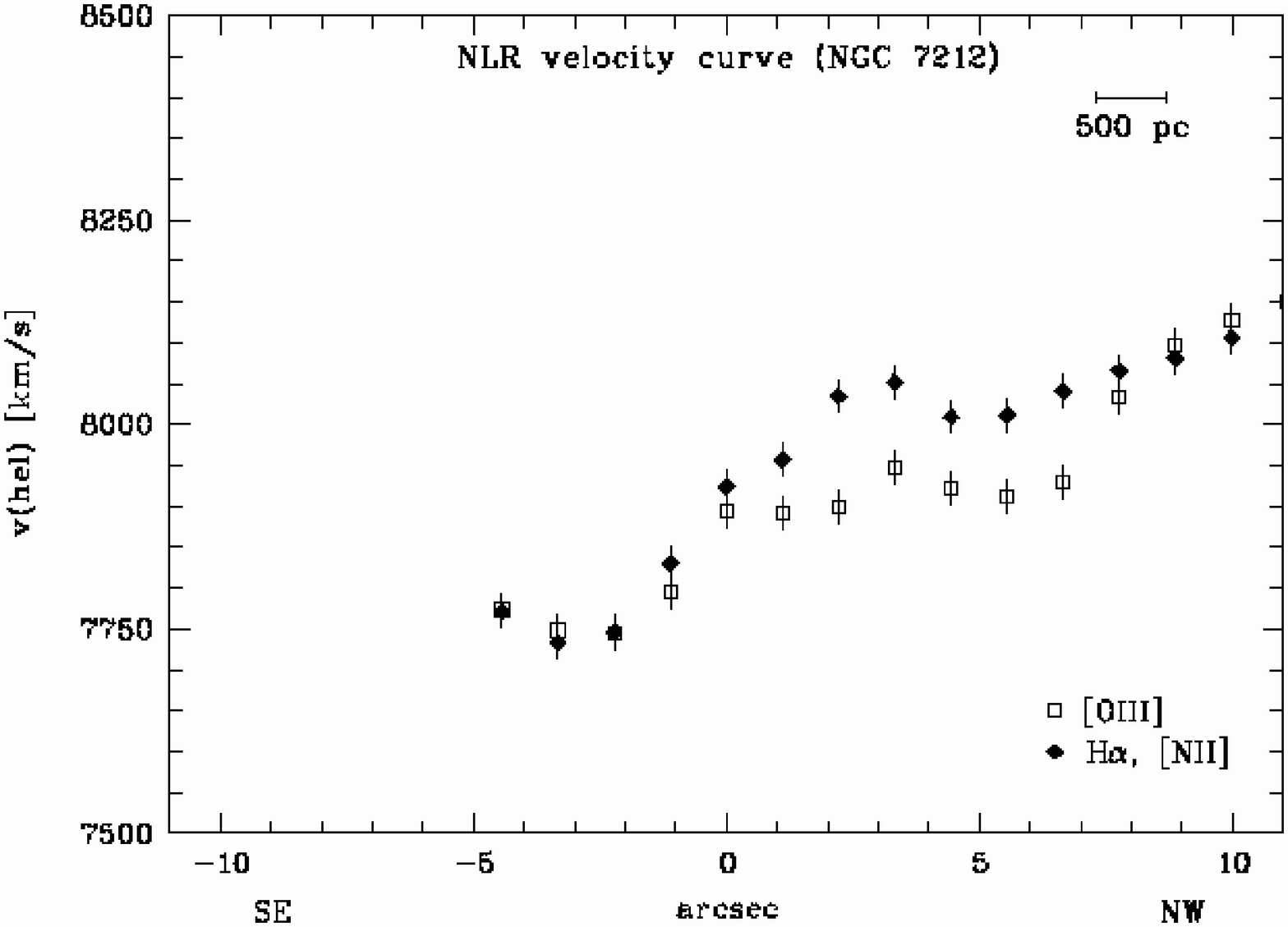}
\includegraphics[width=6.2cm,angle=-90]{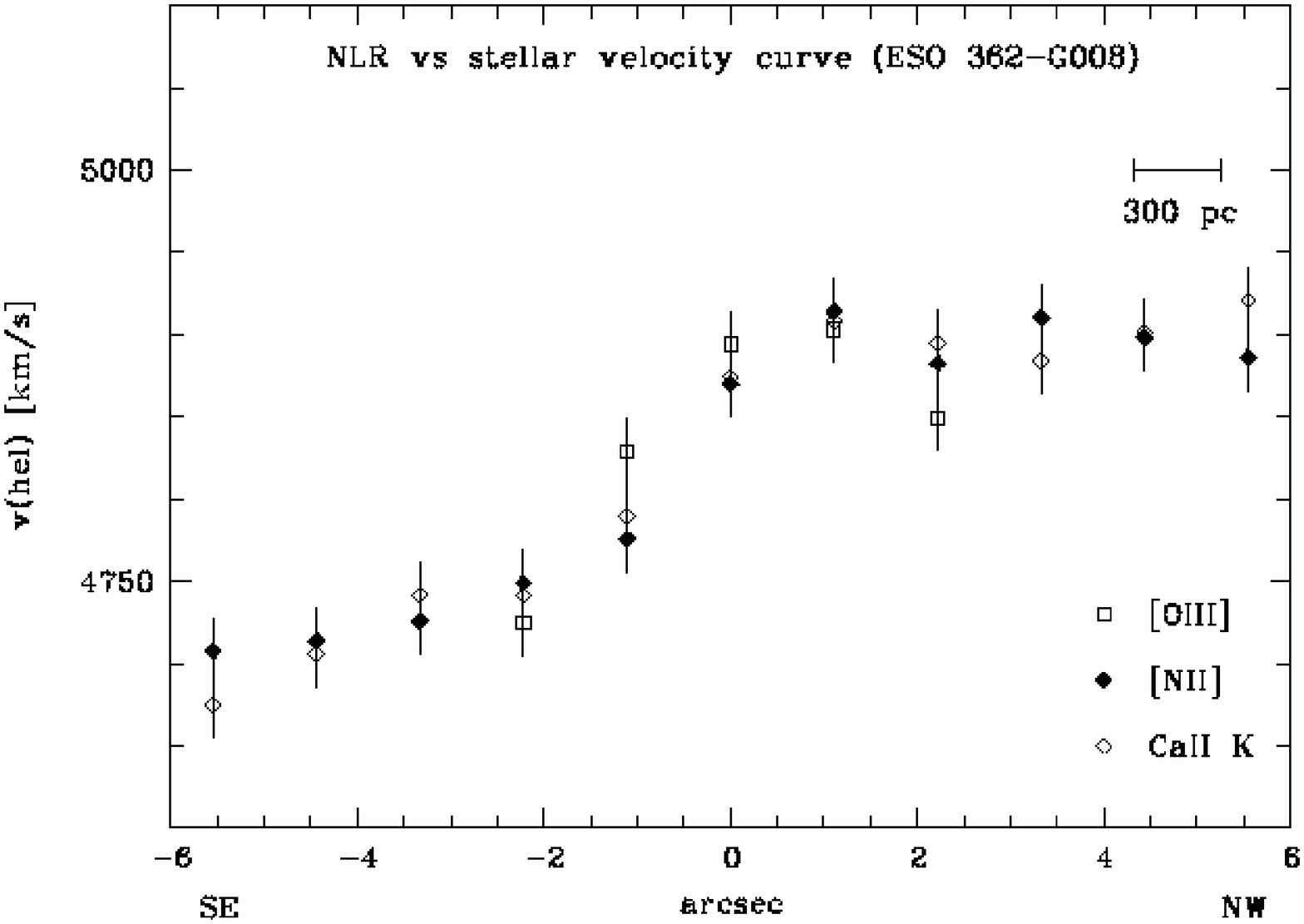}
\includegraphics[width=6.2cm,angle=-90]{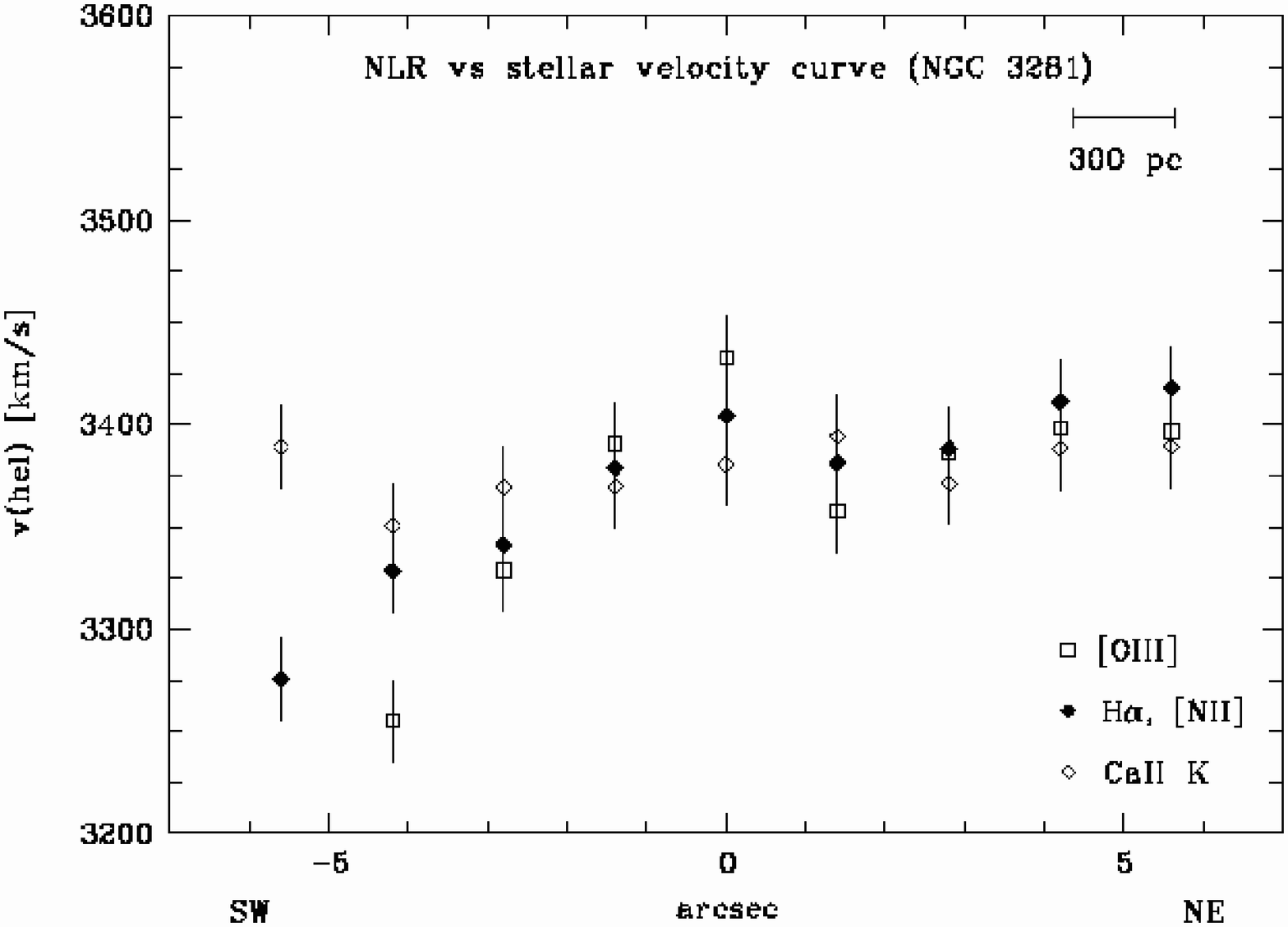}
\includegraphics[width=6.2cm,angle=-90]{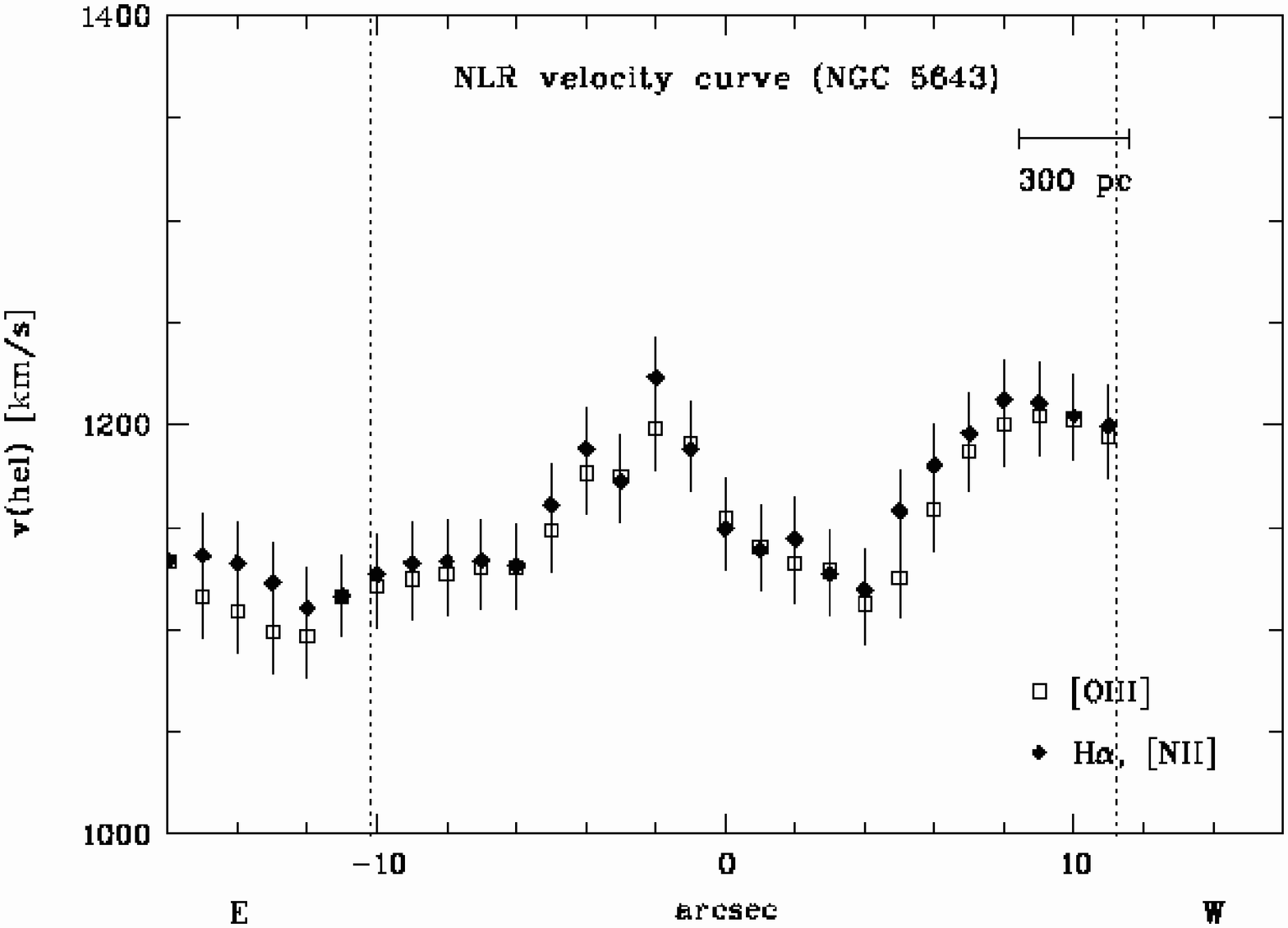}
\caption[]
{\label{vel2} \small
Velocity fields of the Seyfert-2 galaxies IC\,5063, NGC\,7212, 
 ESO\,362-G008, NGC\,3281, and NGC\,5643. The velocities of
the NLR were derived from the average value of the peak wavelengths
of the H$\alpha$ and [\ion{N}{ii}] emission lines (filled diamonds), with the
exception of ESO\,362-G008 where only [\ion{N}{ii}] was used.
The [OIII] velocities are also shown (open squares).
The stellar 
velocities were in most cases determined from the \ion{Ca}{ii}\,K absorption
line ``peak wavelength'' as seen in the ``raw'' spectrum (open diamonds) if
visible at a good S/N with
the exception of IC\,5063 where \ion{Na}{i}\,D was used instead.
For NGC\,5643, the edge of the NLR as determined from the
  diagnostic diagrams is indicated by dotted lines.
}
\end{center}
\end{figure*}

We derived the NLR line-of-sight velocity curve by taking the average of
velocity centroids derived by fitting Gaussians to H$\alpha$ and [\ion{N}{ii}]
emission lines. For comparison, we also show the velocity curve derived
in the same manner from the [\ion{O}{iii}] emission line
which may hint at NLR-jet interaction.
In addition, given the high S/N ratio of our spectra, we
were able to trace the stellar rotation curves from Gaussian fits to
the stellar absorption lines \ion{Ca}{ii} K or \ion{Na}{i} D for three objects
(before subtraction of the stellar template) throughout the whole region
as these lines are not blended with emission lines.
The results are shown in Fig.~\ref{vel2}. 
We estimated the uncertainty in determining
the velocity peaks to $\sim$20\,km\,s$^{-1}$ for both the emission and absorption
lines.
Note that for ESO\,362-G008, the H$\alpha$ line suffers from the 
underlying absorption line which was not
well subtracted in all regions and thus, only [\ion{N}{ii}] was used.
The velocity determined from the [\ion{O}{iii}] line 
is limited to the central 4\arcsec~due to low S/N in the outer parts.

Detailed interpretation of the NLR velocity curves is beyond the scope
of this paper. Even for very simple gas distributions, like a disk with
circular rotation, NLR line-of-sight velocity fields can be quite complex
due to collimation of the radiation field, dust obscuration and
projection effects
whose understanding requires modelling the 3D structure of the NLR
with many degrees of freedom (position angle and inclination of
the gaseous disk, opening angle, inclination and position angle and length
of the ionisation cone, radial density profile, dust distribution,
just to mention the most important ones). Moreover, outflows, random motions
or elliptical streaming due to barred potentials might complicate the
picture.

We will present such a modelling in a subsequent paper.
Here, we limit ourselves to point out that
all the galaxies show large-scale velocity gradients across their NLR.
Based on our preliminary modelling, we believe that to the zeroth order,
all of them can be explained by rotation (though we cannot rule out
outflow components in some cases). The rotation interpretation is
supported by the fact that in two objects for which we measure
the stellar rotation curve (IC\,5063, ESO\,362-G008), this curve has
similar behaviour as that for the gas:
although the stellar velocity gradient is shallower, the NLR gas shows
a component rotating in the same sense as the stars. The shallower slope
of the stellar rotation curve can result from a different line-of-sight
path through stellar and gaseous disks. Also the bulge stars modify the
stellar rotation curve: all our galaxies have large bulges and if both
bulge and disk stars are present along a given line-of-sight, the
velocity centroid of the
resulting absorption lines will be shifted
towards lower values compared to disk stars alone.

\section{Conclusions}
We study high-sensitivity spatially-resolved spectra along the extended
[\ion{O}{iii}] emission of six Seyfert-2 galaxies obtained with the VLT and the NTT. 
To derive the pure emission-line fluxes, we successfully
use the galaxy itself as stellar template to subtract
the underlying absorption lines in four out of six objects.

The nuclear spectra reveal the typical strong NLR emission from oxygen at
different ionisation states, lines from ionised nitrogen and sulphur,
as well as Balmer lines. In most objects, high-excitation iron lines are
additionally seen in the central spectra, originating from the
powerful and hard ionisation source in the centre. 

Plotting line-ratios from our spatially resolved spectra
in diagnostic diagrams, 
we observe a transition of emission-line ratios from the central AGN region
to \ion{H}{ii} region in another Seyfert-2 galaxy (NGC\,5643), in
addition to the Seyfert-2 galaxy NGC\,1386 already discussed in paper I. 
The most probable explanation for 
this transition is that the stellar ionisation field
starts to dominate that of the AGN. 
We are thus able to determine the radius of the NLR 
independent of sensitivity and excluding
[\ion{O}{iii}] contamination from circumnuclear starbursts.
In former spectroscopic studies, the observed [\ion{O}{iii}] has often been
attributed to the extended NLR. We can show that at least part of this
``extended NLR'' emission is actually powered by \ion{H}{ii} regions and
that only the central few arcseconds are indeed gas photoionised by the AGN.

In the other four objects, no such transition is observed but the line ratios
fall in the AGN regime in all three diagnostic diagrams. 
Thus, the determined NLR size (1000-5000\,pc)
is a lower limit, limited by either the S/N of our data or the lack of a
strong surrounding stellar ionisation field.

We derive physical parameters of the NLR such as reddening,
surface brightness, electron density, and ionisation parameter as
a function of projected distance from the nucleus.
The differences between the reddening distributions 
determined from the continuum slope and the Balmer
decrement argue in favour of  dust intrinsic to the NLR
with a varying column density along the line-of-sight.
In most cases, 
both electron density and ionisation parameter decrease with radius.

In all the objects, the gas rotation curve shows a large-scale
velocity gradient suggestive of rotation (though the detailed
modelling was not carried out and some outflow might be present too).
This is an important hint that the NLR gas is distributed in a disk
rather than a sphere, an issue that is still a matter of debate.

We discuss the results for each object (see Appendix).
In individual objects, 
substructures are seen in both the ionisation parameter and
electron density and can often be interpreted as signs of shocks from the
interaction of a radio jet and the NLR gas.

Our results for the five  Seyfert-2 galaxies show that the NLR properties
of the Seyfert-2 galaxy NGC\,1386 in paper I can be considered as prototypical,
putting our conclusions on a larger statistical basis.
We performed a similar study of the NLR of six Seyfert-1 galaxies.
The comparison with the results presented here will be
summarised in \citet{ben06c}.
All results presented here
are described and discussed in detail in \citet{ben05}.

\begin{acknowledgements}
We thank the anonymous referee for valuable comments and suggestions.
N.B. is grateful for financial support by the ``Studienstiftung
des deutschen Volkes''. B.J. acknowledges the support of the Research 
Training Network ``Euro3D-Promoting 3D Spectroscopy in Europe''
(European Commission, Human Potential Network Contract No. 
HPRN-CT-2002-00305) and of the Czech Science Foundation
(grant No. 202/01/D075). M.H. is supported by ``Nordrhein-Westf\"alische
Akademie der Wissenschaften''.
We give special thanks to Christian Leipski for 
providing the NTT spectra of NGC\,5643.
We thank Pierre Ferruit for providing and helping
us with the \texttt{fit/spec} line-fitting tool.
Henrique Schmitt was so kind to provide the continuum-subtracted
HST [\ion{O}{iii}] images of several Seyfert galaxies in this sample.
This research has made use of the NASA/IPAC Extragalactic Database (NED), 
operated by the Jet Propulsion Laboratory, Caltech, under contract with the NASA.
\end{acknowledgements}

\appendix
\section{Comments on Individual Objects}
\label{comments}
We searched the available literature for all objects in our sample and
here summarise the most important results in comparison with our study
(excluding NGC\,1386 which was discussed 
in paper I). Note that the velocity fields will be discussed in detail
when comparing them to those derived from modelling in a subsequent paper.

\subsection{IC\,5063}
\label{ic5063}
IC\,5063 is a Seyfert-2 galaxy with a hidden BLR seen in polarised light
\citep{ing93}. Its radio luminosity is two orders of magnitude greater
\citep{col91} than that typical of nearby Seyferts \citep{ulv84}. Thus,
IC\,5063 is also classified as narrow-line radio galaxy (NLRG).

\citet{ber83} studied the extended ionised gas with long-slit spectroscopy.
Ionised gas is reported out to a distance of 19 kpc
to the north-west of the nucleus ($\sim$60\arcsec~using their adopted distance
to the galaxy and
Hubble constant) and 9 kpc ($\sim$36\arcsec) to the south-east.
\citet{ber83} find both an increasing temperature
and ionisation parameter with increasing distance from the nucleus.  
Under the assumption of one central ionising source,
this implies that the gas density decreases faster than $n_e \propto r^{-2}$.

\citet{wag89} report the detection of an off-nuclear broad emission-line
region at 1\farcs8 north-west of the nucleus. They interpret this emission as
an extreme example of scattered nuclear emission with a very high intrinsic
line emission from the nucleus. The enhanced nuclear
activity, the irregular gas motions, and strong obscuration and scattering
effects are interpreted as IC\,5063 being a recent merger remnant. 

\citet{col91} used high spatial resolution optical and infrared imaging as
well as
optical spectroscopy to study the physical and kinematic conditions of the
ionised gas. They find a highly anisotropic ionising radiation field with a
conical morphology and an opening angle of 50\degr~along a p.a.~of 120\degr.
The cone axis coincides with the photometric  major axis of the galaxy (p.a.~= 
116\degr; RC3).
The ionised gas extends over $d$ $\simeq$ 69\arcsec. Inside the cone, the excitation
conditions are roughly uniform but drop very rapidly outside.
The results of \citet{col91} are similar to that of \citet{ber83}, confirming
a positive excitation gradient outwards from the nucleus. However,
the observed radial dependency of the electron density along p.a.~=
109\degr~suggest $n_{\rm e, obs}$$ \propto
R^{\delta}$ with $\delta = -0.6$. This is not consistent with the requirement of
$\delta$ $<$ -2, needed to explain the increasing ionisation parameter.~They 
suggest that a combined effect of decreasing density and abundance
may explain the increasing excitation outwards from the nucleus.
Several dust lanes are concentrated in the northern side of IC\,5063, running
approximately parallel to the major optical axis.
High-excitation [\ion{Fe}{vii}] and [\ion{Ca}{v}] emission lines are detected
in the central arcseconds.

The 3.6\,cm radio map reveals a linear triple radio structure extending over 4\arcsec, associated
with the NLR \citep{mor98}. At 21\,cm, broad blueshifted \ion{H}{i} absorption is visible, indicating a
fast net outflow.
In the central region, there is clear evidence for
a strong interaction between the radio jet and the ISM.
The shock scenario is supported by HST/NICMOS observations by \citet{kul98}
who find three emission-line regions in [\ion{Fe}{ii}], Pa$\alpha$, and H$_2$
along the major axis aligned with knots seen in radio emission.
The NICMOS data show a very red point source in the nucleus of IC\,5063,
interpreted as dust-obscured active nucleus.

A groundbased [\ion{O}{iii}] image was presented by \citet{mor98}, confirming
the result of \citet{col91} that the high-ionisation line-emitting gas has
an ``X-shaped'' morphology with a basic symmetry axis of
p.a.~$\sim$120\degr. The ionised gas can be traced out to a distance larger
than $r \sim$ 30\arcsec. Although the HST [\ion{O}{iii}] image from \citet{sch03} reveals the same
shape, 
the [\ion{O}{iii}] extension is the fifth part of the groundbased one
($r \sim 6$\arcsec). \citet{sch03} explain the small size by the limited
field-of-view of the linear-ramp filter ($\sim$13\arcsec) which is, however,
twice as large as the observed extension. Thus, it may rather be due to
the low sensitivity of the HST snapshot survey (600 s with the 2.4 m HST) compared to the
groundbased image (1200 s with the 3.6 m ESO telescope).

In our long-slit observations,
we detect [\ion{O}{iii}] emission at a S/N $>$3 out to
a distance of $r \sim$20\arcsec~from the nucleus in both the south-east and
north-west direction (Table~\ref{tablediag}). This is smaller than what has been
reported by \citet{ber83}, but they also find a decrease of line
intensities by a factor of 10-30 at a distance $>$6\arcsec~from the nucleus
and therefore average over a large spatial range to gain a constant S/N. The groundbased
[\ion{O}{iii}] image of \citet{mor98} also reveals a larger extension, but the HST
[\ion{O}{iii}] extension is three times smaller. This discrepancy once again
shows the need for alternative measures of the NLR size rather than 
using the [\ion{O}{iii}] extension alone.

Line ratios at a S/N $>$ 3 can be measured out to a distance of $r \sim
13$\arcsec~from the photometric centre and all fall in the upper right
corner in the three diagnostic diagrams (Fig.~\ref{diag1}). 
Moreover, they show a remarkably small scatter
compared to the AGN-typical line ratios of other galaxies, e.g.~NGC\,5643. 
We classify the corresponding gas as NLR, but we cannot exclude that the NLR
extends even further out where we are not able to measure line ratios to a high
accuracy. 

In the centre, IC\,5063 has the highest [\ion{O}{ii}]/H$\beta$ flux ratio ($F_{\rm dered}
\sim 2.96$) and at the same time the lowest [\ion{O}{iii}]/H$\beta$ ratio ($F_{\rm dered}
\sim 8.03$) of our sample (Table~\ref{lineratio1}). This translates to the lowest
ionisation parameter in the optical nucleus
(Table~\ref{result}). All values are indicative of a less strong
ionisation field in the centre of IC\,5063 compared to the other AGNs in our
sample. 
The central electron temperature is with $T_{\rm e, obs}$ $\sim$ 13865$\pm$1800\,K 
comparable to that of other type 2s in our sample as well as to the central
value measured by \citet{col91} within the errors ($T_{\rm e, obs}$$\sim 15000$\,K). 
High excitation lines such as [\ion{Fe}{vii}] and [\ion{Fe}{x}] are observed
in the central spectra.

IC\,5063 is the only object in our sample where we can follow the
[\ion{O}{iii}]\,$\lambda$4363 emission line at a S/N $>$ 3 
over several arcseconds from the nucleus
(from
$\sim$8\arcsec~north-west to $\sim$5\arcsec~south-east). The resulting temperature ranges
between $T_{\rm e, obs}$ $\sim 15310$\,K to $T_{\rm e, obs}$ $\sim 12245$\,K, without a clear tendency
with radius. The average value of the outer regions
is with $T_{\rm e, obs}$ $\sim$ 13789$\pm$290\,K close to
the central temperature of $T_{\rm e, obs}$ $\sim$13865$\pm$1800\,K.
We cannot confirm the results of \citet{ber83} who reported an increasing
temperature with distance from the nucleus, measured along a different p.a.~(310\degr~compared to our
p.a.~of 115\degr).

The central reddening determined from the Balmer decrement is 
the highest one found in our sample ($E_{B - V}$ $\sim$ 0.89$\pm$0.01\,mag). 
It probably coincides with the red point source seen in
NICMOS images of the nucleus of IC\,5063 \citep{kul98}.
The reddening remains high 1\arcsec~northwest and 
decreases outwards (Fig.~\ref{reddening2}). In the south-east, it reaches a minimum at
$\sim$4\arcsec~and then increases again to a value comparable to the central one
at a distance of 10\arcsec. This increase in the south-east
is not observed in the reddening
determined from the continuum slope relative to the stellar template (Fig.~\ref{reddening2}), 
indicating that it is most likely due to dust intrinsic to the
NLR. Overall, both reddening measures show a comparable reddening distribution, with the
continuum slope reddening covering a smaller range of $\Delta E_{B - V} \sim
0.2$\,mag, while the Balmer decrement yields a reddening distribution with an
amplitude of $\Delta E_{B - V} \sim 1$\,mag. We use the reddening determined
from the H$\alpha$/H$\beta$ ratio to correct the observed line ratios.

The electron-density distribution peaks at the centre 
($n_{\rm e, obs}$ $\sim$ 635$\pm$30\,cm$^{-3}$) and decreases outwards to the low density limit
with two subpeaks at $\sim$5\arcsec~on both sides of the optical nucleus.
This general behaviour is similar to the results of \citet{ber83} and
\citet{col91} with a central electron density comparable to that of
\citet{ber83} ($n_{\rm e, obs}$ $\sim 600$\,cm$^{-3}$) and
slightly higher compared to \citet{col91} ($n_{\rm e, obs}$ $\sim
540$\,cm$^{-3}$ without temperature correction versus their value of $n_{\rm e, obs}$ $\sim 450$\,cm$^{-3}$ along a comparable p.a.~of $\sim$110\degr; their Fig. 9a).

IC\,5063 is besides NGC\,3281 the second object in our sample where the
ionisation parameter does not peak at the optical nucleus. The highest
ionisation is observed instead at 5\farcs5 south-east of the nucleus.
This increase towards the south-east is in agreement with the increasing
ionisation parameter reported by \citet{ber83} and \citet{col91}, but towards
the north-west, we observe a decreasing ionisation parameter.
Interestingly, the position at which the ionisation parameter peaks is the
same at which the surface-brightness distributions start to increase again
outwards. Moreover, it coincides with the south-eastern region of enhanced
electron density. 

\citet{col91} consider combined effects of abundance and density
gradients as the most likely interpretation of these observations.
However, the line ratios in all three diagnostic diagrams show
a very small scatter, arguing against a steep abundance gradient.
Moreover, the slope of decreasing electron density $<$ 2 
needed to explain the increase in excitation conditions
is not observed, neither in their data nor in our data
($n_{\rm e, obs}$ $\propto R^{-1.1}$; Table~\ref{fitden}).
We instead favour a shock scenario from the interaction of the radio jet and the NLR gas.
This is strengthened by the close match between the radio map and the NLR as well as the 
blueshifted \ion{H}{i} absorption 
and the broad profile of the \ion{H}{i} absorption observed by \citet{mor98}.
The shocks may result in a locally increased density and ionise 
the surrounding medium, resulting in the increased
ionisation parameter and surface brightness. 
The shock scenario is supported by the emission-line profiles which show
substructures in the central 9\arcsec~with a blue asymmetry on
the north-west side and a red asymmetry on the south-east side of the centre.
The profiles get significantly broader especially
1-2\arcsec~on both sides of the nucleus.
Unfortunately, our observations are limited by spectral resolution to allow for a detailed
discussion of profile variations.
To probe the shock scenario, we looked for the presence of [\ion{Fe}{x}] emission
[e.g.~\citet{vie89}]:
It is negligible throughout most of the region, with the exception of the
central 2\arcsec~(centre: $F_{\rm dered}
\sim 0.02$, 1\arcsec~north-west: $F_{\rm dered}
\sim 0.003$, 1\arcsec~south-east: $F_{\rm dered}
\sim 0.001$, respectively) and at 3\arcsec~south-east ($F_{\rm dered}
\sim 0.001$), i.e.~close to the region where we observe an increased density.
However, we are limited by both the signal and the low resolution which makes it
difficult to disentangle the weak [\ion{Fe}{x}] from 
[\ion{O}{i}]\,$\lambda$6363\,\AA.

\subsection{NGC\,7212}
\label{ngc7212}
NGC\,7212 is a Seyfert-2 galaxy in a system of three interacting galaxies
\citep{was81}. 
Line emission from ionised gas extending over $\sim$17\farcs5~along both p.a.~of
37\degr~and 127\degr~has been reported by
\citet{dur90} as observed by means of optical long-slit spectroscopy in the
H$\beta$+[\ion{O}{iii}] wavelength range. They find a high excitation value $R$
= $I$([\ion{O}{iii}]\,$\lambda$5007\,\AA+4959\,\AA)/$I$(H$\beta$) = 19 in the nucleus, while the nebulosity is of
variable excitation with $R$ ranging from 5-28. They quote that the large
values are uncertain as H$\beta$ is close to the detection limit.

\citet{tra92} find a broad H$\alpha$ component in the
polarised light of NGC\,7212. However, \citet{tra95} argues that a significant
amount of polarisation is probably not intrinsic to the nucleus of NGC\,7212
but due to transmission through aligned dust grains in the host galaxy.
This is supported by several observational evidences, showing that dust
obscuration plays a significant role in the source. Moreover, the narrow
permitted and forbidden lines also posses a substantial amount of
polarisation.

\citet{tra95} finds a jet-like high-ionisation feature
extending up to 10\arcsec~from the nucleus at a p.a.~of
$\sim$170\degr~in groundbased [\ion{O}{iii}] and H$\alpha$ image, possibly due to
collimated radiation of the nucleus.
This direction coincides with a double radio source on a much smaller
spatial scale discovered by \citet{fal98} 
who compare HST and VLA observations of NGC\,7212.
The continuum image of \citet{fal98} shows multiple dust lanes.
The [\ion{O}{iii}] image exhibits extended emission out to $\sim$3\arcsec~from the
nucleus along p.a.~= 170\degr. The emission is diffuse and composed of several individual knots to
the north and south of the nucleus. 

Along a p.a.~of 170\degr, we observe [\ion{O}{iii}] emission
extending out to 12\arcsec~from the nucleus (Table~\ref{tablediag}), i.e.~four times
larger than the extension seen in the HST image in the same direction. 
It is somewhat smaller than 
the maximum extent observed by \citet{dur90} (p.a.~of 127\degr~and 37\degr). 
The excitation value we observe in the
central spectrum is with $R_{\rm obs} \sim$ 17 comparable high to what has been observed by
\citet{dur90}. It is the highest value in our type-2 sample. 
Note that the [\ion{O}{iii}]/H$\beta$ ratio given in
Table~\ref{lineratio1} does not include the [\ion{O}{iii}]\,$\lambda$4959\,\AA~line
which is one third of the flux of the [\ion{O}{iii}]\,$\lambda$5007\,\AA~line. Thus, one third
of $F_{\rm obs}$ given in Table~\ref{lineratio1} has to be added to gain $R$.
The reddening-corrected value in the centre
is $R_{\rm dered} \sim$ 16 and varies between 6 and 17 in the central
24\arcsec~region, i.e.~it stays high in the whole region which we classify as
NLR. Emission-line ratios at a S/N $>$ 3
were obtained out to $\sim$5\arcsec~south-east of the nucleus and $\sim$10\arcsec~north-west.

The reddening in the centre is rather low ($E_{B - V}$ = 0.33$\pm$0.01\,mag)
and decreases to a value of $\sim0.07$\,mag at 1\arcsec~north-west
of  the nucleus (Fig.~\ref{reddening2}). 
On both sides of this region, it increases and reaches its
maximum value at 4\arcsec~south-east and 7\arcsec~north-west of the
photometric centre ($\Delta E_{B - V}$ $\sim$ 1\,mag). These maxima may be
attributed to dust lines seen in the continuum image by \citet{fal98}.

The surface brightness is highest at 1\arcsec~south-east of the
centre (Fig.~\ref{lum2}). 
Although the highest [\ion{O}{iii}] and H$\alpha$ flux as well as the
highest continuum is observed at 0\arcsec~(this is how we defined the photometric centre),
the reddening-corrected luminosities peak at 1\arcsec~south-east due to the
higher reddening observed in this part. The surface-brightness distributions decrease
outwards and show a secondary maximum at $\sim$6\arcsec~north-west of the
nucleus. In the same region, the highest $R_{\rm dered}$ value of 17 is observed.

Both electron density and ionisation parameter show a slight increase at
$\sim$8\arcsec~north-west of the nucleus.
As the radio maps of NGC\,7212 show a double radio source extending in the
direction of our long-slit observations \citep{fal98},
it is probable that the radio jet
interacts with the NLR, resulting in the observed 
enhanced surface brightness, electron
density, ionisation parameter, and the high
excitation at $\sim$6-8\arcsec~north-west of the nucleus.

\subsection{ESO\,362-G008}
\label{eso362}
Groundbased images of this Seyfert--2 galaxy were studied by
\citet{mul96}, revealing a very red continuum nucleus.
The [\ion{O}{iii}] emission is strongest in the nucleus
and extends out to $\sim$25\arcsec~along the galaxy disk ($\sim$158\degr, see
Fig.~\ref{galaxies2})
and $\sim$10\arcsec~along the galactic minor axis. The regions of high excitation gas are distributed in
a cross-like morphology, with the highest ratio corresponding to an
off-nuclear cloud to the north-east.
\citet{col96} interpret the presence of gas far out of the disk
as sign of a large-scale outflow occurring in ESO\,362-G008.

\citet{fra00} studied the extended NLR of the Seyfert-2 galaxy ESO\,362-G008 in detail to compare it
to that of the Seyfert-1 galaxy MCG\,-05-13-017.  
They use a stellar population template obtained from averaging the
extranuclear spectra. In the
nucleus, a dilution of an intermediate-age burst of star formation is
found. \citet{fra00} 
find no evidence of a featureless continuum contributing more than 5\% in the
near-UV, in agreement with the results of \citet{cid01}.
Emission-line fluxes are measured along p.a.~= 60\degr~out to 14\arcsec~from the
nucleus. The H$\beta$ emission line is too weak to be measured, often not
filling the absorption feature, indicating that the stellar template does not
match the younger stellar population of the nuclear region.
All ratios show a symmetric behaviour on both sides of the
nucleus. The increasing
[\ion{O}{ii}]/[\ion{O}{iii}] ratio indicates a decreasing ionisation parameter.

ESO\,362-G008 is the only object in our sample in which  strong
underlying stellar absorption lines remain even after subtraction of a template
determined in the outer parts of the galaxy (Table~\ref{tabletemplate};
Fig~\ref{figtemplate}). 
This hints the existence of a nuclear starburst as is confirmed by the stellar population
synthesis of \citet{cid98} and the results of \citet{sto00}:
They describe ESO\,362-G008 as relatively evolved nuclear
starburst due to its high order Balmer absorption lines. The continuum is very
red due to a dust lane crossing the nuclear region as seen by \citet{mal98} in
broadband HST images.

As a consequence, our results have to be taken with some caution as the 
H$\beta$ and H$\alpha$ emission are most probably underestimated.
We indeed often see the underlying absorption trough in both lines.
Also \citet{fra00} who use a stellar population template averaged from
extranuclear spectra report dilution of an intermediate-age burst of star
formation, resulting in a mismatch of the stellar template and the younger
stellar population of the nuclear region where the H$\beta$ line is often not
filling the absorption feature.

ESO\,362-G008 is also the object in our sample with the smallest detectable 
[\ion{O}{iii}] extent ($r \sim 4$\arcsec). Moreover, due to the low S/N, the line ratio
study is limited to the central $r \sim 3$\arcsec.
Interestingly, the groundbased image reveals a more than twice as large 
[\ion{O}{iii}] extension (along roughly the same p.a.). \citet{fra00} study
emission lines out to $r \sim$14\arcsec~from the nucleus but along p.a.~=
60\degr~and are also confined to the inner $d \sim 10$\arcsec~at a p.a.~of 160\degr.

We decided not to apply more sophisticated stellar template
corrections due to our limited S/N. 
All line ratios fall in the AGN regime in the diagnostic
diagrams, suggesting that the NLR extends out to at least 3\arcsec~radius from
the centre (Fig.~\ref{diag1}).
The line ratios have rather high values on the x-axis, most
probably due to
the underestimated H$\alpha$ line flux.
Due to the remaining underlying absorption lines in the central spectra, 
only emission-line fluxes from the strongest lines 
[\ion{O}{ii}], H$\beta$, [\ion{O}{iii}],
[\ion{O}{i}], H$\alpha$, [\ion{N}{ii}], and [\ion{S}{ii}] can be derived
(Table~\ref{lineratio1}). 

The reddening determined from the continuum slope relative to the
template shows a similar distribution
as the reddening distribution using the Balmer decrement 
with the highest reddening value in the centre and a slow decrease to the
outer parts (Fig.~\ref{reddening2}). 
Note that the reddening of
the continuum slope shown in Fig.~\ref{reddening2} was set arbitrarily to zero for comparison.
The reddening scatter derived from the continuum slope is small ($\Delta E_{B - V} \sim
0.07$\,mag), while a $\sim$10 times higher range is obtained using the
H$\alpha$/H$\beta$ ratio ($\Delta E_{B - V} \sim 0.6$\,mag). These differences
can be explained first by the relative reddening value that was obtained in 
case of the continuum slope, i.e.~both outer template and central spectra
suffer similar dust extinction which do not reflect in the derived
value. Second, dust may be intrinsic to the NLR. 
We used the reddening determined by the emission-line ratio for correction.

The decrease of the ionisation parameter we observe has already been suggested by \citet{fra00} based
on the increase of the [\ion{O}{ii}]/[\ion{O}{iii}] ratio, at least in the
central $r \sim 5$\arcsec~(their Fig. 11).

\subsection{NGC\,3281}
\label{ngc3281}
\citet{dur88} discovered the extended ionised envelope of
$d \sim$21\arcsec~in NGC\,3281.~They find a large reddening value of
$E_{(B-V)}$ = 0.78\,mag in the nuclear region. The derived electron temperature
equals to 28000 K in the high excitation zone.

NGC\,3281 has been classified as a proto-typical Seyfert-2 galaxy
by \citet{sto92} as it clearly shows features expected
from the unified model: an ionisation cone, heavy obscuration towards
the (hidden) nucleus, a wind outflow along the cone, and emission-line ratios
consistent with photoionisation by a power-law continuum. They
carried out the most extensive study of NGC\,3281 including
direct images in continuum, [\ion{O}{iii}], and H$\alpha$+[\ion{N}{ii}]
as well as long-slit spectroscopy at several slit positions.
The [\ion{O}{iii}] emission resembles the projection of a cone.
The excitation map shows that either the reddening decreases from south to
north or the excitation increases.
The reddening distribution indicates that the nucleus is hidden by a dust
lane. This is strengthened by nuclear spectra which show no dilution by a
featureless continuum.
The stellar population is old and typical of early-type galaxies. 
An S2 template from synthetic spectra by \citet{bic88} was used to subtract
the absorption lines and reddened to match several spectra which where
heavily reddened. 
The stellar population seems not to vary much in the
inner 22\arcsec~$\times$ 10\arcsec~region as suggested from the uniform equivalent
width of absorption lines with the exception of \ion{Na}{i} D. \citet{sto92}
interpret this observation by additional contribution to  \ion{Na}{i} D
from interstellar absorption: Loci of higher \ion{Na}{i} D equivalent widths correspond to
regions with high $E_{(B-V)}$.
The electron densities are found to be highest around the apex of the cone and
decreasing with distance. 
The emission-line ratios are well described by photoionisation models with
varying ionisation parameter. It seems that the ionisation
parameter increases away from the apex along the axis of the cone, possibly
due to a decrease of the density faster than $r^{-2}$.

The HST [\ion{O}{iii}] image resembles the groundbased one \citep{sch03}.
It reveals a conically shaped NLR with opening angle of $\sim$80\degr~towards the
north-east. The emission extends by 6\farcs1 towards the north-south direction and
3\farcs9 along the cone axis.~Compared to the groundbased image, this
extension is less than half the one measured by \citet{sto92}. This
discrepancy is explained by \citet{sch03} by the limited field-of-view of
the linear-ramp filter ($\sim$13\arcsec), but can also be due to less sensitivity of the
800\,s exposure taken with the 2.4\,m HST mirror compared to the 900\,s exposure using
the 4\,m CTIO telescope. The counter ionisation cone is mostly hidden by
the host galaxy disk and only seen as a small blob of emission 4\farcs5 south
of the nucleus. The [\ion{O}{iii}] emission is nearly perpendicular to the
photometric major axis of the host galaxy (p.a.~= 140\degr, RC3).

\citet{vig02} study the broad-band X-ray spectrum of NGC\,3281, revealing
its Compton-thick nature. The nuclear continuum is heavily absorbed (column
density $\sim$ 2 $\cdot$ 10$^{24}$\,cm$^{-2}$).

In our longslit spectroscopy, the [\ion{O}{iii}] line emission at a S/N $>$ 3 can be traced out to a distance of
$r \sim$ 9\arcsec~from the nucleus, i.e.~three times as far as the maximum
radius seen in the HST [\ion{O}{iii}] image of \citet{sch03}.
It is comparable to what has been found by \citet{dur88} and \citet{sto92}.
Line ratios with a S/N $>$ 3 have been measured in the central $r \sim
5$\arcsec, showing values typical for AGN ionisation in all three diagnostic
diagrams. Thus, the NLR extends out to at least $r \sim$ 5\arcsec~from the
optical nucleus. 

The nuclear line intensity ratios relative to H$\beta$ are comparable to those of the
other Seyfert-2 galaxies in our sample (Table~\ref{lineratio1}).
However, NGC\,3281 has the lowest nuclear electron density of the type-2 sample ($n_{\rm e, obs}$
$\sim 540\pm40$\,cm$^{-3}$; Table~\ref{result}). In addition,
the ionisation parameter determined in the nuclear spectrum is rather low.
The central temperature is
also the lowest observed with $T_{\rm e, obs}$ $\sim 13715\pm440$\,K. This is
significantly lower to what has been reported by \citet{dur88} ($T_{\rm e, obs}$ $\sim 28000$\,K).

The similarity in the reddening distribution we determined from  the H$\alpha$/H$\beta$ line ratio 
as well as the continuum indicates that both the continuum and the emission lines are
suffering extinction from foreground dust, e.g.~the dust lane seen south-west
of the nucleus \citep{sto92}.
The surface-brightness distributions fall smoothly with distance from the
centre and show a secondary peak at 3\arcsec~north-east of the nucleus in the
emission lines (Fig.~\ref{lum2}). 
In the outer parts south-west of the photometric centre, 
the H$\alpha$ surface brightness approaches the value of [\ion{O}{iii}],
indicating that the NLR may end somewhere close by. Unfortunately, we are
limited by the S/N to observe a transition towards \ion{H}{ii} regions. 
In Fig.~\ref{density2}, the electron-density
distribution in NGC\,3281 is shown. It peaks at the centre and decreases slowly
outwards down to $n_{\rm e, obs}$ $\sim 110-240$\,cm$^{-3}$, in agreement with the results
of \citet{sto92}.

The ionisation parameter presented in Fig.~\ref{ioni2} reaches the maximum value at 3\arcsec~north-east of the centre,
  coinciding with the secondary peak in the emission-line surface-brightness distribution.
The increase towards the north-east is in
  agreement with the results of \citet{sto92}, but they also find an increase
  towards the south-west which we cannot confirm. It seems that this increase
  starts at a distance $>$5\arcsec~south-west from the nucleus where we do not
  trace the ionisation parameter any more due to limited signal.
 As the electron density
  does not drop off faster than $r^{-2}$  [neither in our observations nor in
  those of \citet{sto92}], this
  straightforward explanation of the observed increasing ionisation parameter
  with distance can be ruled out.
\citet{sto92} suggest that the gas near the apex of
the cone where the reddening is large sees a partially obscured
nuclear ionising source, resulting in a low ionisation parameter.
Comparing \texttt{CLOUDY} 
photoionisation modelling with the observed emission-line
  ratios, they rule out shocks as primary ionisation mechanism. Moreover, no
  extended radio emission is observed which could hint the existence of radio
  jets \citep{sch01}. 
Our observations do not help to further elucidate the
origin of the increased ionisation parameter towards the north-east.

We do not see double-peaked [\ion{O}{iii}] emission within the cone
(i.e.~from the centre towards the north-east), probably
due to the low spectral resolution. Indeed, the profiles are very broad
in the centre and out to $\sim$3\arcsec~north-west.

\subsection{NGC\,5643}
\label{ngc5643}
The Seyfert-2 galaxy NGC\,5643 is a barred spiral galaxy.
It was studied in detail by \citet{sch94} by means of groundbased narrow-band
imaging in [\ion{O}{iii}] and H$\alpha$+[\ion{N}{ii}] and long-slit spectroscopy.
High excitation gas in a bi-conical morphology
is elongated along the bar (p.a.~= 90\degr) out to $\sim$16\arcsec~on each
side of the nucleus and out to $\sim$7\arcsec~perpendicular. 
Further out, the H$\alpha$ image reveals
several \ion{H}{ii} regions on both edges of the bar.
\citet{sch94} speculate that the active nucleus is hidden $\sim$3\arcsec~west of
the continuum peak where both the highest reddening value and the highest
gas density are observed. NGC\,5643 is another Seyfert-2 galaxy fitting the
unified model with the torus obscuring the nucleus and collimating the
ionising radiation leading to the bi-conical morphology. In the central region,
the stellar population is old corresponding to an S3 to S4 template, while in
the \ion{H}{ii} regions, the stellar population is younger
(S6-S7). 
\citet{sch94} did not use a
stellar template to correct for the underlying absorption
lines. Instead, only the H$\alpha$ and H$\beta$ lines were corrected using the
corresponding equivalent widths. Dilution from a blue continuum probably due
to scattered light is found in the inner 10\arcsec.
The ratio [\ion{O}{iii}]/[\ion{O}{ii}] was
used as tracer of the ionisation parameter. It is high throughout the extended
NLR, decreasing more or less abruptly at its edges.
\citet{sch94} found increasing oxygen (O) and nitrogen (N) 
abundances towards the centre where the O/H ratio reaches 
solar and N/H twice the solar value.

HST images in [\ion{O}{iii}] and H$\alpha$+[\ion{N}{ii}] were taken by
\citet{sim97} to study the NLR with a high-resolution of 0\farcs1
(Fig.~\ref{galaxies2}).
The one-sided conical distribution of the
high-excitation gas is clearly seen. The peak of the red continuum emission
coincides with the apex of the cone. The radio structure seen by \citet{mor85}
is closely aligned with the overall shape of emission. 
A more detailed VLA A-array radio map (8.4\,GHz) with radio structure
closely matching the distribution of the NLR gas is presented by
\citet{lei06b}. It shows a diffuse radio jet extended by $\sim$15\arcsec~to both
sides of the centre (while in optical wavelength, the counter cone to the west
is absorbed by dust).

\citet{sim97} suggest that
the absence of a visible counter cone is the consequence of its obscuration by
the galaxy disk. An extended blue continuum region is seen out to 0\farcs9 east of the 
nucleus. It is not clear whether this emission is scattered nuclear continuum
or due to e.g.~a stellar ionising continuum.

NGC\,5643 was
included in the sample of 18 Seyfert galaxy studied by \citet{fra03}. They
present density and reddening as a function of distance from the nucleus as
well as surface-brightness distributions. However, their data have a
significant lower signal-to-noise ratio (S/N)
than our data and, moreover, they
did not take into account the
underlying absorption owing to the contribution of the stellar population
which we show to be important.

We detect [\ion{O}{iii}] emission at a S/N $>$ 3 out to a distance of
$\pm$16\arcsec~from the nucleus (Table~\ref{tablediag}), i.e.~comparable to the
results of \citet{sch94}. However, only the central
$\pm$11\arcsec~originate from the NLR as can be seen in the first and third
diagnostic diagram (Fig.~\ref{diag1}). 
Further out, the emission originates from circumnuclear
\ion{H}{ii} regions, in agreement with the results of \citet{tsv95} who 
list 214 \ion{H}{ii} regions in NGC\,5643, most of them
distributed in a ring-like structure at 20-60\arcsec~from the nucleus,
suggesting that current star formation is occurring in the nearly circular
spiral arms. At a
p.a.~of 90\degr~along the bar, 
the nearest \ion{H}{ii} regions are seen $\sim$10\arcsec~east of the nucleus.
Indeed, we observe a transition between line ratios typical for AGN
ionisation and \ion{H}{ii}-region like ionisation occurring at a distance of
11\arcsec~from the centre, thus determining the size of the NLR to 11\arcsec~($\simeq$
1050\,pc). This size coincides with the region in which \citet{sch94} report
dilution from a blue continuum. 
While \citet{fra03} detect line emission which they classify as extended NLR
out to a distance of $\sim$15-20\arcsec, our analysis shows that the extended
emission beyond 11\arcsec~originate from circumnuclear \ion{H}{ii} regions and
can therefore not be attributed to the NLR.
In addition to the central [\ion{O}{iii}] emission, line-emission 
is found at distances of 47$\pm$3\arcsec~west
from the centre as well as 36$\pm$5\arcsec~east 
and again 73$\pm$2\arcsec~east from the photometric centre, attributable 
to \ion{H}{ii} regions identified by \citet{tsv95} and \citet{sch94}. This is strengthened by the line
ratios which all fall in the \ion{H}{ii}-region regime in the diagnostic
diagrams (open and filled diamonds in Fig.~\ref{diag1}).
When comparing the line ratios of these \ion{H}{ii} regions with those of the
circumnuclear \ion{H}{ii} regions, it is notable, that the ones from the
centre (out to 16\arcsec~east) slowly approaches the outer ones in terms of the
[\ion{O}{iii}]/H$\beta$ ratio. It shows that the transition between NLR and
circumnuclear \ion{H}{ii} regions is not abrupt but that the domi\-nating
ionisation field slowly changes from the central AGN to that of the
circumnuclear stellar one. At 16\arcsec~east (marked with the small letter ``p''
in Fig.~\ref{diag1}), the line ratios are
finally identically with those observed in the \ion{H}{ii} regions in the
spiral arms. This observation strengthens 
our interpretation of the observed transition as true border between NLR and
surrounding \ion{H}{ii} regions. 

The reddening is  highest in the centre
and falls quickly towards the east and also 2\arcsec~to the west where it then
again rises and stays high throughout the western
NLR (Fig.~\ref{reddening2}). The high reddening in
the west may originate from obscuration of dust lanes seen in broadband
NICMOS/WFPC colour maps \citep{sim97,qui99}. The obscuration from dust seems to be
responsible for the one-sided ionisation cone structure seen in the HST
images of \citet{sim97}. 
\citet{sch94} observe the highest reddening at
3\arcsec~west of the continuum peak and speculate that the AGN is hidden
there while we observe the highest reddening value in the centre. 
However, one has to take into account their spatial resolution of 2\arcsec~compared to 
our spatial resolution of $\sim$1\arcsec. Moreover, the coincidence of the highest
reddening with the most luminous spectrum (continuum, H$\alpha$ and [\ion{O}{iii}])
we find is in agreement with the schematic model proposed by \citet{sim97} (their
Fig. 6). 

\begin{figure}  
\begin{center}  
\includegraphics[width=9cm,angle=0]{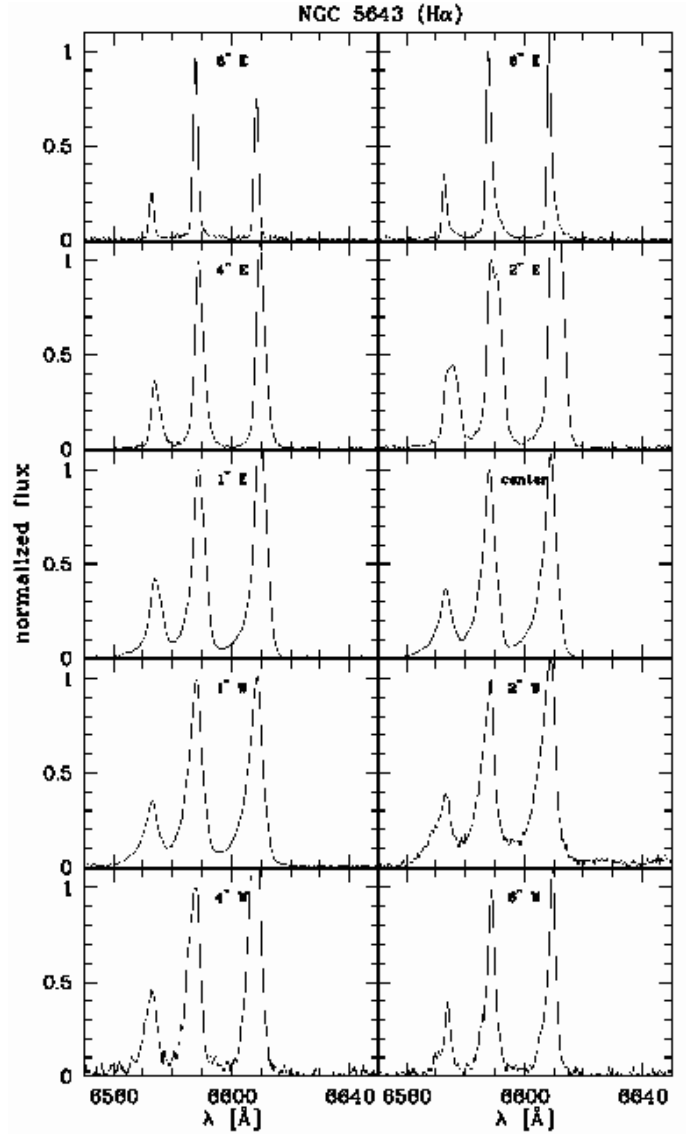}
\caption[H$\alpha$-profile variations in NGC\,5643]
{\label{profile5643} \small
 H$\alpha$ profile variations in NGC\,5643. The peak flux
  of H$\alpha$ is normalised to 1.}
\end{center}  
\end{figure}

The electron density is slightly higher 1\arcsec~west of the photometric
centre than in the centre itself. 
\citet{sch94} also report the highest density at 1\farcs8
west of the nucleus (their first data point west of the photometric centre).
The AGN may indeed reside slightly offset to the west of the [\ion{O}{iii}]
and H$\alpha$ peak as suggested by \citet{sim97}.

Due to the fairly high spectral resolution of the NTT data taken by Christian
Leipski ($\sim$1.5\,\AA~$\simeq$ 90\,km\,s$^{-1}$), we are able to study the emission-line
profiles in detail (Fig.~\ref{profile5643}). 
Interestingly, the velocity field is reflected in the
profiles of H$\beta$, [\ion{O}{iii}], and H$\alpha$:
At the location of the redshifted velocities to the east of the
nucleus, we see red wings in the profiles out to $\sim$8\arcsec. 
The strongest red contribution leading even to a secondary peak is observed
at 2\arcsec~east of the nucleus, i.e.~coinciding with the region
where the highest velocity occurs. From 1\arcsec~east of the nucleus out to
$\sim$7\arcsec~west of the nucleus, a blue wing is observed which reaches its
maximum contribution to the total flux at $\sim3-4$\arcsec~west of the nucleus,
i.e.~coinciding with the maximum blueshifted velocity. 
Along with the profile asymmetries, the observed profiles get also broader. 
\citet{whi82} already reported that the line profiles are identical
within observational uncertainties across the nuclear region and reveal profile
asymmetries such as blue wings.
The pronounced profile substructure is confined to the NLR,
i.e.~the central 22\arcsec~as determined from the diagnostic diagrams.
The observed profile substructure may also reflect the influence of the bar potential.
However, we cannot rule out that both the observed profile substructure and
the red- and blueshifted velocities originate from outflowing gas due to jet/NLR
interactions: Two radio lobes extend in the direction of the bar \citep{mor85}.

\end{document}